\documentclass[12pt,english]{article}
\usepackage{times}
\usepackage[T1]{fontenc}
\usepackage{geometry}
\usepackage{graphicx}
\usepackage{setspace}
\usepackage{amssymb}

\makeatletter

\providecommand{\tabularnewline}{\\}

\usepackage{bm}

\usepackage{babel}
\makeatother
\begin{document}

\section*{\noindent Multi-Scale Single-Bit \emph{RP-EMS} Synthesis for Advanced
Propagation Manipulation through System-by-Design}

\noindent \vfill

\noindent G. Oliveri,$^{(1)}$ \emph{Senior Member, IEEE}, P. Rocca,$^{(1)(2)}$
\emph{Senior Member, IEEE}, M. Salucci,$^{(1)}$ \emph{Senior Member,
IEEE}, D. Erricolo,$^{(4)}$ \emph{Fellow, IEEE}, and A. Massa,$^{(3)(1)(5)}$
\emph{Fellow, IEEE}

\noindent \vfill

\noindent ~

\noindent {\footnotesize $^{(1)}$} \emph{\footnotesize ELEDIA Research
Center} {\footnotesize (}\emph{\footnotesize ELEDIA}{\footnotesize @}\emph{\footnotesize UniTN}
{\footnotesize - University of Trento)}{\footnotesize \par}

\noindent {\footnotesize DICAM - Department of Civil, Environmental,
and Mechanical Engineering}{\footnotesize \par}

\noindent {\footnotesize Via Mesiano 77, 38123 Trento - Italy}{\footnotesize \par}

\noindent \textit{\emph{\footnotesize E-mail:}} {\footnotesize \{}\emph{\footnotesize giacomo.oliveri}{\footnotesize ,}
\emph{\footnotesize paolo.rocca}{\footnotesize ,} \emph{\footnotesize marco.salucci}{\footnotesize ,}
\emph{\footnotesize andrea.massa}{\footnotesize \}@}\emph{\footnotesize unitn.it}{\footnotesize \par}

\noindent {\footnotesize Website:} \emph{\footnotesize www.eledia.org/eledia-unitn}{\footnotesize \par}

\noindent {\footnotesize ~}{\footnotesize \par}

\noindent {\footnotesize $^{(2)}$} \emph{\footnotesize ELEDIA Research
Center} {\footnotesize (}\emph{\footnotesize ELEDIA@XIDIAN} {\footnotesize -
Xidian University)}{\footnotesize \par}

\noindent {\footnotesize P.O. Box 191, No.2 South Tabai Road, 710071
Xi'an, Shaanxi Province - China}{\footnotesize \par}

\noindent {\footnotesize E-mail:} \emph{\footnotesize paolo.rocca@xidian.edu.cn}{\footnotesize \par}

\noindent {\footnotesize Website:} \emph{\footnotesize www.eledia.org/eledia-xidian}{\footnotesize \par}

\noindent ~

\noindent {\footnotesize $^{(3)}$} \emph{\footnotesize ELEDIA Research
Center} {\footnotesize (}\emph{\footnotesize ELEDIA}{\footnotesize @}\emph{\footnotesize UESTC}
{\footnotesize - UESTC)}{\footnotesize \par}

\noindent {\footnotesize School of Electronic Engineering, Chengdu
611731 - China}{\footnotesize \par}

\noindent \textit{\emph{\footnotesize E-mail:}} \emph{\footnotesize andrea.massa@uestc.edu.cn}{\footnotesize \par}

\noindent {\footnotesize Website:} \emph{\footnotesize www.eledia.org/eledia}{\footnotesize -}\emph{\footnotesize uestc}{\footnotesize \par}

\noindent {\footnotesize ~}{\footnotesize \par}

\noindent {\footnotesize $^{(4)}$} \emph{\footnotesize Andrew Electromagnetics
Laboratory} {\footnotesize - University of Illinois Chicago}{\footnotesize \par}

\noindent {\footnotesize Department of Electrical and Computer Engineering,
851 South Morgan Street, Chicago, IL 60607-7053 - USA}{\footnotesize \par}

\noindent {\footnotesize E-mail:} \emph{\footnotesize derric1@uic.edu}{\footnotesize \par}

\noindent {\footnotesize Website:} \emph{\footnotesize https://andrew.lab.uic.edu/}{\footnotesize \par}

\noindent ~

\noindent {\footnotesize $^{(5)}$} \emph{\footnotesize ELEDIA Research
Center} {\footnotesize (}\emph{\footnotesize ELEDIA@TSINGHUA} {\footnotesize -
Tsinghua University)}{\footnotesize \par}

\noindent {\footnotesize 30 Shuangqing Rd, 100084 Haidian, Beijing
- China}{\footnotesize \par}

\noindent {\footnotesize E-mail:} \emph{\footnotesize andrea.massa@tsinghua.edu.cn}{\footnotesize \par}

\noindent {\footnotesize Website:} \emph{\footnotesize www.eledia.org/eledia-tsinghua}{\footnotesize \par}

\noindent \vfill

\noindent \textbf{\emph{This work has been submitted to the IEEE for
possible publication. Copyright may be transferred without notice,
after which this version may no longer be accessible. }}

\noindent \vfill

\newpage
\section*{Multi-Scale Single-Bit \emph{RP-EMS} Synthesis for Advanced Propagation
Manipulation through System-by-Design}

~

~

~

\begin{flushleft}G. Oliveri, P. Rocca, M. Salucci, D. Erricolo, and
A. Massa\end{flushleft}

\vfill

\begin{abstract}
\noindent A new method for synthesizing \emph{Single-Bit Reconfigurable
Passive Electromagnetic Skins} (\emph{1RP-EMS}s) featuring advanced
beam shaping capabilities is proposed. By using single-bit unit cells,
the multi-scale problem of controlling \emph{1RP-EMS}s is formulated
as a two-phase process. First, the macro-scale synthesis of the discrete
surface current that radiates the electromagnetic (\emph{EM}) field
fitting user-designed requirements is performed by means of an innovative
quantized version of the iterative projection method (\emph{QIPM}).
Successively, the meta-atom states of \emph{}the \emph{1RP-EMS} are
optimized with a customized implementation of the \emph{System-by-Design}
paradigm to yield a \emph{1RP-EMS} that supports such a feasible reference
current. A representative set of numerical results is reported to
assess the effectiveness of the proposed approach in designing and
controlling single-bit meta-atom \emph{RP-EMS}s that enable complex
wave manipulations.

\vfill
\end{abstract}
\noindent \textbf{Key words}: Reconfigurable \emph{EM} Skins; \emph{EM}
Holography; Reconfigurable Intelligent Surfaces; Next-Generation Communications;
Iterative Projection Method; System-by-Design; Metamaterials.

\newpage
\section{Introduction and Rationale\label{sec:1 - Introduction}}

\noindent \emph{Electromagnetic Skins} (\emph{EMS}s) are currently
the core of a theoretical, methodological, and practical revolution
within the academic and industrial communities working on wireless
communications \cite{Liaskos 2018}-\cite{Callaghan 2021}. Several
research studies on the foundation, the modeling, the simulation,
the design, and the test of \emph{EMS}s are currently under development
with a strong interdisciplinary effort combining chemistry, physics,
metamaterial science, electromagnetic (\emph{EM}) engineering, telecommunications,
and signal processing expertises \cite{Liaskos 2018}-\cite{Di Renzo 2020}\cite{Tang 2019}\cite{Di Renzo 2020b}.
As a matter of fact, starting from their early conceptualization as
thin metasurfaces able to manipulate the wave propagation beyond Snell's
laws \cite{Yang 2019}, \emph{EMS}s are considered as one of the key
enabling factors of the revolutionary \emph{Smart EM Environment}
(\emph{SEME}) paradigm in wireless communications \cite{Massa 2021}-\cite{Oliveri 2021d}\cite{Rocca 2021}\cite{Benoni 2021}.
Certainly, a multiplicity of methodological and practical challenges
\cite{Di Renzo 2019}-\cite{Massa 2021}\cite{Di Renzo 2020b}\cite{Callaghan 2021}\cite{Yang 2019}\cite{Hodge 2020}
still needs to be addressed to have a full transition from traditional
wireless systems to the \emph{SEME}-enhanced ones. In particular,
the \emph{complexity} associated to the design, the fabrication, the
implementation, the control, and the integration within a wireless
scenario of \emph{EMS}s is the main critical issue. More specifically,
complexity arises at (\emph{i}) the \emph{EMS} design level owing
to the multi-scale nature of its layout that features micro/nano-scale
descriptors combined with meso/macro-scale reflection and communication
properties, (\emph{ii}) the \emph{SEME} level due to the interactions
between the \emph{EMS}s and the large-scale propagation scenario,
and (\emph{iii}) the {}``propagation management'' level because
of the need to fruitfully integrate the \emph{EMS}s in a heterogenous
wireless infrastructure, which includes the base stations, the integrated
access and backhaul (\emph{IAB}) nodes, and the smart repeaters, as
well, to yield measurable performance improvements in the overall
wireless network.

\noindent Within such a framework, the design of planar artificial
materials with advanced propagation management capabilities has been
recently demonstrated for \emph{static passive EM skins} (\emph{SP-EMS})
by exploiting artificial intelligence (\emph{AI}) techniques within
the \emph{System-by-Design} (\emph{SbD}) paradigm \cite{Oliveri 2021c}\cite{Oliveri 2021d}\cite{Massa 2021b}.
Such an approach leverages on the decomposition of the problem at
hand into a source design phase and a subsequent optimization of the
surface descriptors of the \emph{SP-EMS} within the \emph{Generalized
Sheet Transition Condition} (\emph{GSTC}) framework \cite{Oliveri 2021c}\cite{Oliveri 2021d}\cite{Yang 2019}.
Thanks to the modularity of such a synthesis tool and its multi-scale-oriented
nature, the efficient design of wide-aperture \emph{EMS}s that enable
advanced pattern shaping properties has been carried out \cite{Oliveri 2021c}\cite{Oliveri 2021d}
despite the use of extremely simple unit cells.

\noindent Otherwise, \emph{reconfigurable passive EMS}s (\emph{RP-EMS}s)
have been proposed and widely studied to dynamically control the propagation
environment for adaptively improving the communication performance
\cite{Liaskos 2018}-\cite{Di Renzo 2020}\cite{Tang 2019}\cite{Gong 2020}.
Towards this end, \emph{RP-EMS} unit cells needs either analog (e.g.,
varactors/varistor \cite{Liang 2022}\cite{Hodge 2020}\cite{Pitilakis 2021}
and mechanically-tuned sub-parts \cite{Callaghan 2021}) or digitally-controlled
(e.g., p-i-n diodes \cite{Dai 2020}) components. From an applicative
viewpoint, the implementation of a continuous control on each \emph{RP-EMS}
cell can yield to very expensive and complex architectures, thus it
is generally avoided \cite{Kashyap 2020} and the \emph{RP-EMS} analog
states are often discretized using few bits, $B$, per cell \cite{Liang 2022}\cite{Kashyap 2020}
or they are implemented by using binary switches \cite{Dai 2020}.
Therefore, \emph{RP-EMS}s are usually digitally-controlled systems
\cite{Dai 2020}-\cite{Ross 2022} with relatively limited per-cell
degrees-of-freedom (\emph{DoF}s) when compared to \emph{SP-EMS}s \cite{Oliveri 2021c}\cite{Oliveri 2021d}.
A key consequence of such a per-cell constraint, mainly when low-bit
($B\to1$) \emph{RP-EMS} are at hand \cite{Kashyap 2020}, turns out
to be the very limited control of the shape of the reflected beam
\cite{Kashyap 2020}. Thus, the mainstream state-of-the-art literature
on \emph{RP-EMS}s has been concerned with the synthesis of \emph{RP-EMS}s
with {}``simple'' anomalous reflection capabilities and narrow beam
focusing (i.e., pencil beam-like) \cite{Liang 2022}\cite{Dai 2020}-\cite{Ross 2022}.
However, demonstrating more advanced footprint control/shaping with
a digital \emph{RP-EMS} would be of great interest in practice since
it would allow one to efficiently concentrate the reflected power
in arbitrary desired areas (i.e., roads, squares, streets, buildings)
and not just in spots. Unfortunately, the approach derived in \cite{Oliveri 2021c}\cite{Oliveri 2021d}
to design \emph{SP-EMS}s affording shaped footprint patterns cannot
be directly applied to \emph{RP-EMS}s \cite{Oliveri 2021c}\cite{Oliveri 2021d}.
Indeed, the synthesis of the reference surface current, which is performed
in the first step of \cite{Oliveri 2021c}\cite{Oliveri 2021d} and
that exploits the non-uniqueness of the associated inverse source
(\emph{IS}) problem to take advantage of the \emph{non-radiating currents}
(\emph{NRC}s) \cite{Oliveri 2021c}\cite{Salucci 2018}, assumes that
the unit cell of the corresponding \emph{EMS} allows a fine tuning
of the reflection phase \cite{Oliveri 2021c}\cite{Oliveri 2021d}.
By definition, this is actually prevented when dealing with digital
\emph{RP-EMS}s \cite{Kashyap 2020} making the design process ineffective
and potentially unable to fulfil complex coverage requirements\@.

\noindent Dealing with \emph{RP-EMS}s, the objective of this work
is twofold. On the one hand, it is aimed at presenting and validating
an innovative method for the synthesis (i.e., the design and the control)
of high-performance holographic \emph{1RP-EMS}s. On the other hand,
it is devoted to prove that minimum complexity \emph{RP-EMS}s can
be used in \emph{SEME} scenarios to yield complex wave propagation
phenomena despite the coarse tuning of the reflection phase. 

\noindent Starting from the design of a meta-atom of the \emph{RP-EMS}
that features only a single-bit reconfiguration and by generalizing
the theoretical concepts on complex large-scale \emph{EM} wave manipulation
systems \cite{Oliveri 2021c}\cite{Oliveri 2021d}\cite{Oliveri 2015}-\cite{Oliveri 2021b},
the first step of the proposed method for the synthesis of \emph{1RP-EMS}s
deals with the computation of a discrete-phase current that radiates
a field distribution fitting complex footprint patterns. A digital
\emph{SbD}-based \emph{RP-EMS} optimization is then carried out to
set the \emph{1RP-EMS} configuration that supports such a reference
discrete-phase current. Towards this end, suitable \emph{AI} paradigms
for building reliable and computationally-efficient {}``\emph{RP-EMS}
digital twins'' \cite{Oliveri 2021c}\cite{Oliveri 2021d}\cite{Oliveri 2015}-\cite{Oliveri 2021b}
are exploited to properly address the issues related to the multi-scale
complexity of the problem at hand.

\noindent The outline of the paper is as follows. First, the \emph{1RP-EMS}
synthesis problem is formulated (Sect. \ref{sec: 2 - Problem-Formulation}),
then Sect. \ref{sec:3 - Method} details the proposed two-step (i.e.,
design and control) synthesis method. Representative results from
a wide set of numerical experiments are reported for assessment purposes,
while comparisons with state-of-the-art techniques \cite{Oliveri 2021c}\cite{Oliveri 2021d}
are considered (Sect. \ref{sec:4 - Numerical-Analysis-and}). Finally,
some concluding remarks follow (Sect. \ref{sec:5 - Conclusions-and-Remarks}).

\section{\noindent Mathematical Formulation \label{sec: 2 - Problem-Formulation}}

\noindent Let a single-bit \emph{RP-EMS} (\emph{1RP-EMS}) be centered
in the origin of the local coordinate system $\left(x,y,z\right)$
(Fig. 1). The \emph{1RP-EMS} consists of $M\times N$ reconfigurable
binary meta-atoms displaced on a regular grid of cells with sides
$\Delta x$ and $\Delta y$ on a planar region $\Psi_{EMS}$ ($\Psi_{EMS}$
$=$\{$-M\times\frac{\Delta x}{2}\le x\le M\times\frac{\Delta x}{2}$;
$-N\times\frac{\Delta y}{2}\le y\le N\times\frac{\Delta y}{2}$\}).
Each ($m$, $n$)-th ($m=1,...,M$; $n=1,...,N$) meta-atom is defined
by a set of $U$ geometrical/material descriptors $\mathbf{g}\triangleq\left\{ g^{\left(u\right)};\, u=1,...,U\right\} $
and it features, at the $t$-th ($t=1,...,T$) time step, a binary
state $s_{mn}\left(t\right)\in\left\{ 0,1\right\} $. 

\noindent The \emph{1RP-EMS} at the $t$-th ($t=1,...,T$) instant
can be \emph{}univocally identified by the binary \emph{micro-scale}
state \emph{}vector \emph{}$\mathcal{S}\left(t\right)$, $\mathcal{S}\left(t\right)$
$\triangleq$ \{$s_{mn}\left(t\right)$; $m=1,...,M$; $n=1,...,N$\},
and the time-independent (i.e., it is unrealistic to change the atom
layout at each time step) \emph{micro-scale} descriptor vector $\mathbf{g}$.
Otherwise, the \emph{RP-EMS} can be described from an electromagnetic
viewpoint by the \emph{micro-scale} electric/magnetic surface susceptibility
vector $\mathcal{K}\left(t\right)$ ($t=1,...,T$) \emph{}\cite{Oliveri 2021c}\cite{Yang 2019},
whose ($m$, $n$)-th entry ($m=1,...,M$; $n=1,...,N$) is the diagonal
tensor of the electric/magnetic local surface susceptibility of the
($m$, $n$)-th meta-atom, $\overline{\overline{K}}_{mn}\left(t\right)$
{[}$\overline{\overline{K}}_{mn}\left(t\right)=\mathbb{K}\left\{ \mathbf{g};\, s_{mn}\left(t\right)\right\} $
being $\overline{\overline{K}}_{mn}\left(t\right)\triangleq\sum_{q=x,y,z}k_{qq}\left(\mathbf{g};\, s_{mn}\left(t\right)\right)\widehat{q}\widehat{q}${]}.

\noindent According to the \emph{Generalized Sheet Transition Condition}
(\emph{GSTC}) technique \cite{Yang 2019}\cite{Ricoy 1990}\cite{Achouri 2015},
the instantaneous far field pattern, $\overline{E}\left(r,\theta,\varphi;t\right)$,
reflected by the \emph{RP-EMS} when illuminated by a time-harmonic
plane wave at frequency $f$ impinging from the incidence direction
$\left(\theta^{inc},\varphi^{inc}\right)$ and characterized by {}``perpendicular''
and {}``parallel'' complex-valued electric field components $E_{\bot}^{inc}$
and $E_{\parallel}^{inc}$ is a function of the surface susceptibility
vector \emph{$\mathcal{K}$} through the \emph{macro-scale} induced
surface current $\overline{J}$ (i.e., $\overline{E}\left(r,\theta,\varphi;t\right)=\mathbb{F}\left\{ \overline{J}\left(x,y;t\right)\right\} $).
More in detail, it turns out to that \cite{Oliveri 2021c}\cite{Oliveri 2021d}\cite{Yang 2019}\cite{Osipov 2017}\begin{equation}
\begin{array}{r}
\overline{E}\left(r,\theta,\varphi;t\right)=\frac{jk_{0}}{4\pi}\frac{\exp\left(-jk_{0}r\right)}{r}\int_{-\frac{M\Delta x}{2}}^{\frac{M\Delta x}{2}}\int_{-\frac{N\Delta y}{2}}^{\frac{N\Delta y}{2}}\overline{J}\left(x',y';t\right)\times\\
\exp\left[jk_{0}\left(rx'\sin\theta\cos\varphi+ry'\sin\theta\sin\varphi\right)\right]\mathrm{d}x'\mathrm{d}y'\end{array}\label{eq:far field ems}\end{equation}
where the surface current $\overline{J}$ is given by

\noindent \begin{equation}
\overline{J}\left(x,y;t\right)=\widehat{\mathbf{r}}\times\left[\eta_{0}\widehat{\mathbf{r}}\times\overline{J}^{e}\left(x,y;t\right)+\overline{J}^{h}\left(x,y;t\right)\right]\,\,\,\left(x,y\right)\in\Psi_{EMS}\label{eq:decomposition}\end{equation}
where $\overline{J}^{o}$, $o\in\left\{ e,\, h\right\} $, is the
electric/magnetic component of the current induced on the \emph{RP-EMS},
while $k_{0}=2\pi f\sqrt{\varepsilon_{0}\mu_{0}}$ and $\eta_{0}=\sqrt{\frac{\mu_{0}}{\varepsilon_{0}}}$
are the free-space wavenumber and the impedance, respectively, which
depend on the free-space permeability (permittivity) $\mu_{0}$ ($\varepsilon_{0}$).

\noindent Subject to the local periodicity condition, the dependence
of $\overline{J}^{o}$, $o\in\left\{ e,\, h\right\} $, on the entries
of the \emph{micro-scale} electric/magnetic surface susceptibility
vector \emph{$\mathcal{K}$} (i.e., $\overline{J}^{o}\left(x,y;t\right)=\mathbb{G}\left\{ \mathbb{K}^{o}\left\{ \mathbf{g};\, s_{mn}\left(t\right)\right\} ;\overline{E}^{inc}\left(x,y,0;t\right)\right\} $,
$o\in\left\{ e,\, h\right\} $) is made explicit \emph{}in the following
form \cite{Oliveri 2021c}\cite{Oliveri 2021d}\cite{Yang 2019}\cite{Achouri 2015}\begin{equation}
\begin{array}{c}
\overline{J}^{e}\left(x,y;t\right)=\sum_{m=1}^{M}\sum_{n=1}^{N}\left\{ j2\pi f\varepsilon_{0}\left[\overline{\overline{K}}_{mn}^{e}\left(t\right)\cdot\overline{E}_{mn}\left(t\right)\right]_{\tau}-\widehat{\bm{\nu}}\times\nabla_{\tau}\left[\overline{\overline{K}}_{mn}^{h}\left(t\right)\cdot\overline{H}_{mn}\left(t\right)\right]_{\nu}\right\} \Omega_{mn}\left(x,y\right)\\
\overline{J}^{m}\left(x,y;t\right)=\sum_{m=1}^{M}\sum_{n=1}^{N}\left\{ j2\pi f\mu_{0}\left[\overline{\overline{K}}_{mn}^{h}\left(t\right)\cdot\overline{H}_{mn}\left(t\right)\right]_{\tau}+\widehat{\bm{\nu}}\times\nabla_{\tau}\left[\overline{\overline{K}}_{mn}^{e}\left(t\right)\cdot\overline{E}_{mn}\left(t\right)\right]_{\nu}\right\} \Omega_{mn}\left(x,y\right)\end{array}\label{eq:surface currents}\end{equation}
 where $\widehat{\bm{\nu}}$ is the outward normal to $\Psi_{EMS}$,
$\left[\,.\,\right]_{\tau/\nu}$ stands for the tangential/normal
component, and $\Omega_{mn}\left(x,y\right)$ $\triangleq$ \{$1$
if {[}$-\left(m-M-1\right)\times\frac{\Delta x}{2}$ $\le$ $x$ $\le$
$\left(m-M\right)\times\frac{\Delta x}{2}${]} and {[}$-\left(n-N-1\right)\times-\frac{\Delta y}{2}$
$\le$ $y$ $\le$ $\left(n-N\right)\times\frac{\Delta y}{2}${]};
$0$ otherwise\} is the basis function related to the ($m$, $n$)-th
($m=1,...,M$; $n=1,...,N$) cell with support $\Delta\Psi_{EMS}$
($\Delta\Psi_{EMS}\triangleq\Delta x\times\Delta y$), while $\overline{E}_{mn}$
($\overline{H}_{mn}$) is the surface averaged electric (magnetic)
field (see \emph{Appendix}).

\noindent Such a derivation points out that the $t$-th ($t=1,...,T$)
far-field pattern $\overline{E}\left(r,\theta,\varphi;t\right)$ can
be controlled by properly adjusting the $M\times N$ binary entries
of $\mathcal{S}\left(t\right)$, once the \emph{1RP-EMS} is designed
(i.e., $\mathbf{g}$ is set - Sect. \ref{sec:2.1 - RP-EMS Design}).
Accordingly, the problem at hand can be mathematically formulated
as follows

\begin{quotation}
\noindent \textbf{\emph{1RP-EMS Synthesis Problem}} - Find the optimal
setting of the micro-scale descriptor vector, $\mathbf{g}^{opt}$,
and the optimal configuration of $T$ binary \emph{micro-scale} state
\emph{}vectors, \emph{}\{$\mathcal{S}^{opt}\left(t\right)$; $t=1,...,T\}$,
such that\begin{equation}
\Phi\left(\mathbf{g},\mathcal{S}\left(t\right)\right)=\int_{\Psi_{obs}}\Re\left\{ F^{des}\left(\widetilde{x},\widetilde{y},\widetilde{z};t\right)-F\left(\widetilde{x},\widetilde{y},\widetilde{z};t\right)\right\} \mathrm{d}\widetilde{x}\mathrm{d}\widetilde{y}\mathrm{d}\widetilde{z}\label{eq:cost function globale}\end{equation}
is minimized at each $t$-th ($t=1,...,T$) time instant {[}i.e.,
($\mathbf{g}^{opt}$, $\mathcal{S}^{opt}\left(t\right)$) $=$ $\arg$
($\min_{\mathbf{g},\mathcal{S}\left(t\right)}${[}$\Phi\left(\mathbf{g},\mathcal{S}\left(t\right)\right)${]}),
$t=1,...,T${]}.
\end{quotation}
\noindent In (\ref{eq:cost function globale}), $\Re\left\{ \,.\,\right\} $
is the {}``ramp'' function and $F^{des}\left(\widetilde{x},\widetilde{y},\widetilde{z};t\right)$
is the user-defined power pattern footprint at the $t$-th ($t=1,...,T$)
time instant in the observation region $\Psi_{obs}$, $\left(\widetilde{x},\widetilde{y},\widetilde{z}\right)$
being the \emph{RP-EMS} global coordinate system (Fig. 1). Moreover,
the footprint pattern is a function of the reflected far-field pattern
$\overline{E}\left(r,\theta,\varphi;t\right)$, (i.e., $F\left(\widetilde{x},\widetilde{y},\widetilde{z};t\right)=\mathbb{H}\left\{ \overline{E}\left(r,\theta,\varphi;t\right)\right\} $)
and it is given by

\noindent \textbf{\begin{equation}
F\left(\widetilde{x},\widetilde{y},\widetilde{z};t\right)=\left|\overline{E}\left(\sqrt{\widetilde{x}^{2}+\widetilde{y}^{2}+\left(\widetilde{z}-d\right)^{2}},\arctan\frac{\sqrt{\widetilde{y}^{2}+\left(\widetilde{z}-d\right)^{2}}}{\widetilde{x}},\arctan\left(\frac{\widetilde{z}-d}{\widetilde{x}}\right);t\right)\right|^{2}\label{eq:footprint}\end{equation}
}where $d$ is the \emph{1RP-EMS} height over the ground plane (Fig.
1).

\noindent It is worth noticing that, unlike the case of \emph{SP-EMSs},
the synthesis of a \emph{1RP-EMS} cannot be done by minimizing (\ref{eq:cost function globale})
only once since there is a different optimal configuration $\mathcal{S}^{opt}\left(t\right)$
for each $t$-th ($t=1,...,T$) user-defined footprint pattern, $F^{des}\left(\widetilde{x},\widetilde{y},\widetilde{z};t\right)$,
as pointed out in the following expression

\noindent \begin{equation}
\Phi\left(\mathbf{g},\mathcal{S}\left(t\right)\right)=\int_{\Psi_{obs}}\Re\left\{ F^{des}\left(\widetilde{x},\widetilde{y},\widetilde{z};t\right)-\mathbb{H}\left\{ \mathbb{F}\left\{ \mathbb{G}\left\{ \mathbb{K}\left\{ \mathbf{g};\, s_{mn}\left(t\right)\right\} ;\overline{E}^{inc}\left(x,y,0;t\right)\right\} \right\} \right\} \right\} \mathrm{d}\widetilde{x}\mathrm{d}\widetilde{y}\mathrm{d}\widetilde{z}\label{eq:cost esplicita}\end{equation}
where the link between $F^{des}\left(\widetilde{x},\widetilde{y},\widetilde{z};t\right)$
and $\mathcal{S}\left(t\right)$ $\triangleq$ \{$s_{mn}\left(t\right)$;
$m=1,...,M$; $n=1,...,N$\} is made evident. On the other hand, the
$U$ geometrical/material entries of $\mathbf{g}^{opt}$ must be set
once as the optimal trade-off among all $T$ propagation scenarios.

\noindent Furthermore, the problem at hand is more complex than that
of a multi-bit \emph{RP-EMS} and (even) much more than of a \emph{SP-EMS}.
Unlike the \emph{SP} case, the $t$-th ($t=1,...,T$) \emph{micro-scale}
electric/magnetic surface susceptibility vector $\mathcal{K}\left(t\right)$
assumes here only a quantized set of states (i.e., $2^{M\times N}$)
instead of a continuity of values \cite{Oliveri 2021c}\cite{Oliveri 2021d}.
Thus, the macro-scale (reflection) properties of the arising \emph{EMS}
turn out to be more severely constrained than those of a \emph{SP-EMS}
or a multi-bit \emph{RP-EMS}. Consequently, the fulfilment of complex
shaping requirements on the footprint power pattern, as those in \cite{Oliveri 2021c}\cite{Oliveri 2021d},
is certainly more difficult and it may results even physically unfeasible.

\noindent Taking into account these considerations, the {}``\emph{1RP-EMS
Synthesis Problem}'' (\ref{eq:cost function globale}) is then addressed
with a two-step approach where, first, the {}``\emph{1RP-EMS Design
Problem}'' (Sect. \ref{sec:2.1 - RP-EMS Design}) is solved by identifying
the $U$ geometrical/material descriptors of the single-bit meta-atom
(i.e., $\mathbf{g}\leftarrow\mathbf{g}^{opt}$), while the second
step is aimed at setting, at each $t$-th ($t=1,...,T$) time-instant,
the entries of the \emph{micro-scale} state vector \emph{}$\mathcal{S}\left(t\right)$
to fulfil the footprint pattern requirements {[}i.e., $\mathcal{S}\left(t\right)$
$\leftarrow$ $\mathcal{S}^{opt}\left(t\right)${]} ({}``\emph{1RP-EMS
Control Problem}'' - Sect. \ref{sec:2.2 - RP-EMS Control}).

\subsection{\noindent \emph{1RP-EMS} Design Problem\label{sec:2.1 - RP-EMS Design}}

\noindent As for the \emph{1RP-EMS} unit cell design, a key challenge
and preparatory step to enable the footprint pattern control (i.e.,
$F\to F^{des}$) is the choice of a meta-atom structure whose reflection
properties can be suitably modified when its logical state is changed
\cite{Yang 2019}. In principle, an optimal trade-off should be found
by minimizing (\ref{eq:cost esplicita}) with respect to $\mathbf{g}$
across all $T$ user-requirements \{$F^{des}\left(\widetilde{x},\widetilde{y},\widetilde{z};t\right)$;
$t=1,...,T$\}, while, in this paper, a {}``worst case''-strategy
is adopted to yield a more general and flexible implementation. The
design is then carried out by requiring that the \emph{1RP-EMS} meta-atom
supports the widest possible reflection variation to account not only
the $T$ propagation scenarios at hand, but more in general the largest
range of admissible conditions. According to (\ref{eq:surface currents}),
such a guideline corresponds to the maximization of the gap between
the values of the electric/magnetic local surface susceptibility when
switching the status of the generic ($m$, $n$)-th ($m=1,...,M$;
$n=1,...,N$) meta-atom from $s_{mn}\left(t\right)=0$ to $s_{mn}\left(t\right)=1$.
Mathematically, this means to minimize the following cost function\begin{equation}
\begin{array}{l}
\phi\left(\mathbf{g}\right)=\frac{1}{\pi^{2}}\left[\left(\left|\angle\left.\Gamma_{mn}^{\bot\bot}\left(t\right)\right|_{f=f_{0}}^{s_{mn}\left(t\right)=1}-\angle\left.\Gamma_{mn}^{\bot\bot}\left(t\right)\right|_{f=f_{0}}^{s_{mn}\left(t\right)=0}\right|-\pi\right)^{2}+\right.\\
\left.+\left(\left|\left.\Gamma_{mn}^{\parallel\parallel}\left(t\right)\right|_{f=f_{0}}^{s_{mn}\left(t\right)=1}-\angle\left.\Gamma_{mn}^{\parallel\parallel}\left(t\right)\right|_{f=f_{0}}^{s_{mn}\left(t\right)=0}\right|-\pi\right)^{2}\right]\end{array}\label{eq: ELEDIA rising}\end{equation}
to yield the optimal set of the geometrical/material descriptors of
the single-bit meta-atom, $\mathbf{g}^{opt}$ {[}i.e., $\mathbf{g}^{opt}$
$=$ $\arg\left(\min_{\mathbf{g}}\left[\phi\left(\mathbf{g}\right)\right]\right)${]}.
In (\ref{eq: ELEDIA rising}), $f_{0}$ is the central working frequency,
$\angle\cdot$ stands for the phase of the complex argument, and $\Gamma_{mn}^{\bot\bot}\left(t\right)/\Gamma_{mn}^{\parallel\parallel}\left(t\right)$
{[}$\Gamma_{mn}^{\bot\bot}\left(t\right)=\mathbb{Y}^{\bot\bot}\left\{ \mathbf{g};\, s_{mn}\left(t\right)\right\} $
and $\Gamma_{mn}^{\parallel\parallel}\left(t\right)=\mathbb{Y}^{\parallel\parallel}\left\{ \mathbf{g};\, s_{mn}\left(t\right)\right\} ${]}
are the \emph{TE/TM} co-polar components of the reflection tensor
in the ($m$, $n$)-th ($m=1,...,M$; $n=1,...,N$) cell, $\overline{\overline{\Gamma}}_{mn}\left(t\right)$,
while the logical status of the ($m$, $n$)-th cell (i.e., $s_{mn}\left(t\right)\in\left\{ 0,1\right\} $)
is physically implemented by biasing the diodes in the meta-atom layout
(Fig. 2).

\subsection{\noindent \emph{1RP-EMS} Control Problem\label{sec:2.2 - RP-EMS Control}}

\noindent Once the \emph{1RP-EMS} has been designed by setting $\mathbf{g}^{opt}$,
the computation of $\mathcal{S}^{opt}\left(t\right)$ should be performed
by minimizing the constrained ($\mathbf{g}\equiv\mathbf{g}^{opt}$)
version of (\ref{eq:cost function globale})\begin{equation}
\Phi\left(\mathbf{g}^{opt},\mathcal{S}\left(t\right)\right)=\int_{\Psi_{obs}}\Re\left\{ F^{des}\left(\widetilde{x},\widetilde{y},\widetilde{z};t\right)-\mathbb{H}\left\{ \mathbb{F}\left\{ \mathbb{G}\left\{ \mathbb{K}\left\{ \mathbf{g}^{opt};\, s_{mn}\left(t\right)\right\} ;\overline{E}^{inc}\left(x,y,0;t\right)\right\} \right\} \right\} \right\} \mathrm{d}\widetilde{x}\mathrm{d}\widetilde{y}\mathrm{d}\widetilde{z}\label{Lui}\end{equation}
{[}i.e., $\mathcal{S}^{opt}\left(t\right)$ $=$ $\arg\left(\min_{\mathcal{S}}\left[\Phi\left(\mathbf{g}^{opt},\mathcal{S}\left(t\right)\right)\right]\right)$,
which directly relates the state vector $\mathcal{S}\left(t\right)$
with the footprint target $F^{des}\left(\widetilde{x},\widetilde{y},\widetilde{z};t\right)$.
However, when dealing with \emph{aperiodic} wave manipulation devices
\cite{Oliveri 2021c}\cite{Encinar 2004}-\cite{Pozar 1999}, such
a single-phase solution approach is usually avoided in favour of splitting
the problem at hand into two parts. The former phase ({}``\emph{Reference
Current Computation}'') addresses a \emph{macro-scale} objective
that consists in the computation of an ideal equivalent surface current
$\overline{J}^{opt}\left(x,y;t\right)$ that affords the desired footprint
pattern $F^{des}\left(\widetilde{x},\widetilde{y},\widetilde{z};t\right)$,
which is coded into the following macro-scale cost function\begin{equation}
\Phi\left(\overline{J}\left(x,y;t\right)\right)=\int_{\Psi_{obs}}\Re\left\{ F^{des}\left(\widetilde{x},\widetilde{y},\widetilde{z};t\right)-\mathbb{H}\left\{ \mathbb{F}\left\{ \overline{J}\left(x,y;t\right)\right\} \right\} \right\} \mathrm{d}\widetilde{x}\mathrm{d}\widetilde{y}\mathrm{d}\widetilde{z},\label{Macro cost-funct}\end{equation}
to be minimized\begin{equation}
\overline{J}^{opt}\left(x,y;t\right)=\arg\left(\min_{\overline{J}\left(x,y\right)}\left[\Phi\left(\overline{J}\left(x,y;t\right)\right)\right]\right).\label{Gi}\end{equation}
The second (\emph{microscale}) phase ({}``\emph{1RP-EMS Configuration}'')
\cite{Oliveri 2021c}\cite{Encinar 2004}-\cite{Pozar 1999} is devoted
to choose the meta-atoms configuration $\mathcal{S}^{opt}\left(t\right)$
that supports the reference current $\overline{J}^{opt}\left(x,y;t\right)$
by solving the following optimization problem\begin{equation}
\mathcal{S}^{opt}\left(t\right)=\arg\left(\min_{\mathcal{S}}\left[\psi\left(\mathcal{S}\left(t\right)\right)\right]\right),\label{sssvi}\end{equation}
where\begin{equation}
\psi\left(\mathcal{S}\left(t\right)\right)\triangleq\frac{\left\Vert \overline{J}^{opt}\left(x,y;t\right)-\mathbb{G}\left\{ \mathbb{K}\left\{ \mathbf{g};\, s_{mn}\left(t\right)\right\} ;\overline{E}^{inc}\left(x,y,0;t\right)\right\} \right\Vert }{\left\Vert \overline{J}^{opt}\left(x,y;t\right)\right\Vert }.\label{micro cost-funct}\end{equation}
This two-phase process exploits the fast Fourier relation between
currents and patterns (\ref{eq:far field ems}), which results in
very efficient implementations for large apertures \cite{Oliveri 2021c}\cite{Encinar 2004}-\cite{Pozar 1999},
as well. Moreover, the arising currents can be re-used to design \emph{EMS}
arrangements with different unit cells \cite{Pozar 1999}. Furthermore,
the micro-scale synthesis step does not involve here the optimization
of $\overline{\overline{K}}_{mn}$ to achieve ideal susceptibility
distributions (which may yield, even in the \emph{SP-EMS} case \cite{Oliveri 2021c}\cite{Oliveri 2021d},
to non-feasible anisotropy requirements on the cell), but it is aimed
at setting the ($m$, $n$)-th ($m=1,...,M$; $n=1,...,N$) atom state
$s_{mn}\left(t\right)$ that locally minimizes the mismatch with the
target surface current.

\noindent On the other hand, it has to be noticed that in principle
the problem at hand, whatever the solution approach (direct or two-phases),
requires the phase of the wave reflected by the meta-atoms to vary
over continuous intervals \cite{Oliveri 2021c}\cite{Encinar 2004}-\cite{Pozar 1999}.
This is clearly not true when dealing with $B$-bits \emph{RP-EMS}s,
since each meta-atom can only assume $2^{B}$ states for each $t$-th
($t=1,...,T$) time instant. Such a limitation is even more critical
for \emph{1RP-EMSs} ($B=1$). Moreover, despite the two-phase decomposition,
the multi-scale and quantized nature of the \emph{1RP-EMS Control
Problem} still yields to a solution space with a size (i.e., $2^{M\times N}$)
that grows exponentially with the \emph{RP-EMS} aperture. 

\noindent To take into account these pros \& cons, a dedicated strategy
needs to be implemented (Sect. \ref{sec:3 - Method}).

\section{\noindent Solution Method \label{sec:3 - Method}}

\noindent While the {}``\emph{1RP-EMS} \emph{Design}'' problem (Sect.
\ref{sec:2.1 - RP-EMS Design}) is a quite standard real-variable
optimization problem to be addressed with a standard optimization
tool, the \emph{{}``1RP-EMS} \emph{Control}'' one (Sect. \ref{sec:2.2 - RP-EMS Control})
turns out to be a new challenge. As a matter of fact, the most intuitive
strategy for solving this latter would be that of exploiting the methodology
discussed in \cite{Oliveri 2021c} by simply replacing the model of
the local susceptibility dyadics of the \emph{SP-EMS} with that of
the reconfigurable single-bit meta-atom at hand. However, such an
approach has a fundamental drawback when applied to the \emph{1RP-EMS}
control. By ignoring the quantized nature of the \emph{1RP-EMS} surface
currents in the {}``\emph{Reference Current Computation}'' (\ref{Gi}),
there may not to be an implementable current distribution, $\overline{J}$
($\overline{J}\left(x,y;t\right)=\mathbb{G}\left\{ \mathbb{K}\left\{ \mathbf{g};\, s_{mn}\left(t\right)\right\} ;\overline{E}^{inc}\left(x,y,0;t\right)\right\} $),
that approximates the synthesized reference current $\overline{J}^{opt}$,
regardless of the approach to configure the \emph{1RP-EMS} (\ref{sssvi}).
Therefore, an innovative method is proposed (Sect. \ref{sub:3.1 - QIPM-Based-Reference-Current})
to compute a {}``feasible'' ideal equivalent surface current $\overline{J}^{opt}$
that affords the desired footprint pattern $F^{des}$ , while the
approach used in \cite{Oliveri 2021c} for the design of an \emph{SP-EMS}
is customized here to control the \emph{1RP-EMS} (\ref{sub:3.2 - 1RP-EMS-Configuration-Method}).

\subsection{\noindent \emph{QIPM}-Based Reference Current Computation \label{sub:3.1 - QIPM-Based-Reference-Current}}

\noindent In order to define a {}``feasible'' reference current,
a quantized version of the iterative projection method (\emph{QIPM})
is derived.

\noindent Let $\mathcal{C}$ be the {}``\emph{1RP-EMS Current Space}{}``
composed by the whole set of the \emph{1RP-EMS} admissible surface
currents having the following mathematical form\begin{equation}
\overline{J}\left(x,y;t\right)=\sum_{m=1}^{M}\sum_{n=1}^{N}\alpha_{mn}\left(t\right)\exp\left[j\chi_{mn}\left(t\right)\right]\Omega_{mn}\left(x,y\right)\widehat{\iota}\label{eq:feasibility currents}\end{equation}
where $\widehat{\iota}$ denotes the current polarization, while $\alpha_{mn}\left(t\right)$
{[}$\alpha_{mn}\left(t\right)=\mathbb{A}\left\{ s_{mn}\left(t\right)\right\} ${]}
and $\chi_{mn}\left(t\right)$ {[}$\chi_{mn}\left(t\right)=\mathbb{X}\left\{ s_{mn}\left(t\right)\right\} ${]}
are the values of the locally-controlled magnitude and phase of the
surface current that belong to the discrete (two-elements) alphabets
$\mathcal{A}$ and $\mathcal{X}$, respectively. The elements of $\mathcal{A}$
and $\mathcal{X}$ are the magnitude and the phase of the current
that each meta-atom can support when configured in one of its binary
states, $s_{mn}\left(t\right)\in\left\{ 0,1\right\} $, ($\mathcal{A}$
$\triangleq$\{$\mathbb{A}\left\{ s_{mn}\left(t\right)=0\right\} $
, $\mathbb{A}\left\{ s_{mn}\left(t\right)=1\right\} $\} and $\mathcal{X}$
$\triangleq$ \{$\mathbb{X}\left\{ s_{mn}\left(t\right)=0\right\} $,
$\mathbb{X}\left\{ s_{mn}\left(t\right)=1\right\} $\}).

\noindent Starting from a random initialization of the discrete coefficients
$\alpha_{mn}^{\left(p\right)}\left(t\right)$ and $\chi_{mn}^{\left(p\right)}\left(t\right)$
($m=1,...,M$; $n=1,...,N$), whose values are randomly drawn from
$\mathcal{A}$ and $\mathcal{X}$, the \emph{QIPM} generates a succession
of $P$ trial current distributions, \{$\overline{J}^{\left(p\right)};$
$p=1,...,p$\}. First, the footprint pattern $F^{\left(p\right)}\left(\widetilde{x},\widetilde{y},\widetilde{z};t\right)$
afforded by $\overline{J}^{\left(p\right)}$ is computed (\ref{eq:far field ems})(\ref{eq:footprint}).
It is then projected into the corresponding feasibility space through
the projection operator $R^{\left(p\right)}$ {[}$R^{\left(p\right)}\left(\widetilde{x},\widetilde{y},\widetilde{z};t\right)$
$=$ $\mathbb{R}$ \{$F^{\left(p\right)}\left(\widetilde{x},\widetilde{y},\widetilde{z};t\right)$,
$F^{des}\left(\widetilde{x},\widetilde{y},\widetilde{z};t\right)$\}{]}\begin{equation}
R^{\left(p\right)}\left(\widetilde{x},\widetilde{y},\widetilde{z};t\right)=\left\{ \begin{array}{ll}
F^{des}\left(\widetilde{x},\widetilde{y},\widetilde{z};t\right) & if\, F^{\left(p\right)}\left(\widetilde{x},\widetilde{y},\widetilde{z};t\right)<F^{des}\left(\widetilde{x},\widetilde{y},\widetilde{z};t\right)\\
F^{\left(p\right)}\left(\widetilde{x},\widetilde{y},\widetilde{z};t\right) & \mathrm{otherwise}.\end{array}\right.\label{eq:pattern projection}\end{equation}
The \emph{QIPM} convergence is checked and the iterations are stopped
if either $p=P$ or if the index $\Xi^{\left(p\right)}\left(t\right)$
($\Xi^{\left(p\right)}\left(t\right)\triangleq\frac{\int_{\Psi_{obs}}\left|R^{\left(p\right)}\left(\widetilde{x},\widetilde{y},\widetilde{z};t\right)-F^{\left(p\right)}\left(\widetilde{x},\widetilde{y},\widetilde{z};t\right)\right|\mathrm{d}\widetilde{x}\mathrm{d}\widetilde{y}\mathrm{d}\widetilde{z}}{\int_{\Psi_{obs}}\left|F^{\left(p\right)}\left(\widetilde{x},\widetilde{y},\widetilde{z};t\right)\right|\mathrm{d}\widetilde{x}\mathrm{d}\widetilde{y}\mathrm{d}\widetilde{z}}$)
complies with the convergence condition $\Xi^{\left(p\right)}\left(t\right)\leq\Xi^{th}$.
If this holds true, the reference current is set to the $p$-th estimate,
$\overline{J}^{opt}=\overline{J}^{\left(p\right)}$. Otherwise, the
minimum norm current, $\overline{J}_{MN}^{\left(p\right)}$, corresponding
to $R^{\left(p\right)}\left(\widetilde{x},\widetilde{y},\widetilde{z};t\right)$
is retrieved by means of the truncated singular value decomposition
\cite{Oliveri 2021c}\cite{Salucci 2018}

\noindent \begin{equation}
\overline{J}_{MN}^{\left(p\right)}=\mathbb{F}^{-1}\left\{ \mathbb{H}^{-1}\left\{ R^{\left(p\right)}\left(\widetilde{x},\widetilde{y},\widetilde{z};t\right)\right\} \right\} .\label{eq:}\end{equation}
The quantization of the minimum norm current is subsequently carried
out by approximating it with the closest element of $\mathcal{C}$
($\overline{J}^{\left(p+1\right)}\approx\overline{J}_{MN}^{\left(p\right)}$,
$\overline{J}^{\left(p+1\right)}\in\mathcal{C}$). More in detail,
the amplitude and the phase coefficients of $\overline{J}^{\left(p+1\right)}$
are determined by minimizing the mismatch cost function\begin{equation}
\rho\left(\alpha_{mn}\left(t\right),\,\chi_{mn}\left(t\right)\right)=\left\{ \frac{\left\Vert \sum_{m=1}^{M}\sum_{n=1}^{N}\alpha_{mn}\left(t\right)\exp\left[j\chi_{mn}\left(t\right)\right]\Omega_{mn}\left(x,y\right)\widehat{\iota}-\overline{J}_{MN}^{\left(p\right)}\right\Vert ^{2}}{\left\Vert \overline{J}_{MN}^{\left(p\right)}\right\Vert ^{2}}\right\} \label{eq:projection current}\end{equation}
$\left\Vert \cdot\right\Vert $ being the $\ell_{2}$ norm {[}i.e.,
$\left(\alpha_{mn}^{\left(p+1\right)}\left(t\right),\,\chi_{mn}^{\left(p+1\right)}\left(t\right)\right)=\arg\min_{\chi_{mn}\left(t\right)\in\mathcal{X}}^{\alpha_{mn}\left(t\right)\in\mathcal{A}}\left\{ \rho\left(\alpha_{mn}\left(t\right),\,\chi_{mn}\left(t\right)\right)\right\} ${]},
they are then substituted in (\ref{eq:feasibility currents}) to yield
$\overline{J}^{\left(p+1\right)}$. The iteration index is then updated
($p\leftarrow p+1$) and the entire \emph{QIPM} process is restarted
from the footprint pattern computation.

\noindent It is worth pointing out that, unlike state-of-the-art approaches
\cite{Oliveri 2021c}\cite{Oliveri 2021d}, the \emph{}operation in
(\ref{eq:projection current}) outputs an estimated current $\overline{J}^{\left(p+1\right)}$
that fulfils the feasibility condition, thus it is assured that the
current distribution determined at the convergence, $\overline{J}^{opt}$,
can be surely implemented with a \emph{1RP-EMS} layout.

\subsection{\noindent \emph{1RP-EMS} Configuration Method\label{sub:3.2 - 1RP-EMS-Configuration-Method}}

\noindent By following the guidelines in \cite{Oliveri 2021c}, but
here customized to a binary control problem, a \emph{SbD}-based optimization
is carried out to identify the \emph{1RP-EMS} discrete micro-scale
status $\mathcal{S}^{opt}\left(t\right)$ of $M\times N$ binary entries.
Towards this end, a set of $L$ trial \emph{1RP-EMS} configurations\begin{equation}
\left\langle \mathcal{S}\left(t\right)\right\rangle \triangleq\left\{ \mathcal{S}_{l}\left(t\right);\, l=1,...,L\right\} \label{eq:trial solution sequence}\end{equation}
is iteratively processed until either the number of \emph{SbD} iterations
reaches the maximum value $I$ ($i=I$, $i$ being the iteration index)
or the feasible reference current distribution $\overline{J}^{opt}$,
computed in Sect. \ref{sub:3.1 - QIPM-Based-Reference-Current}, is
matched (\ref{micro cost-funct}) {[}i.e., $\psi\left(\mathcal{S}^{opt}\left(t\right)\right)\le\psi^{th}$,
$\mathcal{S}^{opt}\left(t\right)$ $=$ $\arg\left(\min_{l,i}\left[\psi\left(\mathcal{S}_{l}^{\left(i\right)}\left(t\right)\right)\right]\right)$,
$\psi^{th}$ being a user-defined convergence threshold{]}.

\noindent Starting from a random initial configuration, $\left\langle \mathcal{S}^{\left(i\right)}\left(t\right)\right\rangle _{i=0}$,
each $i$-th ($i=1,...,I$) iteration consists of the following operations:

\begin{itemize}
\item \noindent \emph{1RP-EMS} \emph{Surrogate Modeling} - The set of $L$
\emph{micro-scale} electric/magnetic surface susceptibility vectors,
$\left\langle \mathcal{K}^{\left(i\right)}\left(t\right)\right\rangle $
($\left\langle \mathcal{K}\left(t\right)\right\rangle \triangleq\left\{ \mathcal{K}_{l}\left(t\right);\, l=1,...,L\right\} $),
is predicted with an \emph{AI}-based technique, featuring an \emph{Ordinary
Kriging} implementation, according to the most recent trends in the
surrogate modeling of wave manipulating devices \cite{Oliveri 2020}\cite{Salucci 2018c}.
For each $l$-th entry of $\left\langle \mathcal{K}^{\left(i\right)}\left(t\right)\right\rangle $,
the diagonal tensor of the electric/magnetic local surface susceptibility
of the ($m$, $n$)-th ($m=1,...,M$; $n=1,...,N$) meta-atom, $\overline{\overline{K}}_{mn}\left(t\right)$,
is approximated with its digital-twin (\emph{DT}), $\overline{\overline{K}}_{mn}\left(t\right)\approx\overline{\overline{K}}_{mn}^{DT}\left(t\right)$
($\overline{\overline{K}}_{mn}^{DT}\left(t\right)\triangleq\mathbb{K}^{DT}\left\{ \mathbf{g};\, s_{mn}\left(t\right)\right\} $),
which is off-line trained starting from $V$ full-wave evaluations
of the meta-atom response \{{[}$\mathbf{g}_{v}$, $s_{mn}^{v}\left(t\right)$;
$\mathbb{K}\left\{ \mathbf{g}_{v};\, s_{mn}^{v}\left(t\right)\right\} ${]};
$v=1,...,V$\} \cite{Oliveri 2020}\cite{Salucci 2018c};
\item \noindent \emph{Surface Current} \emph{Computation} - The distribution
of the surface current $\overline{J}_{l}\left(x,y;t\right)$ induced
on the $l$-th ($l=1,...,L$) \emph{1RP-EMS}, which is modeled with
the surrogate susceptibility vector $\mathcal{K}_{l}^{DT}\left(t\right)$,
is computed by setting $\overline{\overline{K}}_{mn}\left(t\right)=\overline{\overline{K}}_{mn}^{DT}\left(t\right)$
in (\ref{eq:surface currents});
\item \noindent \emph{Surface Current} \emph{Fitness Evaluation} - The mismatch
between $\overline{J}_{l}^{\left(i\right)}$ ($l=1,...,L$) and $\overline{J}^{opt}$
is quantified by calculating the value of the micro-scale cost function
(\ref{micro cost-funct}), $\psi\left(\mathcal{S}_{l}^{\left(i\right)}\left(t\right)\right)$; 
\item \emph{Guess Current} \emph{Update} - A new set of \emph{1RP-EMS} states,
$\left\langle \mathcal{S}^{\left(i+1\right)}\left(t\right)\right\rangle $,
is generated by applying the Genetic-Algorithm (\emph{GA}) operators
\cite{Rocca 2009} to the previous guesses, $\left\langle \mathcal{S}^{\left(i\right)}\left(t\right)\right\rangle $,
according their fitness values, $\left\langle \psi^{\left(i\right)}\left(t\right)\right\rangle $
($\left\langle \psi^{\left(i\right)}\left(t\right)\right\rangle $
$\triangleq$ \{$\psi\left(\mathcal{S}_{l}^{\left(i\right)}\left(t\right)\right)$;
$l=1,...,L$). Unlike \cite{Oliveri 2021c}\cite{Oliveri 2021d},
a \emph{GA}-based optimization is performed due to the binary \emph{DoF}s
of the problem at hand.
\end{itemize}

\section{\noindent Numerical Results\label{sec:4 - Numerical-Analysis-and}}

\noindent This section is aimed at illustrating the synthesis process
of \emph{1RP-EMS}s described in Sect. \ref{sec:3 - Method} as well
as at demonstrating its effectiveness and potentialities. Towards
this end, the design of the single-bit meta-atom is first presented
along with the full-wave validation of its properties (Sect. \ref{sub:4.1 - Meta-Atom-Design-and}).
Afterwards, the \emph{1RP-EMS} control is assessed through a selected
set of numerical experiments (Sect. \ref{sub:4.2 - Multi-Scale-Single-Bit-RIS}).
For the full-wave modeling of both the meta-atom and the finite \emph{1RP-EMS}
layouts, the \emph{Ansys HFSS} \cite{HFSS 2021} \emph{EM} simulator
has been used.

\subsection{\noindent Single-Bit Meta-Atom Design and Validation\label{sub:4.1 - Meta-Atom-Design-and}}

\noindent Since a key objective of this work is to prove that it is
possible to achieve advanced beam shaping properties with minimum-complexity
\emph{RP-EMS}s, the design of the single-bit meta-atom has been carried
out according to Sect. \ref{sec:2.1 - RP-EMS Design} by also taking
into account the following constraints: (\emph{i}) the meta-atom features
a single-layer geometry to minimize the fabrication complexity; (\emph{ii})
the single-bit \emph{}($B=1$) reconfigurability of the \emph{RP-EMS}
unit cell is obtained by applying a single bias voltage; (\emph{iii})
the shape of the layout of the printed cell is very regular to keep
its \emph{EM} behavior independent on the accuracy of the fabrication
process; (\emph{iv}) the \emph{1RP-EMS} structure works whatever the
polarization of the incident field. 

\noindent The unit cell in \cite{Yang 2017} has been then considered
as reference model. It consists of a simple square patch (Fig. 2)
with two edges connected to the ground plane through two p-i-n diodes
{[}green rectangles - Fig. 2(\emph{a}){]} and two vias {[}yellow circles
- Fig. 2(\emph{a}){]}. By applying a bias voltage at the center of
the patch, the diodes can be either both set to the {}``ON'' {[}$s_{mn}\left(t\right)=1${]}
or both to the {}``OFF'' {[}$s_{mn}\left(t\right)=0${]} states
to implement the single-bit-per-atom reconfigurability. 

\noindent To operate at the central frequency of the sub-6GHz $n78$
band \cite{Ciydem 2020} (i.e., $f_{0}=3.5$ {[}GHz{]}), such a reference
model has been tuned by considering a Rogers RO4350 ($\varepsilon_{r}=3.66$,
$\tan\delta=4.0\times10^{-3}$) substrate with thickness of $1.524\times10^{-3}$
{[}m{]} that includes $3.5\times10^{-5}$ {[}m{]}-thick metallizations
and the MACOM MADP-000907-14020 diodes. The values of the $\mathbf{g}^{opt}$
entries are listed in Tab. I, while the CAD models of the unit cell
and of the switching device are shown in Fig. 2(\emph{b}) and Fig.
2(\emph{c}), respectively. 

\noindent The reflection performance of the optimized meta-atom are
illustrated in Fig. 3 for the broadside incidence. More in detail,
the plots of the phase {[}Fig. 3(\emph{a}){]} and the magnitude {[}Fig.
3(\emph{b}){]} of the \emph{TE/TM} components of the local reflection
tensor $\overline{\overline{\Gamma}}_{mn}\left(t\right)$ indicate
that such a meta-atom supports a $\approx180$ {[}deg{]} phase difference
between the {}``ON'' {[}$s_{mn}\left(t\right)=1${]} and the {}``OFF''
{[}$s_{mn}\left(t\right)=0${]} states at $f_{0}$ {[}Fig. 3(\emph{a}){]}.
Thanks to the symmetry of the layout, the arising unit-cell is insensitive
to the polarization {[}Fig. 3(\emph{a}){]}. Moreover, the losses are
limited {[}$<4$ {[}dB{]} - Fig. 3(\emph{b}){]} and the cross-polarization
level is low {[}$<-18$ {[}dB{]} - Fig. 3(\emph{b}){]} within the
whole frequency band.

\noindent It is finally worthwhile to remark that, while the successive
control step (Sect. \ref{sec:2.2 - RP-EMS Control}) has been performed
in this paper with the single-bit cell in Fig. 2, the proposed approach
for configuring the \emph{1RP-EMS} can be adopted regardless of the
working frequency, the number of bits per cell, $B$, and the meta-atom
complexity \cite{Oliveri 2015b}.

\subsection{\noindent Single-Bit \emph{RP-EMS} Control\label{sub:4.2 - Multi-Scale-Single-Bit-RIS}}

\noindent To assess the features and the potentialities of the \emph{1RP-EMS}
control method in Sects. \ref{sub:3.1 - QIPM-Based-Reference-Current}-\ref{sub:3.2 - 1RP-EMS-Configuration-Method},
different \emph{EMS} apertures and target radiation performance have
been analyzed by considering a \emph{SEME} scenario where a \emph{1RP-EMS}
is placed at $d=5$ {[}m{]} over the ground (Fig. 1), it is illuminated
by a base station located along the \emph{EMS} broadside direction
{[}i.e., $\left(\theta^{inc},\varphi^{inc}\right)=\left(0,0\right)$
{[}deg{]} $\to$ $\widehat{\mathbf{e}}_{\bot}=\widehat{\mathbf{y}}$
and $\widehat{\mathbf{e}}_{\parallel}=\widehat{\mathbf{x}}${]}, and
it is equipped with a slant $+45$ {[}deg{]} linearly polarized antenna
(i.e., $E_{\bot}^{inc}=E_{\parallel}^{inc}=1$). As for the calibration
setup of the \emph{1RP-EMS} control, the following values have been
chosen according to the guidelines in \cite{Oliveri 2021c}\cite{Oliveri 2020}:
$V=2\times10^{4}$, $P=10^{2}$, $L=20$, $\Xi^{th}=10^{-4}$, $\psi^{th}=10^{-3}$,
and $I=10^{4}$.

\noindent The first experiment is aimed at configuring a $M\times N=10\times10$
\emph{1RP-EMS} to maximize the reflected power in a square $\Psi_{cov}$
of size $10\times10$ {[}$\mathrm{m}^{2}${]} located in the global
coordinate system (Fig. 1) at $\left(\widetilde{x},\widetilde{y},\widetilde{z}\right)=\left(25,30,0\right)$
{[}m{]}, which corresponds to set the desired footprint pattern as
follows\begin{equation}
F^{des}\left(\widetilde{x},\widetilde{y},\widetilde{z};t\right)=\left\{ \begin{array}{ll}
-10\,[dB] & \left(\widetilde{x},\widetilde{y},\widetilde{z}\right)\in\Psi_{cov}\\
-50\,[dB] & \left(\widetilde{x},\widetilde{y},\widetilde{z}\right)\notin\Psi_{cov}\end{array}\right.\label{eq: Desired footprint (1)}\end{equation}
with $t=T=1$.

\noindent Figure 4(\emph{a}) shows the behaviour of the macro-scale
cost function (\ref{Macro cost-funct}) $\Phi^{\left(p\right)}$ during
the \emph{QIPM}-based process ($p=1,...,P$) for the synthesis of
the reference surface current in comparison with that of the \emph{IPM}
technique \cite{Oliveri 2021c}. As expected, the \emph{QIPM} does
not outperform the \emph{IPM} in terms of footprint pattern matching
(i.e., $\Phi_{QIPM}^{\left(p\right)}>\Phi_{IPM}^{\left(p\right)}$,
$p=1,...,P$) since the former is a constrained version of the latter
owing to the binary nature of the meta-atoms and the quantization
of the arising current distribution. Indeed, unlike the smoothly varying
phase distribution of the \emph{IPM} {[}Fig. 4(\emph{b}){]} that ignores
any limitation to the phase control, the profile of the phase distribution
of the \emph{QIPM} current turns out to be binarized {[}Fig. 4(\emph{c}){]}.
Such an apparent drawback {[}Fig. 4(\emph{a}){]} is actually a fundamental
advantage of the \emph{QIPM} when dealing with the subsequent \emph{SbD}-driven
micro-scale state optimization (Fig. 5). As a matter of fact, the
plot of the local error $\sigma\left(m,n\right)$ ($m=1,...,M$; $n=1,...,N$)\begin{equation}
\sigma\left(m,n;t\right)\triangleq\angle\overline{J}^{opt}\left(x_{m},y_{n};t\right)-\angle\overline{J}^{*}\left(x_{m},y_{n};t\right)\label{eq:}\end{equation}
{[}$\overline{J}^{*}\left(x,y;t\right)=\mathbb{G}\left\{ \mathbb{K}\left\{ \mathbf{g};\, s_{mn}^{opt}\left(t\right)\right\} ;\overline{E}^{inc}\left(x,y,0;t\right)\right\} ${]}
in approximating the phase of the reference current distribution with
the \emph{1RP-EMS} in Fig. 5(\emph{b}) points out that it is more
difficult to match the \emph{IPM}-synthesized one, while the mismatch
reduces in the \emph{QIPM} case {[}Fig. 5(\emph{a}){]} (i.e., $3\le\sigma^{IPM}\left(m,n\right)\le121$
{[}deg{]} vs. $0.2\le\sigma^{QIPM}\left(m,n\right)\le0.45$ {[}deg{]}),
as one can visually notice by comparing the phase profiles of the
reference and the synthesized currents {[}$\overline{J}_{IPM}^{opt}\left(x,y;t\right)$
- Fig. 4(\emph{b}) vs. $\overline{J}_{IPM}^{*}\left(x,y;t\right)$
- Fig. 5(\emph{c}); $\overline{J}_{QIPM}^{opt}\left(x,y;t\right)$
- Fig. 4(\emph{c}) vs. $\overline{J}_{QIPM}^{*}\left(x,y;t\right)$
- Fig. 5(\emph{d}){]}.

\noindent In order to analyze the impact of those results on the coverage
performance, the plots of the analytically-computed {[}Figs. 6(\emph{c})-6(\emph{d}){]}
and the \emph{HFSS}-simulated {[}Figs. 6(\emph{e})-6(\emph{f}){]}
footprint patterns generated by the \emph{IPM} {[}Fig. 6(\emph{a}){]}
and the \emph{QIPM} {[}Fig. 6(\emph{b}){]} \emph{1RP-EMS} in an observation
region $\Psi_{obs}$ of $75\times60$ {[}$\mathrm{m}^{2}${]} located
in front of the \emph{RP-EMS} are reported.

\noindent Despite the relatively small \emph{EMS} aperture and its
very limited (binary) reconfiguration capabilities, the \emph{QIPM}
configuration {[}Fig. 6(\emph{b}){]} of the \emph{1RP-EMS} focuses
the reflected beam in the desired coverage region $\Psi_{cov}$ (i.e.,
along a non-Snell direction) better than the \emph{IPM} one {[}Fig.
6(\emph{a}){]} with a lower number of sidelobes {[}Fig. 6(\emph{c})
vs. Fig. 6(\emph{d}){]}. Moreover, the close fitting between analytically-computed
{[}Figs. 6(\emph{c})-6(\emph{d}){]} and \emph{HFSS}-simulated {[}Figs.
6(\emph{e})-6(\emph{f}){]} patterns proves the accuracy of the analytic
prediction of the reflection/focusing properties of the \emph{1RP-EMS}
layout despite the finite \emph{EMS} aperture and the intrinsic approximations
of the analytical model. Such an outcome, which is also in line with
the conclusions drawn for the \emph{SP-EMS} case \cite{Oliveri 2021c}\cite{Oliveri 2021d},
further confirms the reliability of the proposed multi-scale design
without the need of recurring to expensive full-wave simulations in
the on-line synthesis process as well as its effectiveness to control
the macro-scale wave manipulation properties of \emph{1RP-EMS}s.

\noindent When increasing the \emph{EMS} size $\Psi_{EMS}$ ($M=N=10$
$\to$ $M=N=30$) by keeping the same target coverage region $\Psi_{obs}$
and footprint requirements (\ref{eq: Desired footprint (1)}), similar
considerations to those of the first numerical experiment hold true.
For the sake of completeness and analogously to the $M\times N=10\times10$
case, Figures 7-9 illustrate the process for configuring the \emph{1RP-EMS}
by also comparing the \emph{QIPM}-based approach with the \emph{IPM}
one. Figure 7 deals with the current synthesis, while Figure 8 is
concerned with the configuration of the \emph{1RP-EMS}, and Figure
9 gives the radiated footprint patterns. More in detail, Figure 7(\emph{a})
shows the iterative \emph{QIPM}/\emph{IPM} minimization of the macro-scale
cost function (\ref{Macro cost-funct}) to define the reference (phase)
current profiles in Figs. 7(\emph{b})-7(\emph{c}) that are approximated
{[}Figs. 7(\emph{d})-7(\emph{e}){]} by the \emph{SbD}-optimized setups
{[}Figs. 8(\emph{b})-8(\emph{c}){]} of the \emph{1RP-EMS} in Fig.
8(\emph{a}) to afford the footprint patterns in Figs. 9(\emph{a})-9(\emph{b}).
Once again, the constrained nature of the \emph{QIPM} solution {[}Fig.
7(\emph{c}){]} allows one to better configure {[}Fig. 8(\emph{c}){]}
the single-bit \emph{RP-EMS} {[}Fig. 8(\emph{a}){]} for more faithfully
fulfilling the target coverage {[}Fig. 9(\emph{b}){]}. The improved
focusing performance of the \emph{QIPM}-based control are quantified
by the value of the \emph{footprint coverage index} $\gamma$,

\noindent \begin{equation}
\gamma\triangleq\frac{W_{cov}}{W_{ext}}\label{eq:}\end{equation}
where $W_{\Psi}\triangleq\frac{1}{2\eta_{0}}\int_{\Psi}F\left(\widetilde{x},\widetilde{y},\widetilde{z};t\right)\mathrm{d}\widetilde{x}\mathrm{d}\widetilde{y}\mathrm{d}\widetilde{z}$
is the power reflected in the $\Psi$ region and $\Psi_{ext}=\Psi_{obs}-\Psi_{cov}$,
which is equal to $\gamma^{QIPM}\approx4.3\times10^{-1}$, while $\gamma^{IPM}\approx3.6\times10^{-1}$
{[}Fig. 10(\emph{a}){]}.

\noindent In order to give the interested readers a more exhaustive
picture of the advantages of using the \emph{QIPM} approach instead
of the \emph{IPM} one when dealing with discrete \emph{RP-EMS}s, Figure
10(\emph{a}) compares the behavior of $\gamma^{IPM}$ and $\gamma^{QIPM}$
versus the size of the \emph{1RP-EMS} aperture by reporting the relative
index $\Delta\gamma$ ($\Delta\gamma\triangleq\frac{\gamma^{QIPM}-\gamma^{IPM}}{\gamma^{IPM}}$),
as well. As it can be observed, the proposed method (Sect. \ref{sec:3 - Method})
always determines a configuration of the same \emph{1RP-EMS} that
better focuses the reflected power towards the coverage region $\Psi_{cov}$
(i.e., $\gamma^{QIPM}>\gamma^{IPM}$) with a non-negligible improvement
of the power efficiency {[}i.e., $8$ \% $\le\Delta\gamma\le$ $30$
\% - Fig. 10(\emph{a}){]} also when wide apertures are at hand.

\noindent For illustrative purposes, the synthesized ON/OFF configurations
of the \emph{1RP-EMS} {[}Figs. 10(\emph{b})-10(\emph{e}){]} and the
corresponding footprint patterns (Fig. 11) when $M\times N=50\times50$
{[}Figs. 10(\emph{b})-10(\emph{c}) and Figs. 11(\emph{a})-11(\emph{b}){]}
and $M\times N=200\times200$ {[}Figs. 10(\emph{d})-10(\emph{e}) and
Figs. 11(\emph{c})-11(\emph{d}){]} are reported, as well. Despite
the exponentially increasing complexity of the optimization problem
at hand (\ref{sssvi}) owing to the widening of the discrete solution
space {[}i.e., $8.4\times10^{270}$ ($M\times N=30\times30$), $3.7\times10^{752}$
($M\times N=50\times50$), and $1.5\times10^{12041}$ ($M\times N=200\times200$)
binary configurations, $\mathcal{S}\left(t\right)${]}, the control
method in Sect. \ref{sub:3.2 - 1RP-EMS-Configuration-Method} turns
out to be very effective in finding the optimal \emph{1RP-EMS} configuration
$\left.\mathcal{S}^{opt}\left(t\right)\right\rfloor _{t=T=1}$ whatever
the size of $\Psi_{EMS}$, thus improving the beam focusing capabilities
of the \emph{1RP-EMS} {[}Fig. 10(\emph{a}){]} by fully exploiting
the aperture enlargement {[}Fig. 6(\emph{d}) ($M\times N=10\times10$)
vs. Fig. 9(\emph{b}) ($M\times N=30\times30$) vs. Fig. 11(\emph{b})
($M\times N=50\times50$) vs. Fig. 11(\emph{d}) ($M\times N=200\times200$){]}.

\noindent The next numerical experiment is concerned with the case
of a \emph{Multi-Static Reconfigurability} ($t\in\left\{ t_{1},\, t_{2}\right\} $;
$t=1,...,T$) and it deals with the installation of a \emph{1RP-EMS}
to alternatively target the wireless coverage of Piazza della Signoria
or Piazzale degli Uffizi in Florence (Italy) (i.e., one of the most
frequented urban areas in Europe) that consist of a L-shaped wider
square, $\Psi_{cov}^{\left(1\right)}$, and a narrow adjacent site,
$\Psi_{cov}^{\left(2\right)}$, where the entrance to the Uffizi museum
is located {[}Fig. 12(\emph{a}){]}. The \emph{1RP-EMS}, which has
been assumed to be placed at $d=15$ {[}m{]} on the building in Fig.
12(\emph{b}), is requested to switch between the {}``Signoria+Uffizi''
coverage (i.e., $\left.F^{des}\left(\widetilde{x},\widetilde{y},\widetilde{z};t\right)\right\rfloor _{t=t_{1}}$
as in (\ref{eq: Desired footprint (1)}) by setting $\Psi_{cov}=\Psi_{cov}^{\left(1\right)})$
and the {}``Uffizi'' coverage (i.e., $\left.F^{des}\left(\widetilde{x},\widetilde{y},\widetilde{z};t\right)\right\rfloor _{t=t_{2}}$
as in (\ref{eq: Desired footprint (1)}) by setting $\Psi_{cov}=\Psi_{cov}^{\left(2\right)})$. 

\noindent The configurations of the ON/OFF states of a $M\times N=30\times30$
layout (i.e., $\Psi_{EMS}\approx1.3\times1.3$ {[}$m^{2}${]}), which
afford the $T=2$ footprint patterns in Figs. 13(\emph{c})-13(\emph{d}),
are reported in Figs. 13(\emph{a})-13(\emph{b}). As it can be observed,
there are few similarities between the two control maps, $\left.\mathcal{S}^{opt}\left(t\right)\right\rfloor _{t=t_{1}}$
and $\left.\mathcal{S}^{opt}\left(t\right)\right\rfloor _{t=t_{2}}$,
even though the coverage regions at hand, \{$\Psi_{cov}^{\left(c\right)}$,
($c=1,..,C$; $C=2$)\}, partially overlap {[}Fig. 12(\emph{a}){]}.
Such a behavior is not unexpected due to both the strong non-linearity
of the control problem at hand (\ref{sssvi}) and the binary nature
of the control \emph{DoF}s. 

\noindent Concerning the distribution of the radiated power pattern,
Figure 14 confirms the effectiveness of the synthesized layouts in
fulfilling the coverage requirements, the footprint patterns faithfully
overlapping the area of interest, despite the irregular geometries
of the regions-of-interest and the relatively limited number of reconfigurable
states {[}$\le1$ {[}Kbit{]} - Fig. 8(\emph{a}){]}. On the other hand,
it is also worth noticing that, although the wireless coverage of
the {}``Signoria+Uffizi'' area, $\Psi_{cov}^{\left(1\right)}$,
is a more challenging problem than that of the {}``Uffizi'' site,
$\Psi_{cov}^{\left(2\right)}$, since it requires the \emph{1RP-EMS}
focuses the reflected power also at very low elevation angles with
respect to its location {[}Fig. 13(\emph{c}){]}, the amount of power
reflected by the \emph{1RP-EMS} is kept almost unaltered (i.e., $\frac{W_{cov}^{\left(2\right)}}{W_{cov}^{\left(1\right)}}\approx0.78$).

\noindent The ability to afford more elaborated/over-constrained footprints
by controlling a single-bit \emph{RP-EMS} has been the assessed with
the synthesis of a $M\times N=100\times100$ \emph{1RP-EMS} devoted
to manipulate the reflected power for matching the {}``ELEDIA''
logo pattern {[}Fig. 15(\emph{b}){]}. The plot of the full-wave simulated
footprint pattern, $\left.F^{opt}\left(\widetilde{x},\widetilde{y},\widetilde{z};t\right)\right\rfloor _{t=1}$,
in an observation region $\Psi_{obs}^{\left(c\right)}$ of extension
$80\times40$ {[}$\mathrm{m}^{2}${]} {[}Fig. 15(\emph{b}){]} proves
the reliability of the \emph{EMS} configuration $\left.\mathcal{S}^{opt}\left(t\right)\right\rfloor _{t=T=1}$
in Fig. 15(\emph{a}) to match the coverage requirements on a complex
region $\Psi_{cov}$. The readers are suggested to notice that this
test case has been already successfully addressed in \cite{Oliveri 2021c}
with \emph{SP-EMS}s, but here the \emph{DoF}s are far less than those
available in the \emph{SP-EMS} case \cite{Oliveri 2021c}.

\noindent Finally, the numerical assessment ends with a test on the
performance of the proposed \emph{1RP-EMS} synthesis method in a scenario
that needs a \emph{Dynamic} \emph{Multi-Beam} ($C=3$) \emph{Reconfigurability}.
More in detail, the problem at hand is that of $C=3$ users, each
occupying a coverage region $\Psi_{cov}^{\left(c\right)}$ ($c=1,...,C$)
of size $10\times10$ {[}$\mathrm{m}^{2}${]}, that move at different
speeds in different directions as sketched in Fig. 16. By still considering
the $M\times N=100\times100$ \emph{1RP-EMS} aperture, the plots of
the footprint pattern {[}Figs. 17(\emph{e})-17(\emph{h}){]} radiated
by the corresponding ON/OFF configuration of the \emph{EMS} {[}Figs.
17(\emph{a})-17(\emph{d}){]} in $T=4$ subsequent time-instants confirm
that such a technological solution fits the users' needs without installing
multiple \emph{RP-EMS}s or using multi-bit-per-atom reconfiguration
schemes.

\noindent From a computational perspective and to give some insights
on the burden for dynamically managing an \emph{EMS}-driven wireless
planning, let us consider that a $t$-th ($t=1,...,T$) reconfiguration
of the $M\times N=100\times100$ \emph{1RP-EMS} for the last scenario
(Fig. 16) required less than $0.2$ {[}s{]} to a non-optimized MATLAB
implementation of the control algorithm (Sects. \ref{sub:3.1 - QIPM-Based-Reference-Current}-\ref{sub:3.2 - 1RP-EMS-Configuration-Method})
running on a standard laptop equipped with a single-core $1.6$ GHz
CPU. Such a quite impressive result has been obtained thanks to the
profitable integration of the \emph{QIPM} strategy (Sect. \ref{sub:3.1 - QIPM-Based-Reference-Current})
and the \emph{SbD}-based binary optimization (Sect. \ref{sub:3.2 - 1RP-EMS-Configuration-Method}).
Moreover, to the best of our knowledge on the state-of-the-art literature
on \emph{EMS}s, it turns out that the proposed \emph{EMS} implementation/control
can be properly considered as a suitable candidate/tool for the real-time
coverage of time-varying wireless scenarios.

\section{\noindent Conclusions\label{sec:5 - Conclusions-and-Remarks}}

\noindent An innovative method for the synthesis of \emph{RP-EMS}s,
based on single-bit meta-atoms able to support advanced propagation
manipulation features in \emph{SEME} scenarios, has been proposed.
The arising multi-scale optimization problem has been addressed by
means of a two-step approach starting from the design of a meta-atom
that features only a single-bit reconfiguration. First, a discrete-phase
current, which radiates a field distribution fitting complex user-defined
requirements on the footprint pattern, has been computed. Then, a
digital \emph{SbD}-based optimization has been carried out to set
the binary configuration of the \emph{RP-EMS} atoms that supports
such a reference discrete-phase current.

\noindent To the best of the authors' knowledge on the state-of-the
art literature, the main theoretical and methodological advancements
of this work lie in:

\begin{itemize}
\item the assessment that \emph{RP-EMS} architectures featuring single-bit
meta-atoms can allow complex wave manipulations without the need of
continuous phase variations \cite{Oliveri 2021c}\cite{Oliveri 2021d};
\item the derivation of an approach for the control of the \emph{RP-EMS}
to afford complex footprint shapes and not only pencil beams \cite{Liang 2022};
\item the non-trivial extension of the synthesis paradigm, adopted so far
to synthesize static reflectarrays and \emph{SP-EMS}s \cite{Oliveri 2021c}\cite{Oliveri 2021d}\cite{Salucci 2018},
to minimum-complexity \emph{RP-EMS}s by deriving a computationally
effective reconfiguration method.
\end{itemize}
\noindent From the numerical validation, the following outcomes can
be drawn:

\begin{itemize}
\item \noindent the \emph{QIPM}-based approach for the definition of the
reference currents significantly improves the coverage efficiency
with respect to state-of-the-art techniques \cite{Oliveri 2021c}
regardless of the \emph{1RP-EMS} aperture at hand {[}Fig. 10(\emph{a}){]};
\item \noindent despite the minimum complexity of the meta-atoms ($B=1$),
the synthesized \emph{RP-EMS}s feature advanced wave manipulation
properties in realistic scenarios (Fig. 14) as well as in very complex
{}``demonstrative'' cases (e.g., Fig. 15);
\item the proposed method turns out to be an enabling tool for multi-beam
reconfiguration and/or independent user-tracking through \emph{1RP-EMS}
layouts (Fig. 16).
\end{itemize}
\noindent Future works, beyond the scope of this manuscript, will
be aimed at assessing the performance of the proposed method when
using multi-bit meta-atoms and or different meta-atom geometries.

\section*{\noindent Acknowledgements}

\noindent This work benefited from the networking activities carried
out within the Project {}``Cloaking Metasurfaces for a New Generation
of Intelligent Antenna Systems (MANTLES)'' (Grant No. 2017BHFZKH)
funded by the Italian Ministry of Education, University, and Research
under the PRIN2017 Pmrogram (CUP: E64I19000560001). Moreover, it benefited
from the networking activities carried out within the Project {}``SPEED''
(Grant No. 61721001) funded by National Science Foundation of China
under the Chang-Jiang Visiting Professorship Program, the Project
'Inversion Design Method of Structural Factors of Conformal Load-bearing
Antenna Structure based on Desired EM Performance Interval' (Grant
no. 2017HZJXSZ) funded by the National Natural Science Foundation
of China, and the Project 'Research on Uncertainty Factors and Propagation
Mechanism of Conformal Loab-bearing Antenna Structure' (Grant No.
2021JZD-003) funded by the Department of Science and Technology of
Shaanxi Province within the Program Natural Science Basic Research
Plan in Shaanxi Province. A. Massa wishes to thank E. Vico for her
never-ending inspiration, support, guidance, and help.

\section*{\noindent Appendix}

\subsection*{\noindent Expression of $\overline{E}_{mn}\left(t\right)$ and $\overline{H}_{mn}\left(t\right)$}

The surface averaged fields $\overline{E}_{mn}\left(t\right)$ and
$\overline{H}_{mn}\left(t\right)$ can be expressed as \cite{Oliveri 2021c}\cite{Oliveri 2021d}\cite{Yang 2019}\begin{equation}
\overline{E}_{mn}\left(t\right)=\frac{\int_{-\frac{M\Delta x}{2}}^{\frac{M\Delta x}{2}}\int_{-\frac{N\Delta y}{2}}^{\frac{N\Delta y}{2}}\left\{ \overline{\overline{1}}+\overline{\overline{\Gamma}}_{mn}\left(t\right)\right\} \cdot\overline{E}^{inc}\left(x,y,0\right)\Omega_{mn}\left(x,y\right)\mathrm{d}x\mathrm{d}y}{2\times\Delta x\times\Delta y},\label{eq:field average}\end{equation}
and\begin{equation}
\overline{H}_{mn}\left(t\right)=\frac{\int_{-\frac{M\Delta x}{2}}^{\frac{M\Delta x}{2}}\int_{-\frac{N\Delta y}{2}}^{\frac{N\Delta y}{2}}\left\{ \mathbf{k}^{inc}\times\overline{E}^{inc}\left(x,y,0\right)+\mathbf{k}^{ref}\times\overline{\overline{\Gamma}}_{mn}\left(t\right)\cdot\overline{E}^{inc}\left(x,y,0\right)\right\} \Omega_{mn}\left(x,y\right)\mathrm{d}x\mathrm{d}y}{2\times\Delta x\times\Delta y\times\eta_{0}\times k_{0}}\label{eq:field H average}\end{equation}
respectively, where $\overline{\overline{\Gamma}}_{mn}\left(t\right)$
is the local reflection tensor in the ($m$, $n$)-th cell \cite{Oliveri 2021c}\cite{Oliveri 2021d}\cite{Yang 2019}
{[}$\overline{\overline{\Gamma}}_{mn}\left(t\right)=\mathbb{Y}\left\{ \mathbf{g};\, s_{mn}\left(t\right)\right\} ${]},
$\overline{E}^{inc}$ is the incident electric field \cite{Osipov 2017}\begin{equation}
\overline{E}^{inc}\left(x,y,z\right)\triangleq\left(E_{\bot}^{inc}\widehat{\mathbf{e}}_{\bot}+E_{\parallel}^{inc}\widehat{\mathbf{e}}_{\parallel}\right)\exp\left[-j\mathbf{k}^{inc}\cdot\left(x\widehat{\mathbf{x}}+y\widehat{\mathbf{y}}+z\widehat{\mathbf{z}}\right)\right],\label{eq:incident wave}\end{equation}
where $\mathbf{k}^{inc}$ is the incident wave vector ($\mathbf{k}^{inc}$
$\triangleq$ $-k_{0}$ {[}$\sin\left(\theta^{inc}\right)\cos\left(\varphi^{inc}\right)\widehat{\mathbf{x}}$
$+$ $\sin\left(\theta^{inc}\right)\sin\left(\varphi^{inc}\right)\widehat{\mathbf{y}}$
$+$ $\cos\left(\theta^{inc}\right)\widehat{\mathbf{z}}${]}), $\mathbf{k}^{ref}$
is the corresponding reflected wave vector according to standard plane
wave theory \cite{Yang 2019}, and $\widehat{\mathbf{e}}_{\bot}=\frac{\mathbf{k}^{inc}\times\widehat{\bm{\nu}}}{\left|\mathbf{k}^{inc}\times\widehat{\bm{\nu}}\right|}$
and $\widehat{\mathbf{e}}_{\parallel}=\frac{\widehat{\mathbf{e}}_{\bot}\times\mathbf{k}^{inc}}{\left|\widehat{\mathbf{e}}_{\bot}\times\mathbf{k}^{inc}\right|}$
are the {}``perpendicular'' and {}``parallel'' unit vectors (i.e.,
\emph{TE} and \emph{TM} modes) \cite{Oliveri 2021c}\cite{Oliveri 2021d}\cite{Yang 2019}.

\section*{FIGURE CAPTIONS}

\begin{itemize}
\item \textbf{Figure 1.} \emph{Problem geometry}. Sketch of the smart \emph{EM}
environment scenario.
\item \textbf{Figure 2.} \emph{1RP-EMS Design} - Unit cell geometry: (\emph{a})
top view and 3D CAD model of (\emph{b}) the meta-atom and zoom on
(\emph{c}) the switching device.
\item \textbf{Figure 3.} \emph{1RP-EMS Design} - Plots of (\emph{a}) the
phase and (\emph{b}) the magnitude of the TE/TM components of the
local reflection tensor $\overline{\overline{\Gamma}}_{mn}$ versus
the frequency in correspondence with the {}``ON'' ($s_{mn}=1$)
and the {}``OFF'' ($s_{mn}=0$) states.
\item \textbf{Figure 4.} \emph{1RP-EMS Control} (\emph{{}``Square'' Footprint},
$M=N=10$, $d=5$ {[}m{]}; $P=10^{2}$) - Behaviour of (\emph{a})
the macro-scale cost $\Phi^{\left(p\right)}$ versus the iteration
index ($p=1,...,P$) and plot of (\emph{b})(\emph{c}) the phase of
the reference current, \{$\overline{J}^{opt}\left(x_{m},y_{n};t\right)$;
($m=1,...,M$; $n=1,...,N$)\}, synthesized with (\emph{b}) the \emph{IPM},
$\overline{J}_{IPM}^{opt}$, \emph{}and (\emph{c}) the \emph{QIPM},
$\overline{J}_{QIPM}^{opt}$.
\item \textbf{Figure 5.} \emph{1RP-EMS Control} (\emph{{}``Square'' Footprint},
$M=N=10$, $d=5$ {[}m{]}) - Plots of (\emph{a}) the distribution
of the local error $\sigma_{\ell}$ {[}$\ell=n+N\times\left(m-1\right)$;
($m=1,...,M$; $n=1,...,N$){]} and of (\emph{c})(\emph{d}) the phase
of the current, \{$\overline{J}^{*}\left(x_{m},y_{n};t\right)$; ($m=1,...,M$;
$n=1,...,N$)\}, generated by the \emph{1RP-EMS} (\emph{b}) configured
with (\emph{c}) the \emph{IPM}-based approach, $\overline{J}_{IPM}^{*}$,
or (\emph{d}) the \emph{QIPM} one, $\overline{J}_{QIPM}^{*}$.
\item \textbf{Figure 6.} \emph{1RP-EMS Control} (\emph{{}``Square'' Footprint},
$M=N=10$, $d=5$ {[}m{]}) - Plots of (\emph{a})(\emph{b}) the ON/OFF
states and the corresponding (\emph{c})(\emph{d}) analytically-computed
and (\emph{e})(\emph{f}) HFSS-simulated footprint patterns of (\emph{a})
the \emph{IPM} and (\emph{b}) the \emph{QIPM} \emph{1RP-EMS}.
\item \textbf{Figure 7.} \emph{1RP-EMS Control} (\emph{{}``Square'' Footprint},
$M=N=30$, $P=10^{2}$) - Behaviour of (\emph{a}) the macro-scale
cost $\Phi^{\left(p\right)}$ versus the iteration index ($p=1,...,P$)
and plots of the phase of (\emph{b})(\emph{c}) the reference current,
\{$\overline{J}^{opt}\left(x_{m},y_{n};t\right)$; ($m=1,...,M$;
$n=1,...,N$)\}, and of (\emph{d})(\emph{e}) the current, \{$\overline{J}^{*}\left(x_{m},y_{n};t\right)$;
($m=1,...,M$; $n=1,...,N$)\}, generated by the configured \emph{1RP-EMS}
in {[}Fig. 8(a){]} when applying (\emph{b})(\emph{d}) the \emph{IPM}-based
approach or (\emph{c})(\emph{e}) the \emph{QIPM} one.
\item \textbf{Figure 8.} \emph{1RP-EMS Control} (\emph{{}``Square'' Footprint},
$M=N=30$, $d=5$ {[}m{]}) - Plot of (\emph{a}) the 3D model of the
\emph{RP-EMS} and map of (\emph{b})(\emph{c}) the ON/OFF states of
the \emph{1RP-EMS} yielded with (\emph{b}) the \emph{IPM}, $\mathcal{S}_{IPM}^{opt}$,
and (\emph{d}) the \emph{QIPM}, $\mathcal{S}_{QIPM}^{opt}$, approaches.
\item \textbf{Figure 9.} \emph{1RP-EMS Control} (\emph{{}``Square'' Footprint},
$M=N=30$, $d=5$ {[}m{]}) - Plots of the HFSS-simulated footprint
pattern radiated by (\emph{a}) the \emph{IPM} and (\emph{b}) the \emph{QIPM}
\emph{1RP-EMS}.
\item \textbf{Figure 10.} \emph{1RP-EMS Control} (\emph{{}``Square'' Footprint},
$d=5$ {[}m{]}) - Behaviours of (\emph{a}) the total, $\gamma$, and
the relative, $\Delta\gamma$, coverage indexes versus the \emph{1RP-EMS}
size and maps of (\emph{b})-(\emph{e}) the ON/OFF configurations of
the \emph{1RP-EMS} synthesized with (\emph{b})(\emph{d}) the \emph{IPM}
and (\emph{c})(\emph{e}) the \emph{QIPM} when (\emph{b})(\emph{c})
$M\times N=50\times50$ and (\emph{d})(\emph{e}) $M\times N=200\times200$.
\item \textbf{Figure 11.} \emph{1RP-EMS Control} (\emph{{}``Square'' Footprint},
$d=5$ {[}m{]}) - Plots of the HFSS-simulated footprint pattern radiated
by (\emph{a})(\emph{c}) the \emph{IPM} and (\emph{b})(\emph{d}) the
\emph{QIPM} \emph{1RP-EMS} when (\emph{a})(\emph{b}) $M\times N=50\times50$
and (\emph{c})(\emph{d}) $M\times N=200\times200$.
\item \textbf{Figure 12.} \emph{1RP-EMS Control} (\emph{{}``Signoria+Uffizi''
and {}``Uffizi'' Footprints}, $M=N=30$, $d=15$ {[}m{]}) - View
of (\emph{a}) the scenario and of (\emph{b}) the location of the \emph{1RP-EMS}.
\item \textbf{Figure 13.} \emph{1RP-EMS Control} (\emph{{}``Signoria+Uffizi''
and {}``Uffizi'' Footprints}, $M=N=30$, $d=15$ {[}m{]}; \emph{QIPM})
- Plots of (\emph{a})(\emph{b}) the ON/OFF states of the \emph{1RP-EMS}
and (\emph{c})(\emph{d}) the corresponding HFSS-simulated footprint
patterns when focusing on (\emph{a})(\emph{c}) the {}``Signoria+Uffizi''
area and (\emph{b})(\emph{d}) the {}``Uffizi'' area.
\item \textbf{Figure 14.} \emph{1RP-EMS Control} (\emph{{}``Signoria+Uffizi''
and {}``Uffizi'' Footprints}, $M=N=30$, $d=15$ {[}m{]}; \emph{QIPM})
- Coverage check when dealing with (\emph{a}) the {}``Signoria+Uffizi''
and (\emph{b}) the {}``Uffizi'' coverage scenarios.
\item \textbf{Figure 15.} \emph{1RP-EMS Control} (\emph{{}``ELEDIA'' Footprint},
$M=N=100$, $d=15$ {[}m{]}; \emph{QIPM}) - Plots of (\emph{a}) the
ON/OFF states of the \emph{1RP-EMS} and (\emph{b}) the corresponding
HFSS-simulated footprint pattern.
\item \textbf{Figure 16.} \emph{1RP-EMS Control} (\emph{Multi-Beam Footprint},
$M=N=100$, $d=5$ {[}m{]}; \emph{QIPM}) - Users' trajectories.
\item \textbf{Figure 17.} \emph{1RP-EMS Control} (\emph{Multi-Beam Footprint},
$M=N=100$, $d=5$ {[}m{]}; \emph{QIPM}) - Plots of (\emph{a})-(\emph{d})
the ON/OFF states of the \emph{1RP-EMS} and (\emph{e})-(\emph{h})
the corresponding HFSS-simulated footprint patterns radiated at (\emph{a})(\emph{e})
$t=1$, (\emph{b})(\emph{f}) $t=2$, (\emph{c})(\emph{g}) $t=3$,
and (\emph{d})(\emph{h}) $t=T$ ($T=4$).
\end{itemize}

\section*{TABLE CAPTIONS}

\begin{itemize}
\item \textbf{Table 1.} \emph{1RP-EMS Design} - Geometrical descriptors.
\end{itemize}
~

\newpage
\begin{center}~\vfill\end{center}

\begin{center}\includegraphics[%
  clip,
  width=0.95\columnwidth,
  keepaspectratio]{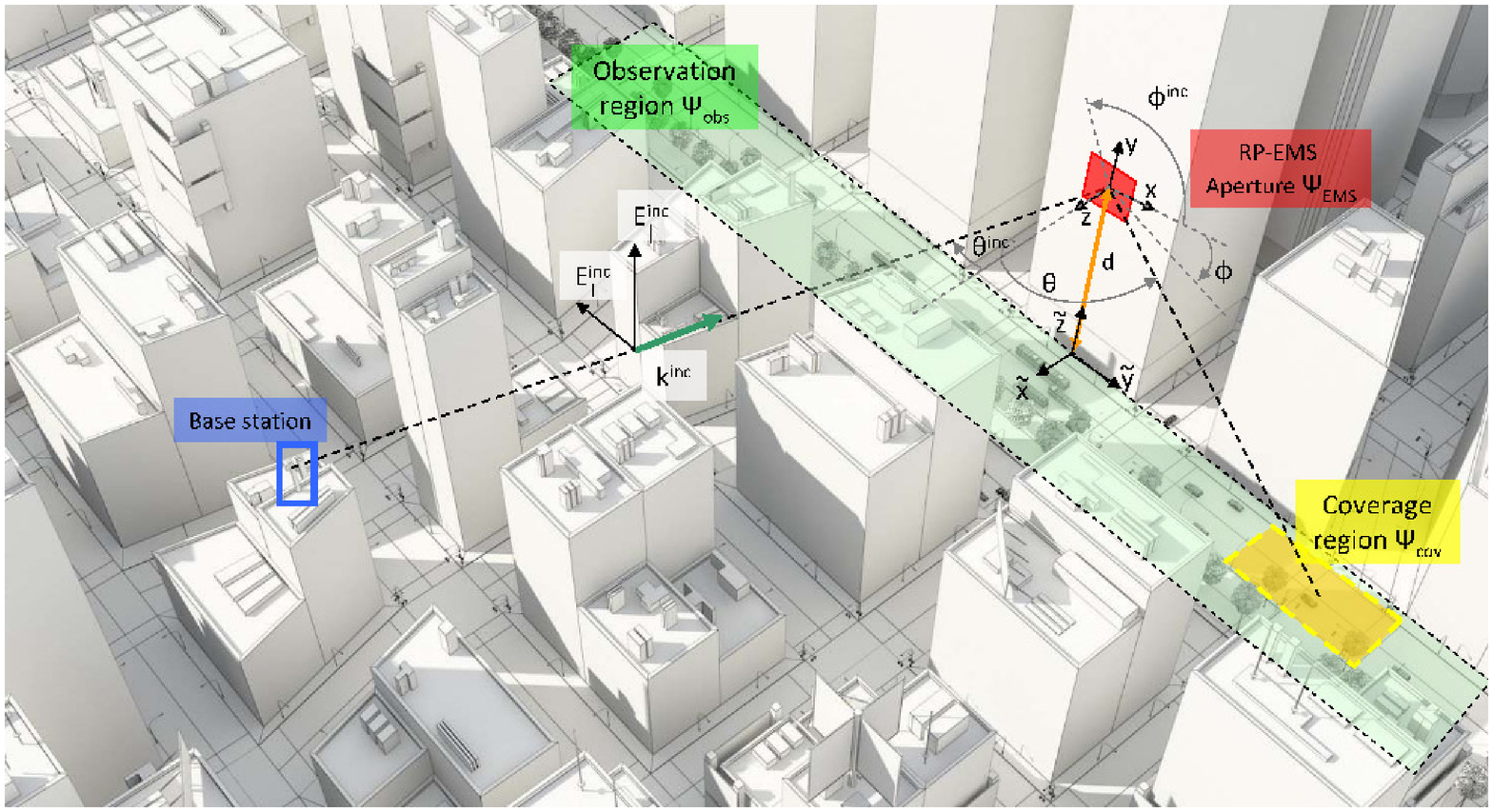}\end{center}

\begin{center}~\vfill\end{center}

\begin{center}\textbf{Fig. 1 - G. Oliveri et} \textbf{\emph{al.}}\textbf{,}
{}``Multi-Scale Single-Bit \emph{RP-EMS} Synthesis for ...''\end{center}

\newpage
\begin{center}\begin{tabular}{c}
\includegraphics[%
  clip,
  width=0.60\columnwidth,
  keepaspectratio]{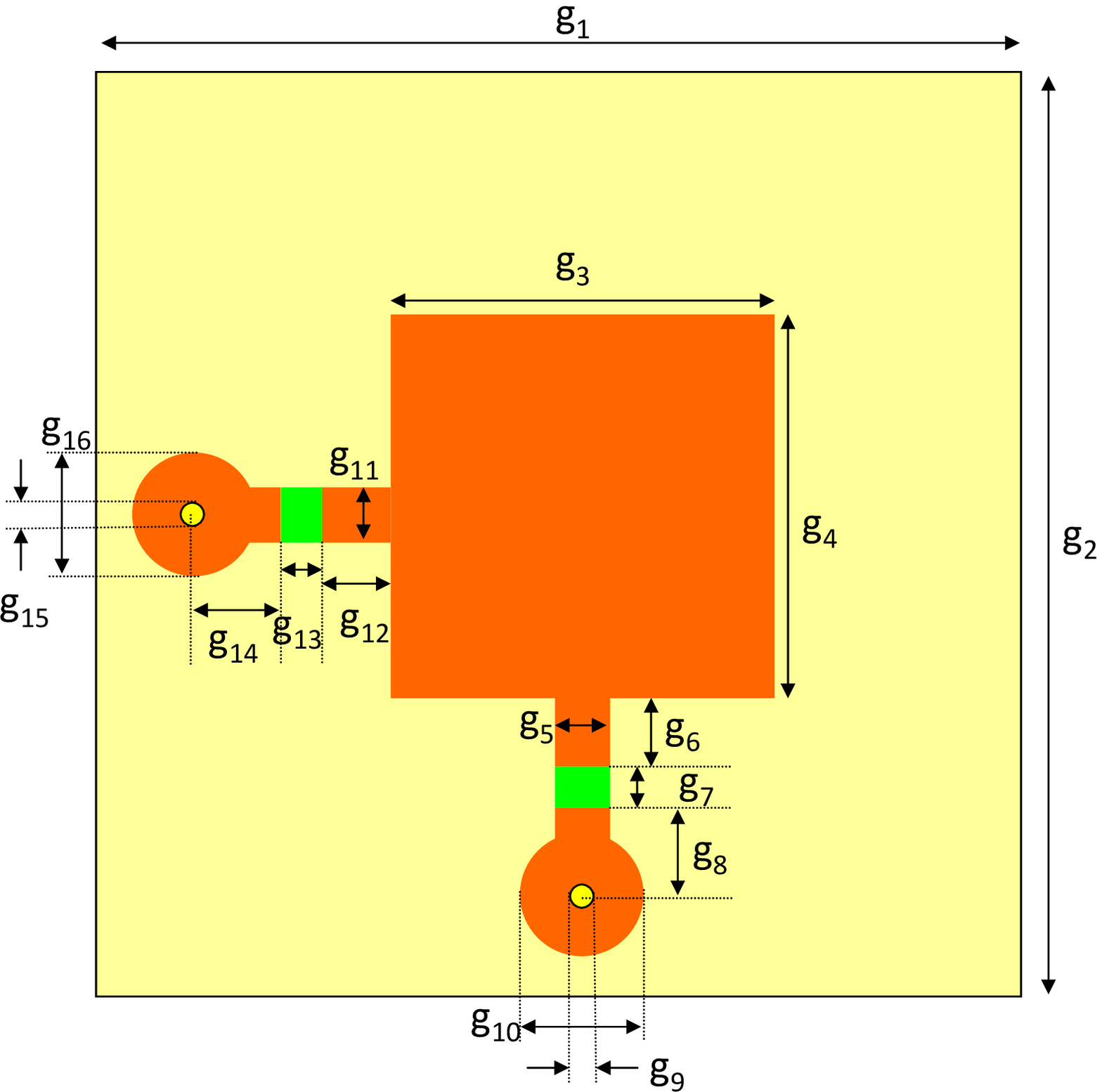}\tabularnewline
(\emph{a})\tabularnewline
\includegraphics[%
  clip,
  width=0.55\columnwidth,
  keepaspectratio]{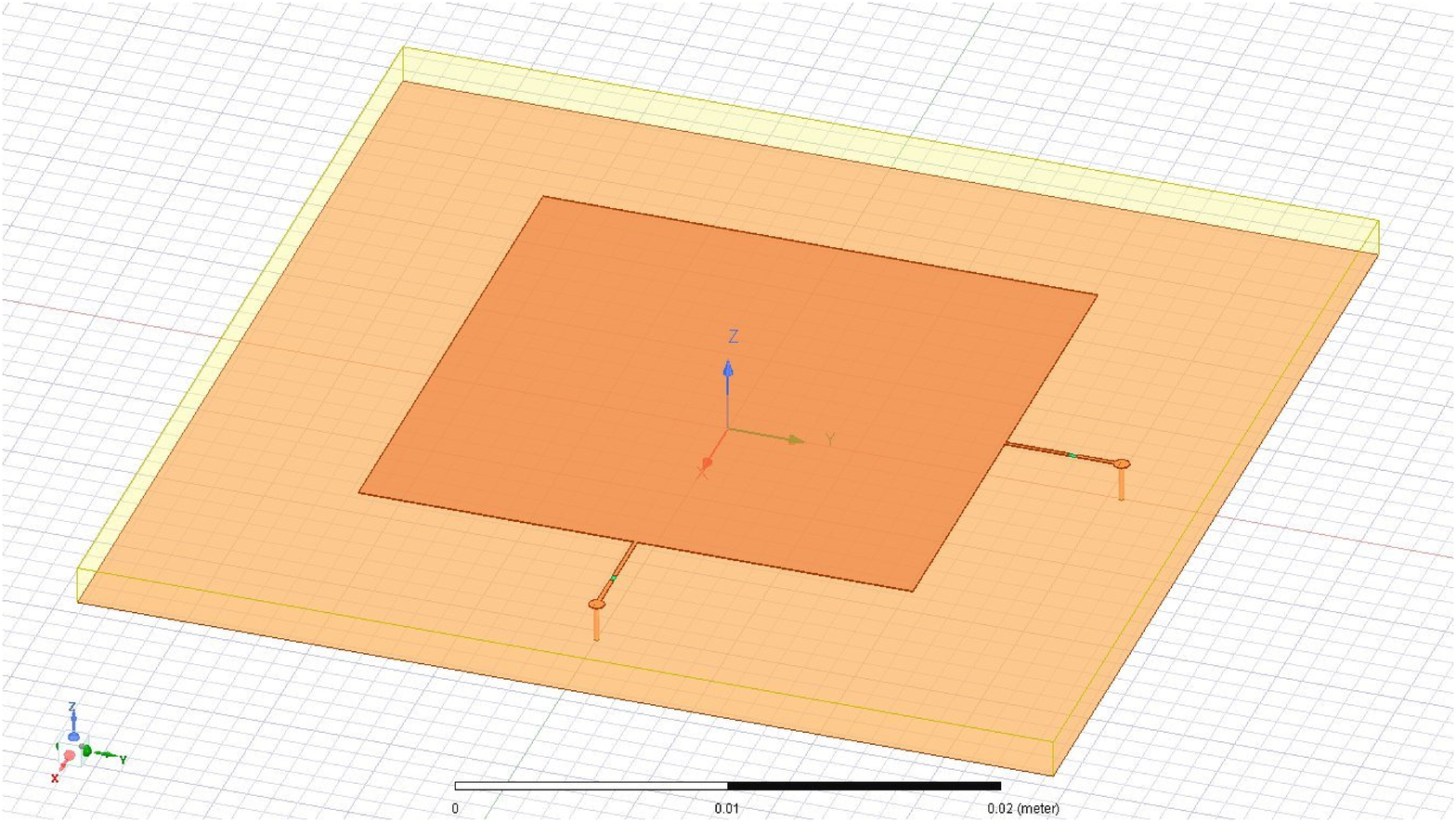}\tabularnewline
(\emph{b})\tabularnewline
\includegraphics[%
  clip,
  width=0.45\columnwidth,
  keepaspectratio]{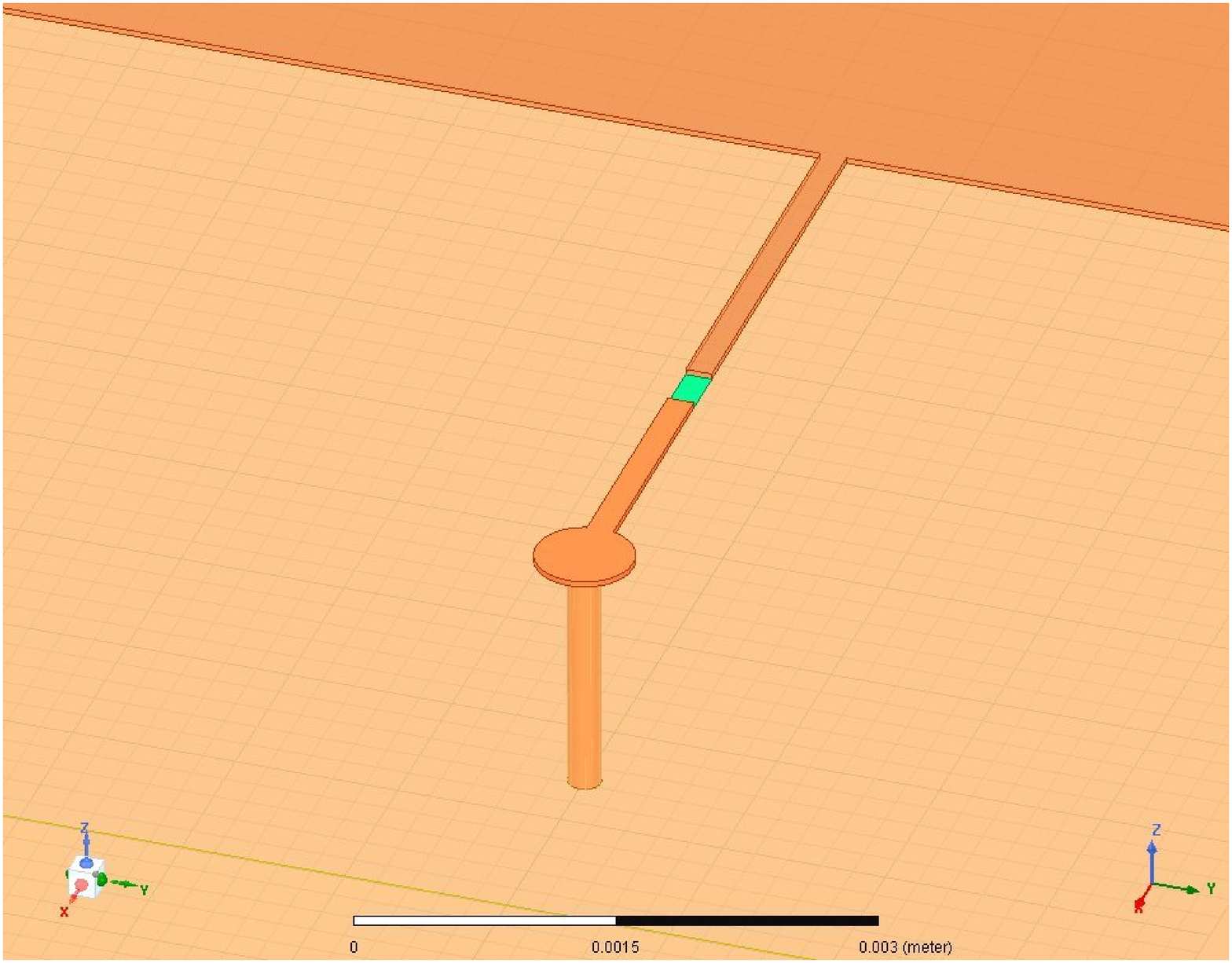}\tabularnewline
(\emph{c})\tabularnewline
\end{tabular}\end{center}

\begin{center}\textbf{Fig. 2 - G. Oliveri et} \textbf{\emph{al.}}\textbf{,}
{}``Multi-Scale Single-Bit \emph{RP-EMS} Synthesis for ...''\end{center}

\newpage
\begin{center}~\vfill\end{center}

\begin{center}\begin{tabular}{c}
\includegraphics[%
  clip,
  width=0.90\columnwidth,
  keepaspectratio]{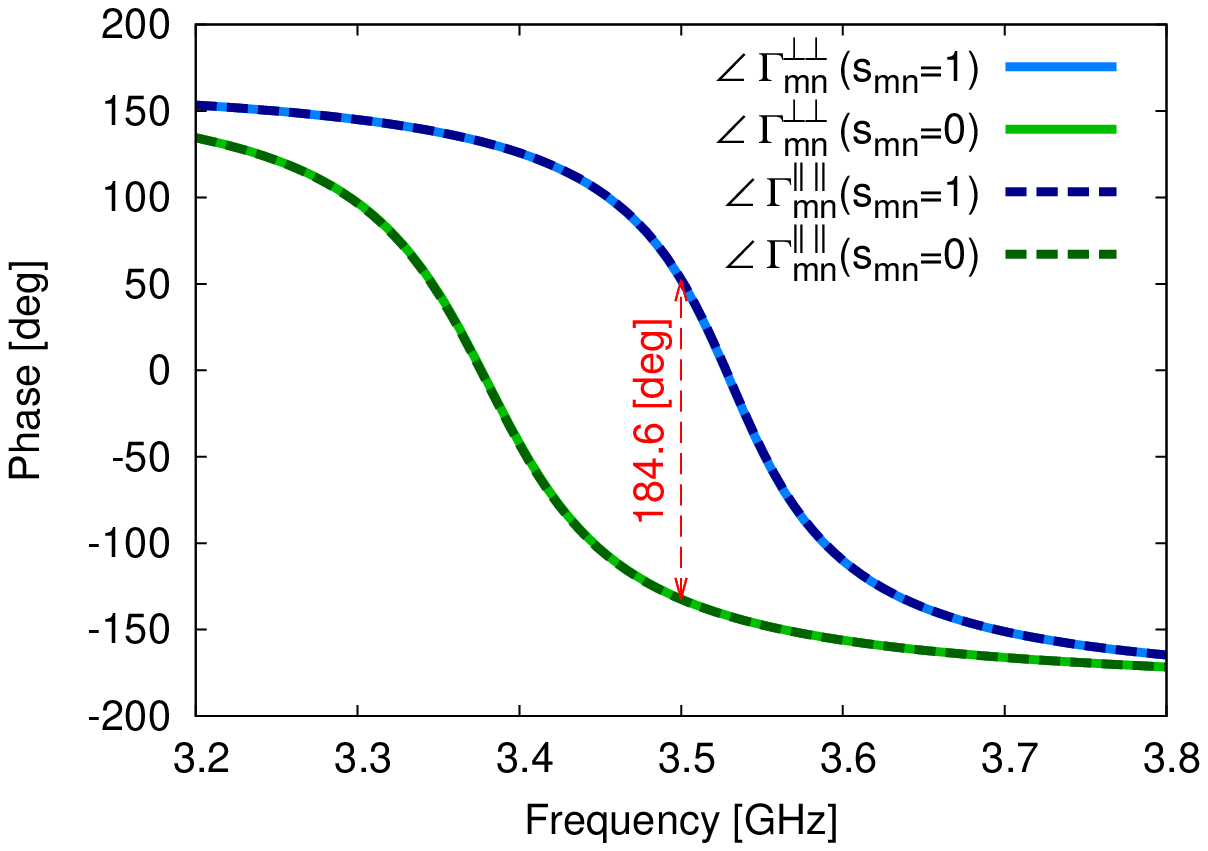}\tabularnewline
(\emph{a})\tabularnewline
\includegraphics[%
  clip,
  width=0.90\columnwidth,
  keepaspectratio]{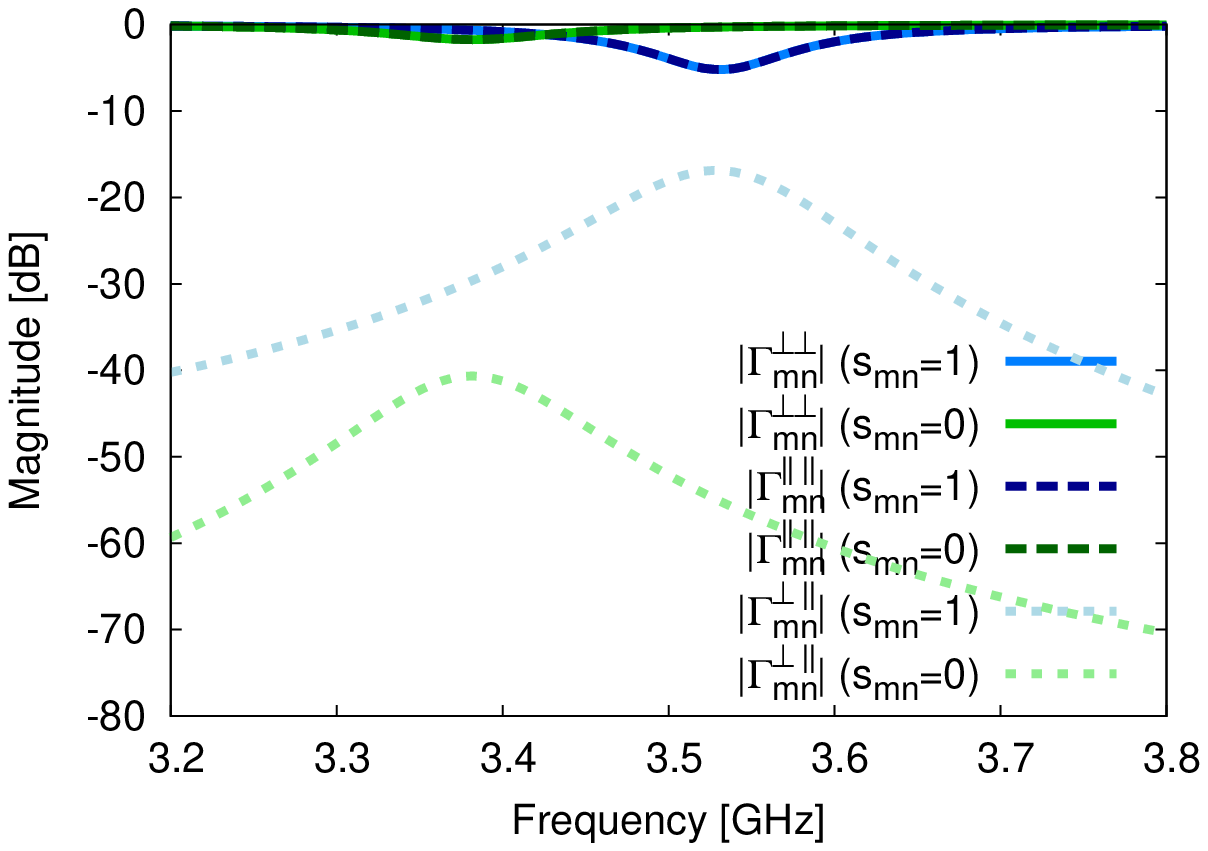}\tabularnewline
(\emph{b})\tabularnewline
\end{tabular}\end{center}

\begin{center}~\vfill\end{center}

\begin{center}\textbf{Fig. 3 - G. Oliveri et} \textbf{\emph{al.}}\textbf{,}
{}``Multi-Scale Single-Bit \emph{RP-EMS} Synthesis for ...''\end{center}

\newpage
\begin{center}~\vfill\end{center}

\begin{center}\begin{tabular}{cc}
\multicolumn{2}{c}{\includegraphics[%
  clip,
  width=0.95\columnwidth,
  keepaspectratio]{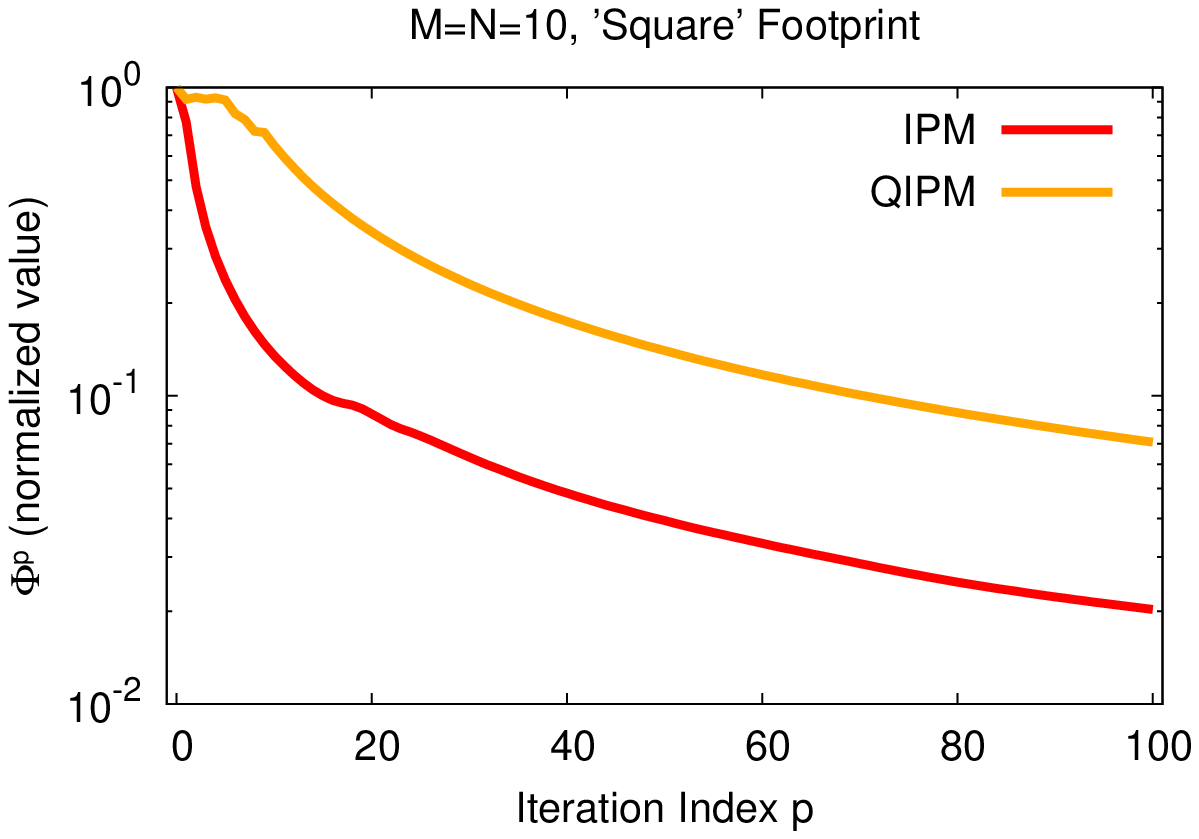}}\tabularnewline
\multicolumn{2}{c}{(\emph{a})}\tabularnewline
\includegraphics[%
  clip,
  width=0.48\columnwidth,
  keepaspectratio]{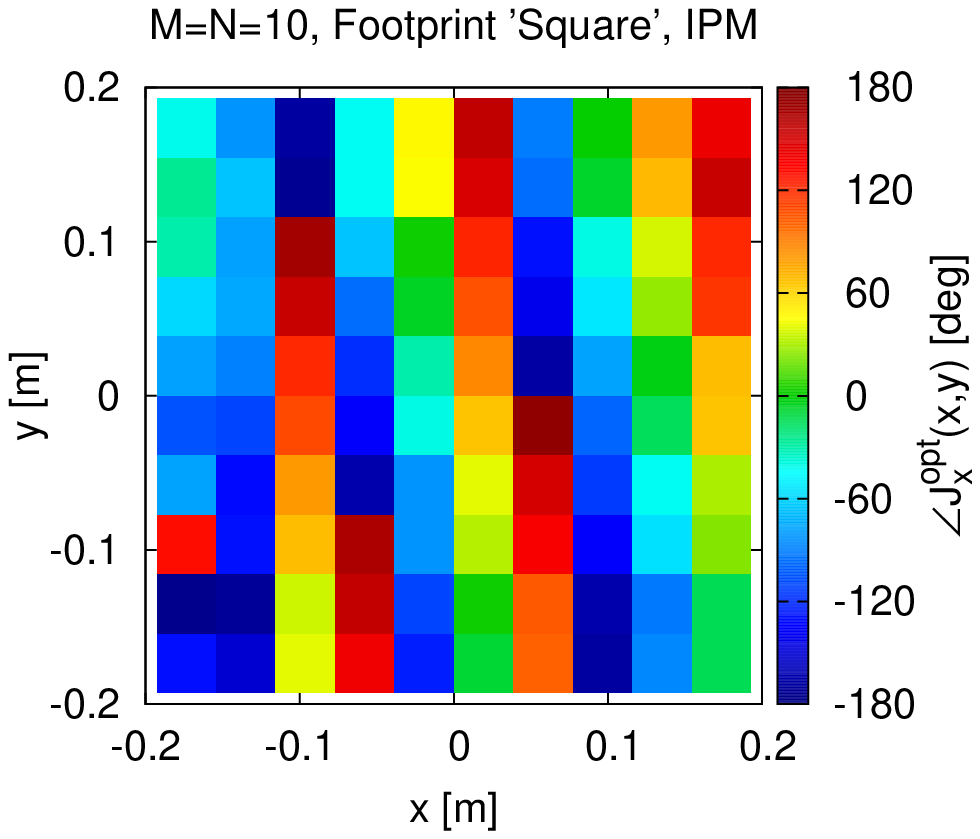}&
\includegraphics[%
  clip,
  width=0.48\columnwidth,
  keepaspectratio]{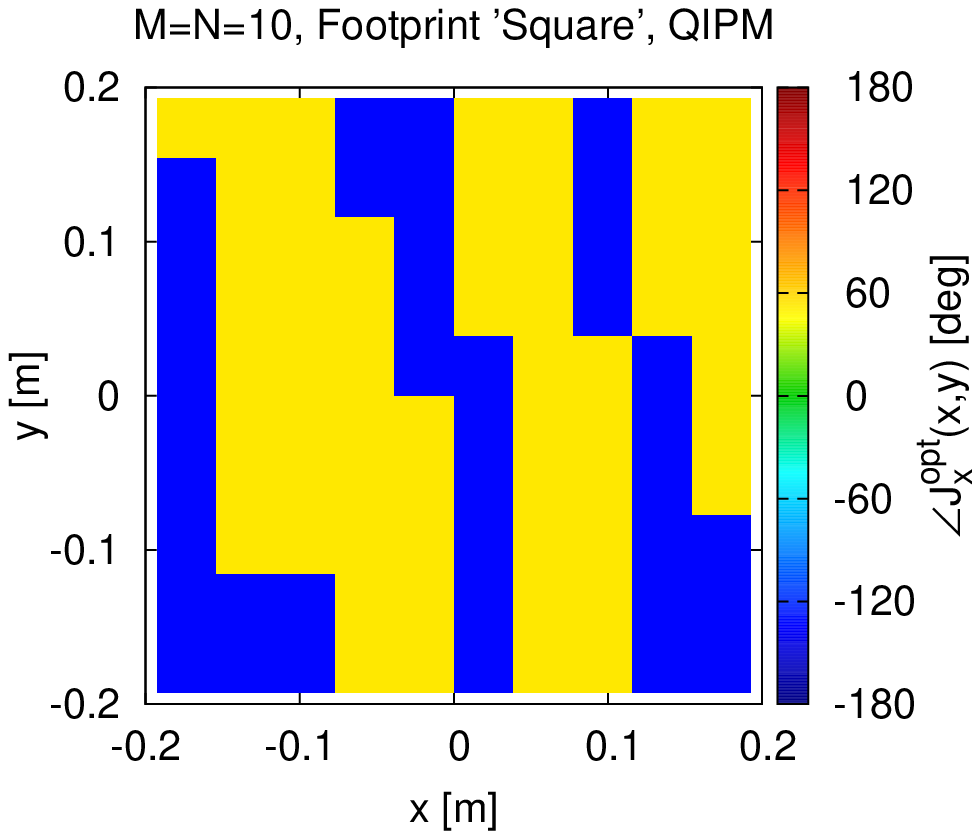}\tabularnewline
(\emph{b})&
(\emph{c})\tabularnewline
\end{tabular}\end{center}

\begin{center}~\vfill\end{center}

\begin{center}\textbf{Fig. 4 - G. Oliveri et} \textbf{\emph{al.}}\textbf{,}
{}``Multi-Scale Single-Bit \emph{RP-EMS} Synthesis for ...''\end{center}

\newpage
\begin{center}\begin{tabular}{cc}
\multicolumn{2}{c}{\includegraphics[%
  clip,
  width=0.50\columnwidth,
  keepaspectratio]{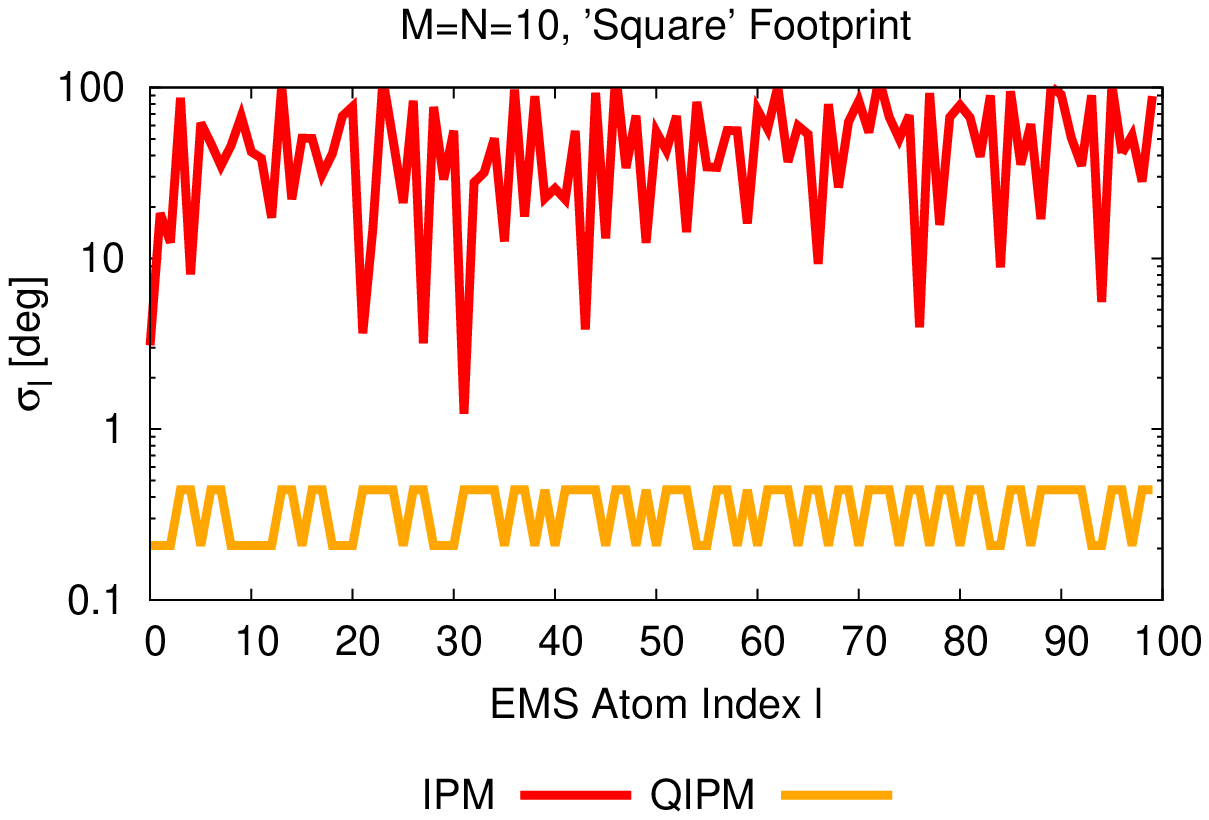}}\tabularnewline
\multicolumn{2}{c}{(\emph{a})}\tabularnewline
\multicolumn{2}{c}{\includegraphics[%
  clip,
  width=0.45\columnwidth,
  keepaspectratio]{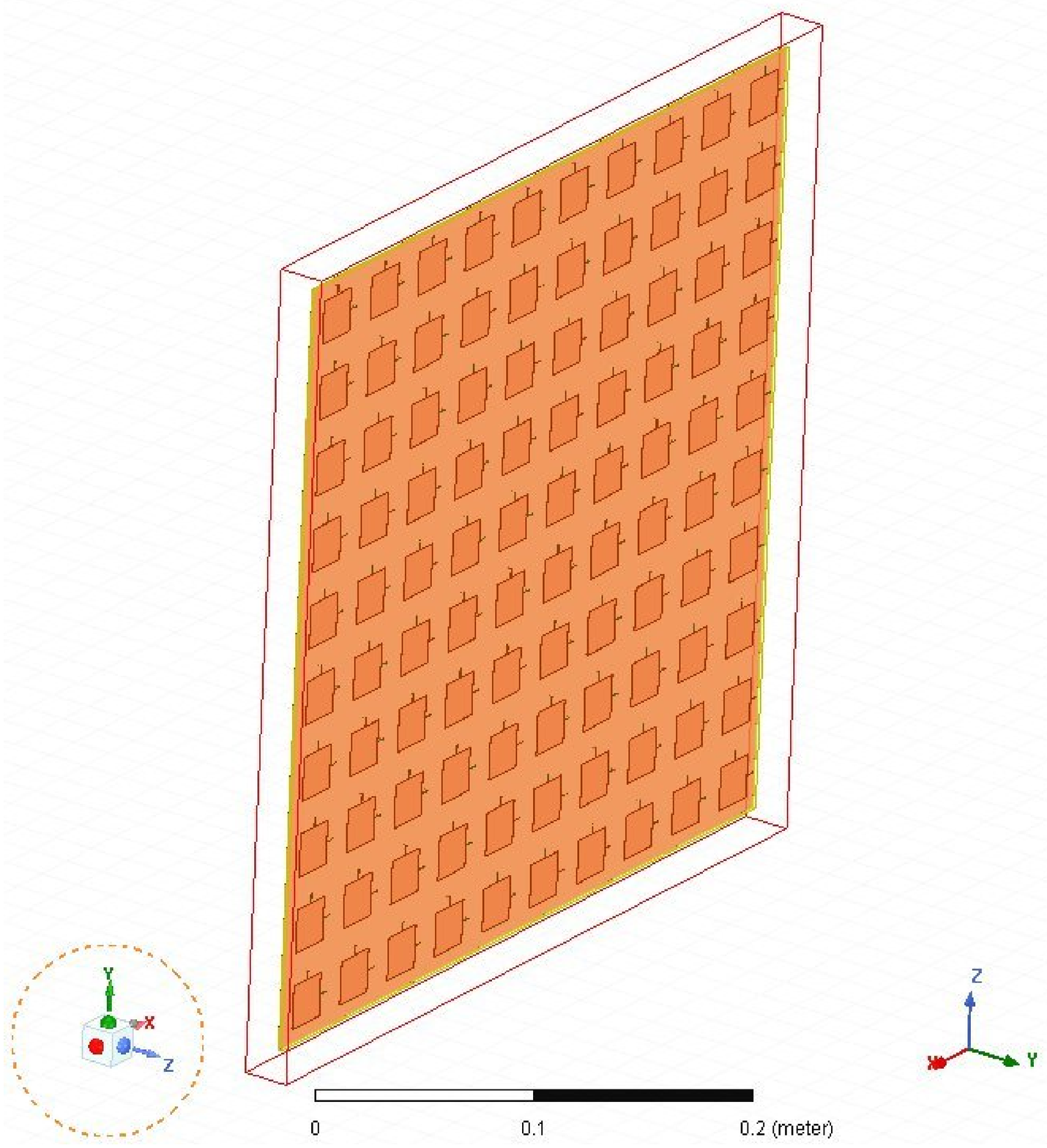}}\tabularnewline
\multicolumn{2}{c}{(\emph{b})}\tabularnewline
\includegraphics[%
  clip,
  width=0.48\columnwidth,
  keepaspectratio]{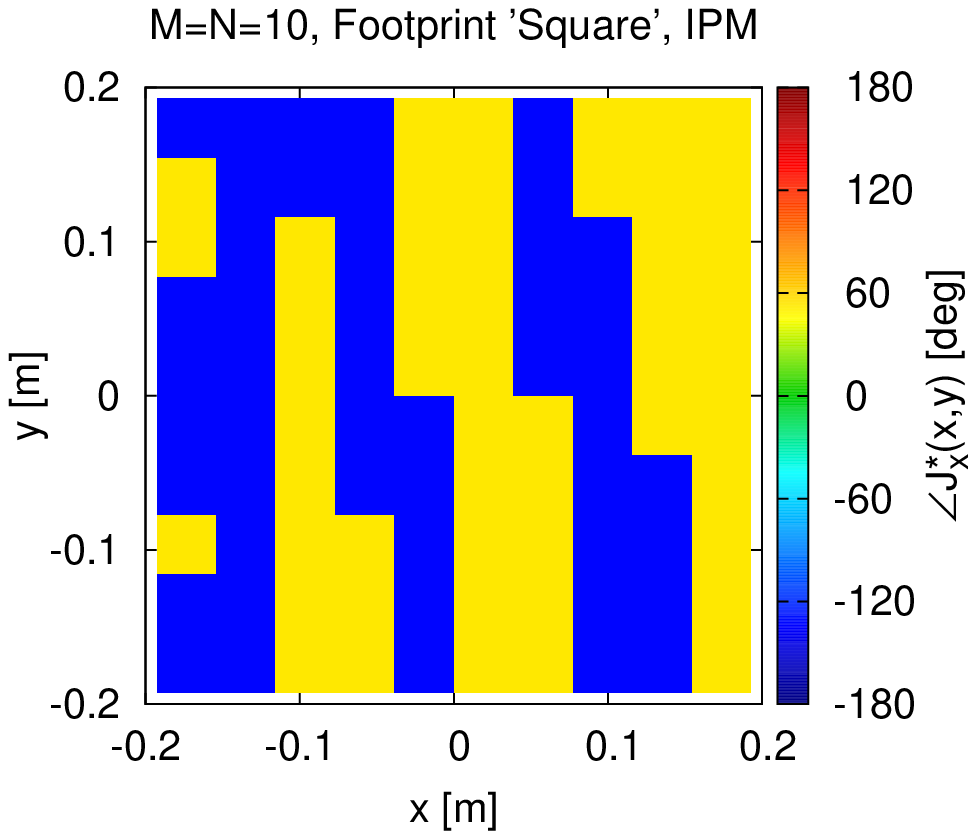}&
\includegraphics[%
  clip,
  width=0.48\columnwidth,
  keepaspectratio]{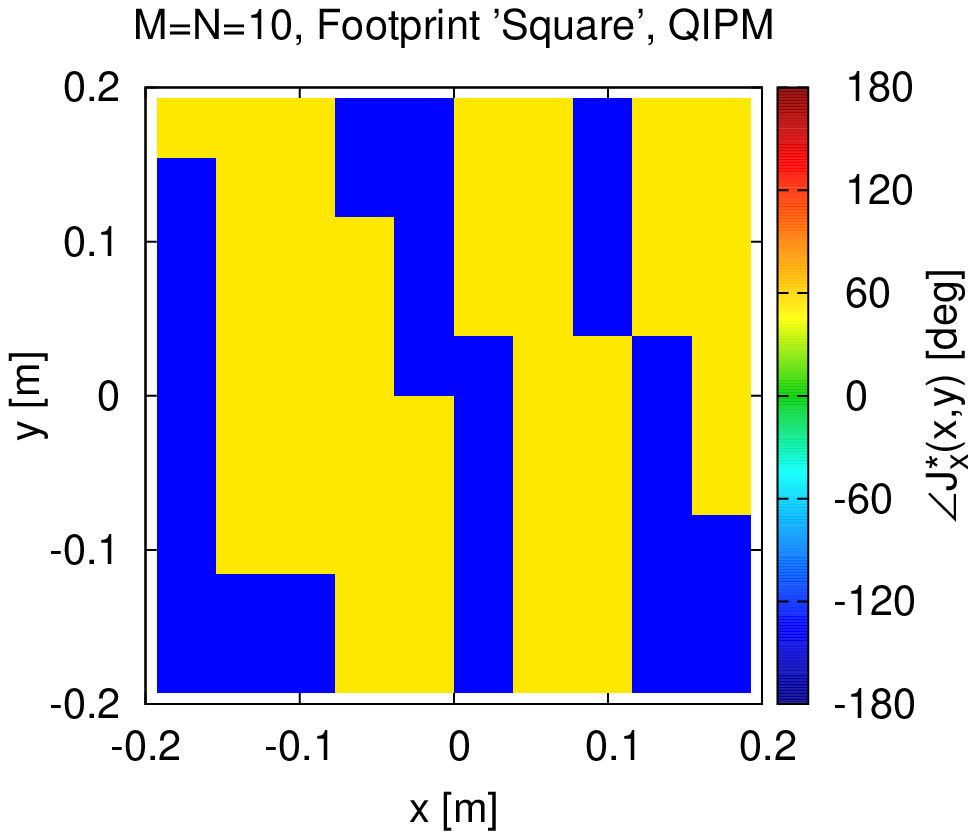}\tabularnewline
(\emph{c})&
(\emph{d})\tabularnewline
\end{tabular}\end{center}

\begin{center}\textbf{Fig. 5 - G. Oliveri et} \textbf{\emph{al.}}\textbf{,}
{}``Multi-Scale Single-Bit \emph{RP-EMS} Synthesis for ...''\end{center}

\newpage
\begin{center}~\vfill\end{center}

\begin{center}\begin{tabular}{cc}
\includegraphics[%
  clip,
  width=0.43\columnwidth,
  keepaspectratio]{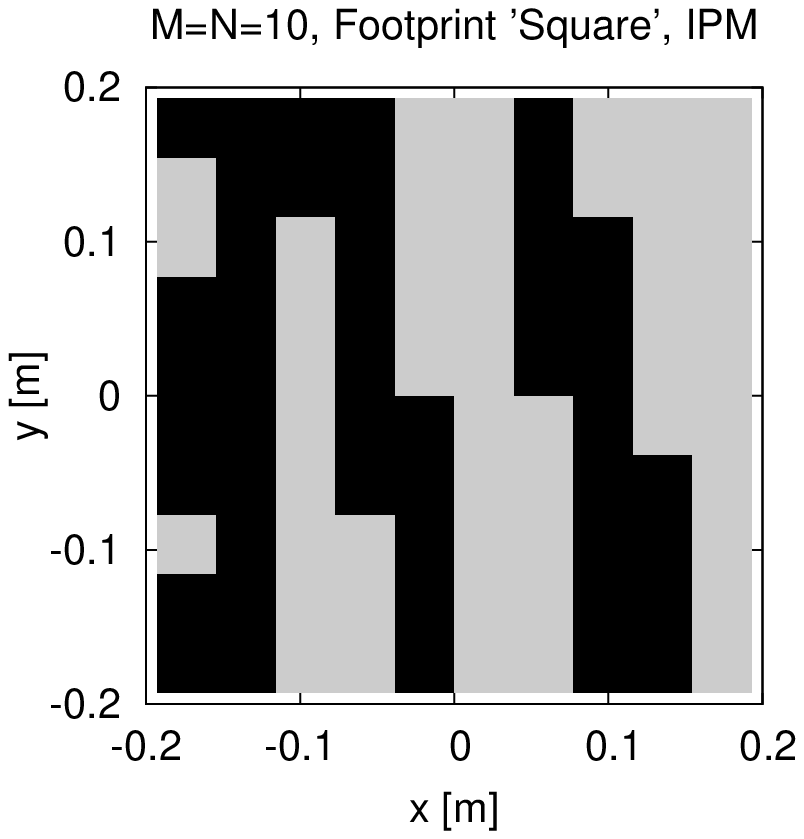}&
\includegraphics[%
  clip,
  width=0.43\columnwidth,
  keepaspectratio]{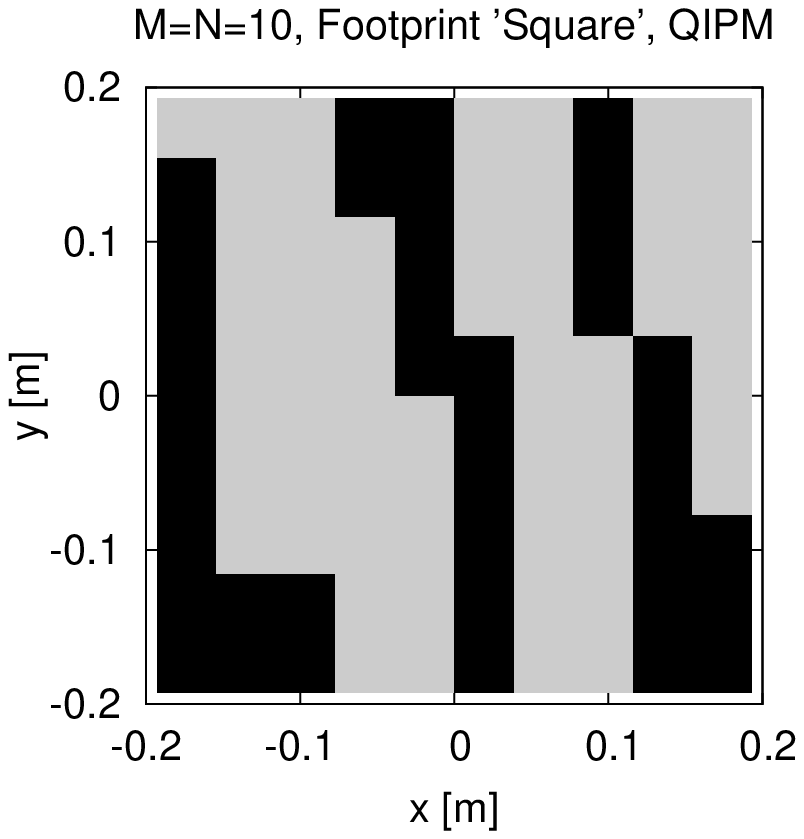}\tabularnewline
(\emph{a})&
(\emph{b})\tabularnewline
\includegraphics[%
  clip,
  width=0.48\columnwidth,
  keepaspectratio]{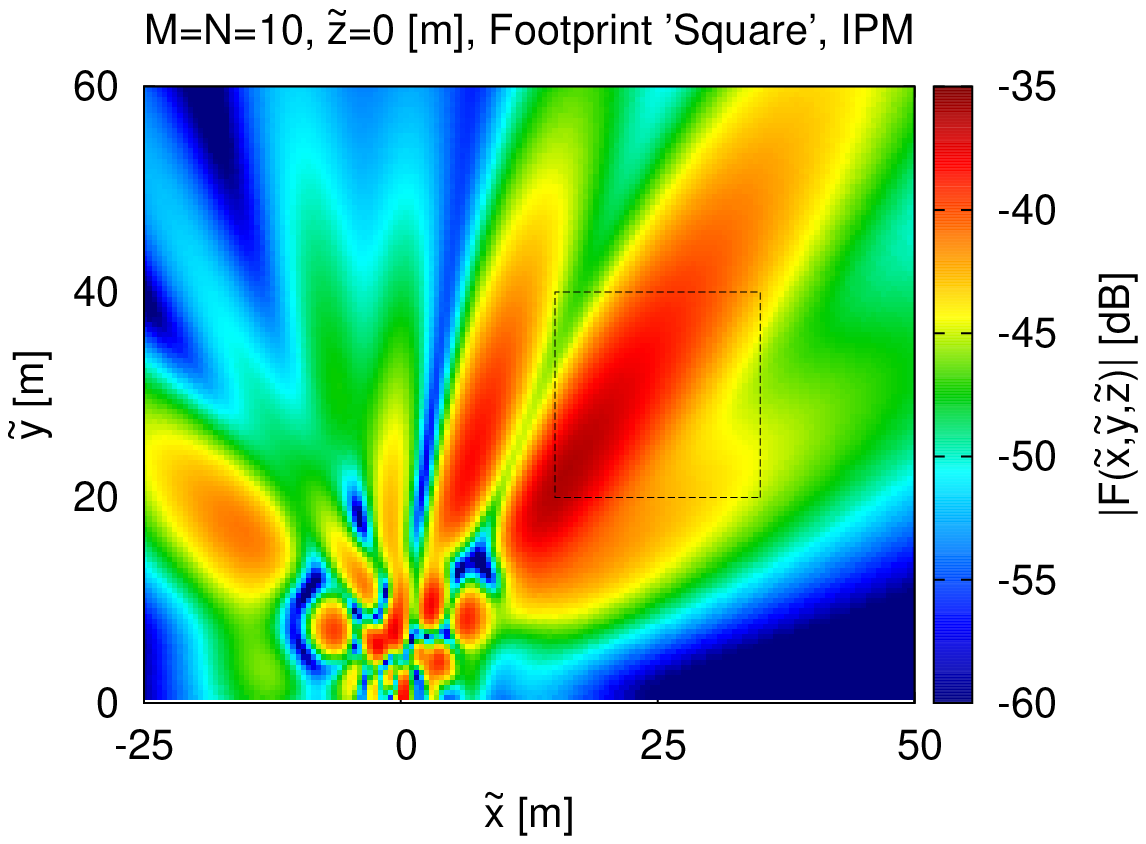}&
\includegraphics[%
  clip,
  width=0.48\columnwidth,
  keepaspectratio]{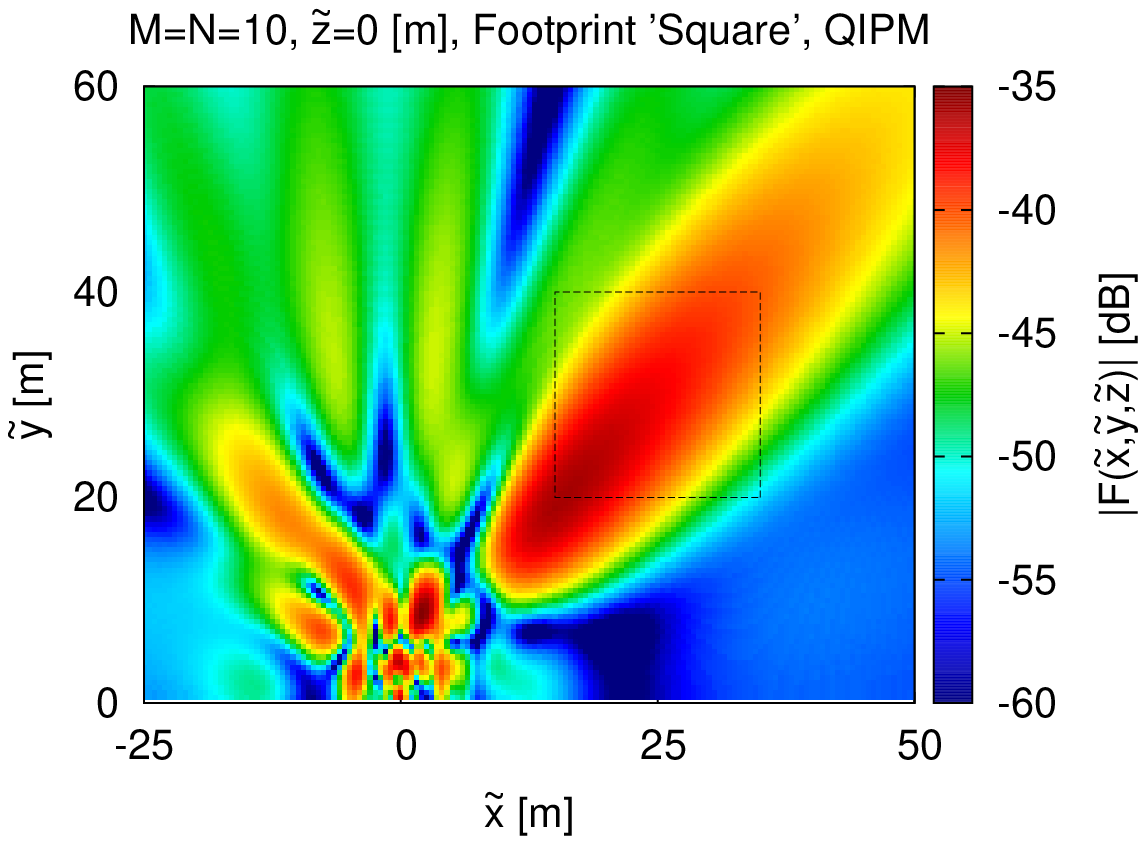}\tabularnewline
(\emph{c})&
(\emph{d})\tabularnewline
\includegraphics[%
  clip,
  width=0.48\columnwidth,
  keepaspectratio]{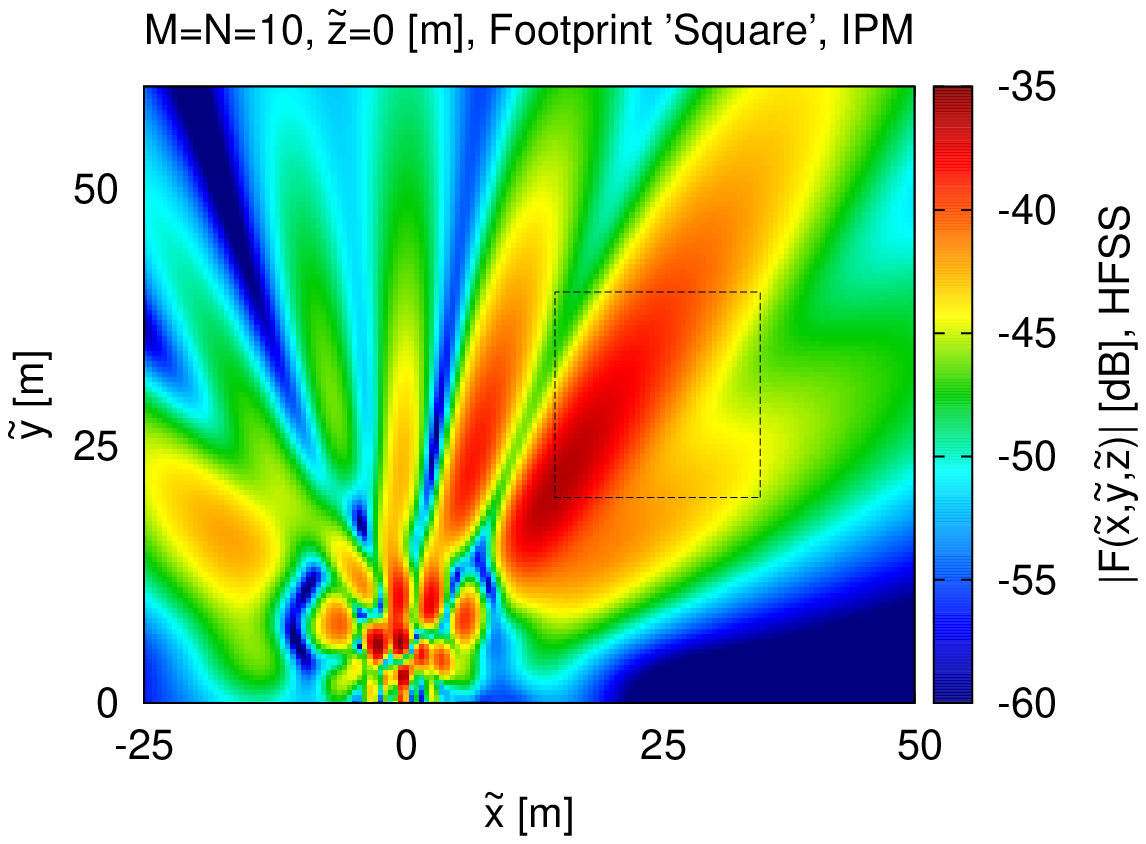}&
\includegraphics[%
  clip,
  width=0.48\columnwidth,
  keepaspectratio]{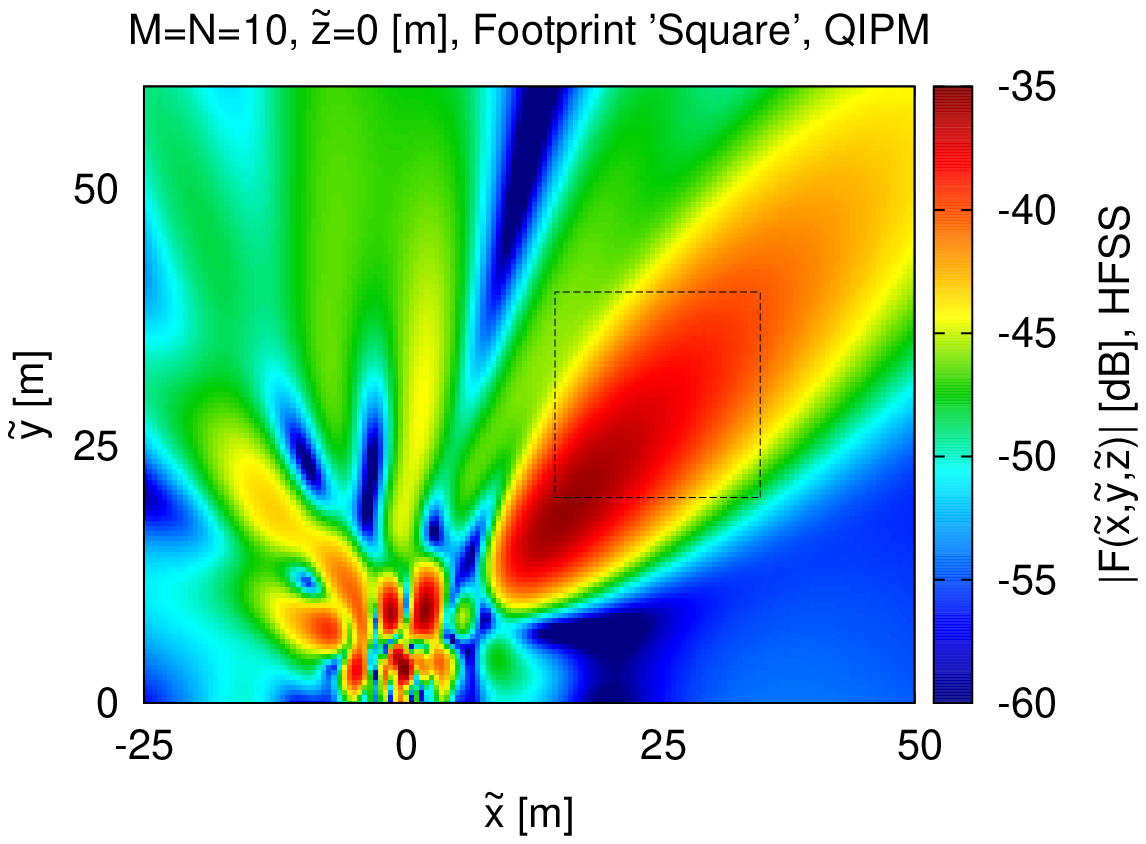}\tabularnewline
(\emph{e})&
(\emph{f})\tabularnewline
\end{tabular}\end{center}

\begin{center}~\vfill\end{center}

\begin{center}\textbf{Fig. 6 - G. Oliveri et} \textbf{\emph{al.}}\textbf{,}
{}``Multi-Scale Single-Bit \emph{RP-EMS} Synthesis for ...''\end{center}

\newpage
\begin{center}\begin{tabular}{cc}
\multicolumn{2}{c}{\includegraphics[%
  clip,
  width=0.65\columnwidth,
  keepaspectratio]{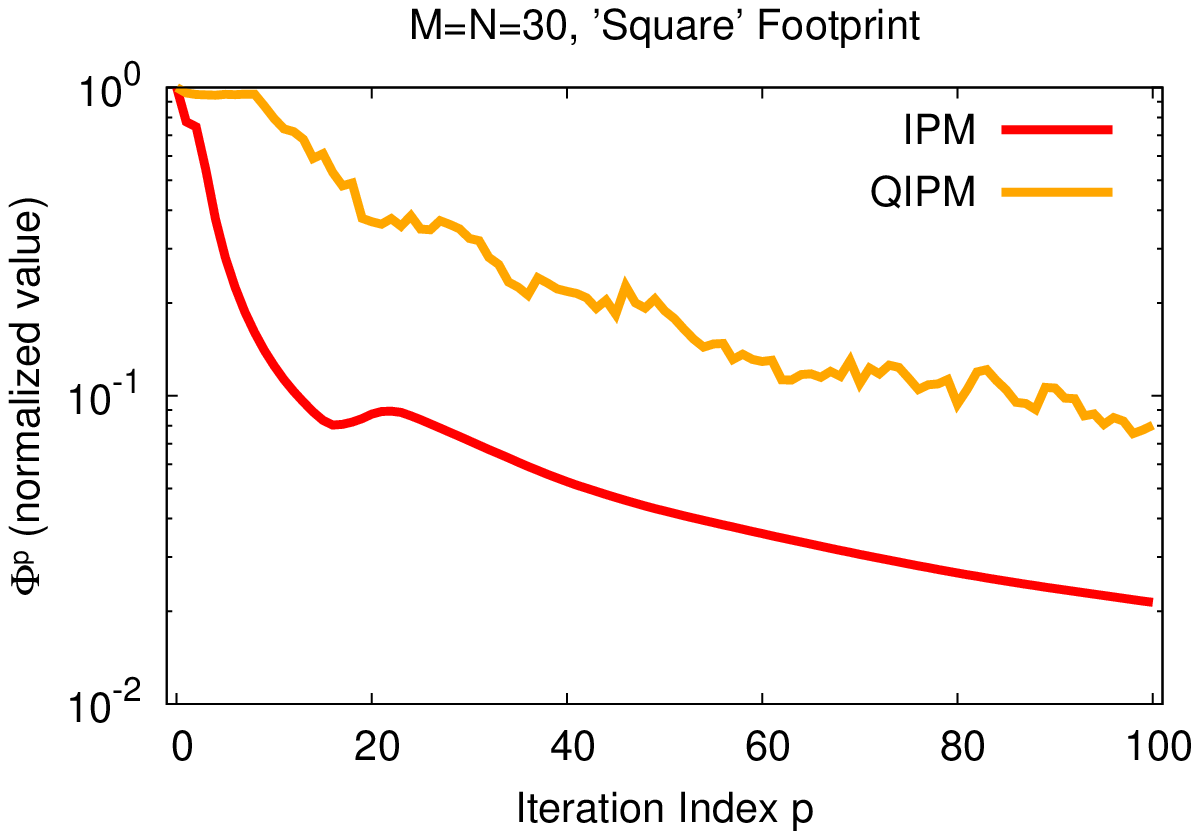}}\tabularnewline
\multicolumn{2}{c}{(\emph{a})}\tabularnewline
\includegraphics[%
  clip,
  width=0.48\columnwidth,
  keepaspectratio]{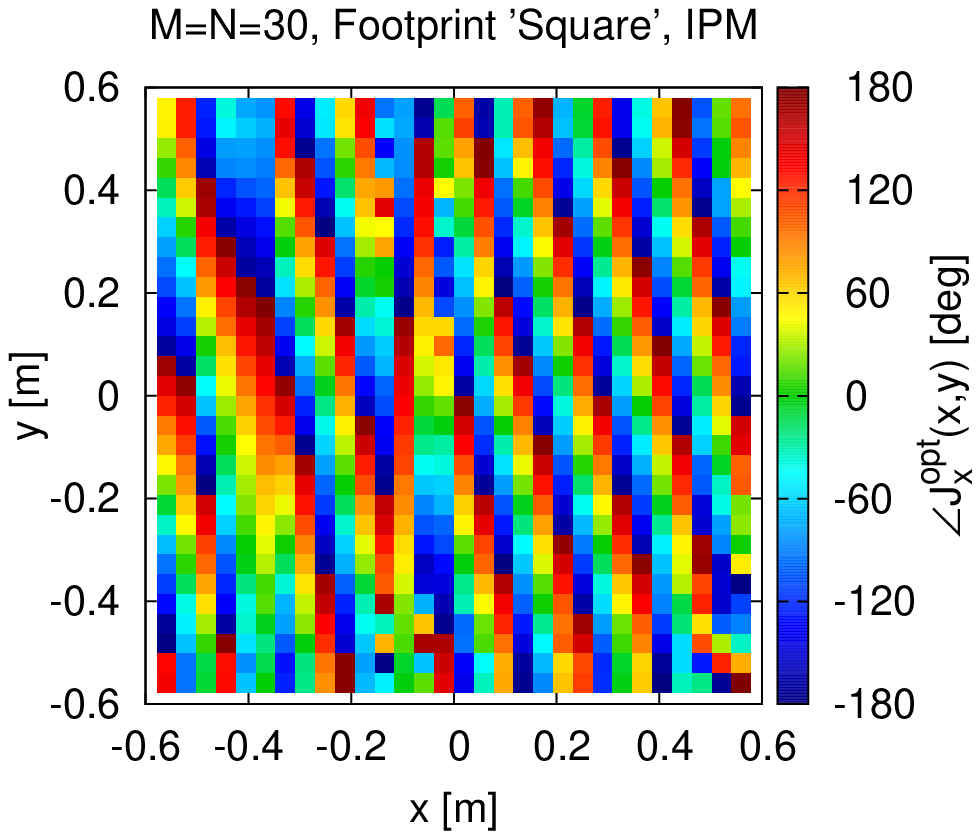}&
\includegraphics[%
  clip,
  width=0.48\columnwidth,
  keepaspectratio]{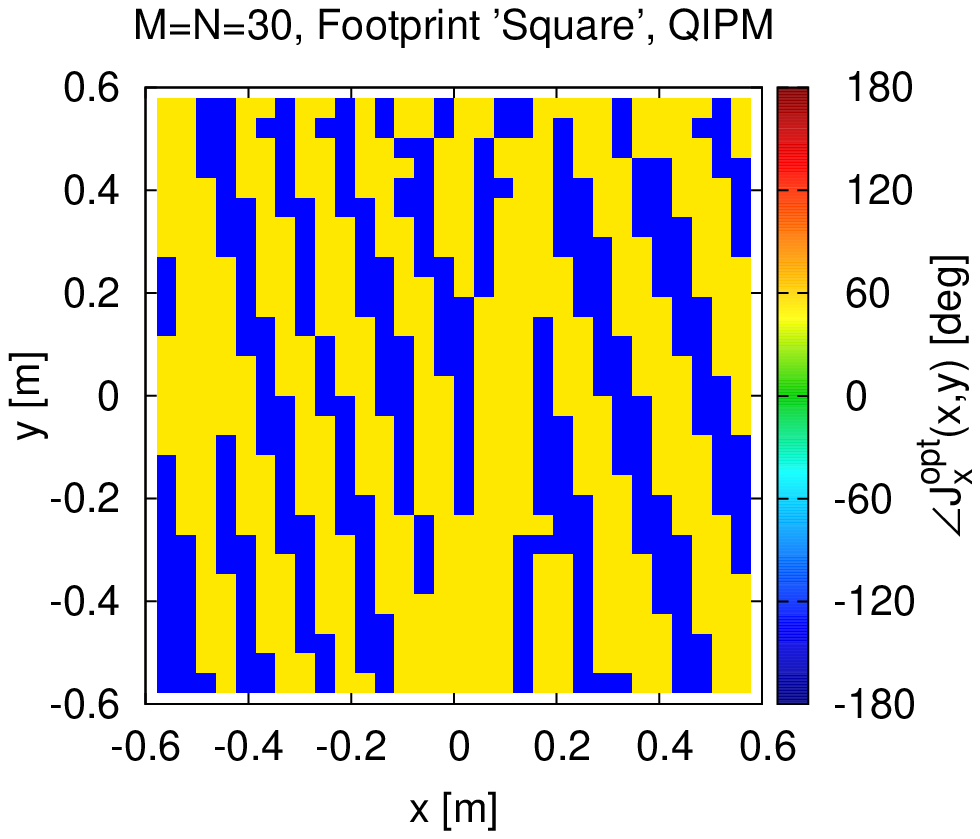}\tabularnewline
(\emph{b})&
(\emph{c})\tabularnewline
\includegraphics[%
  clip,
  width=0.48\columnwidth,
  keepaspectratio]{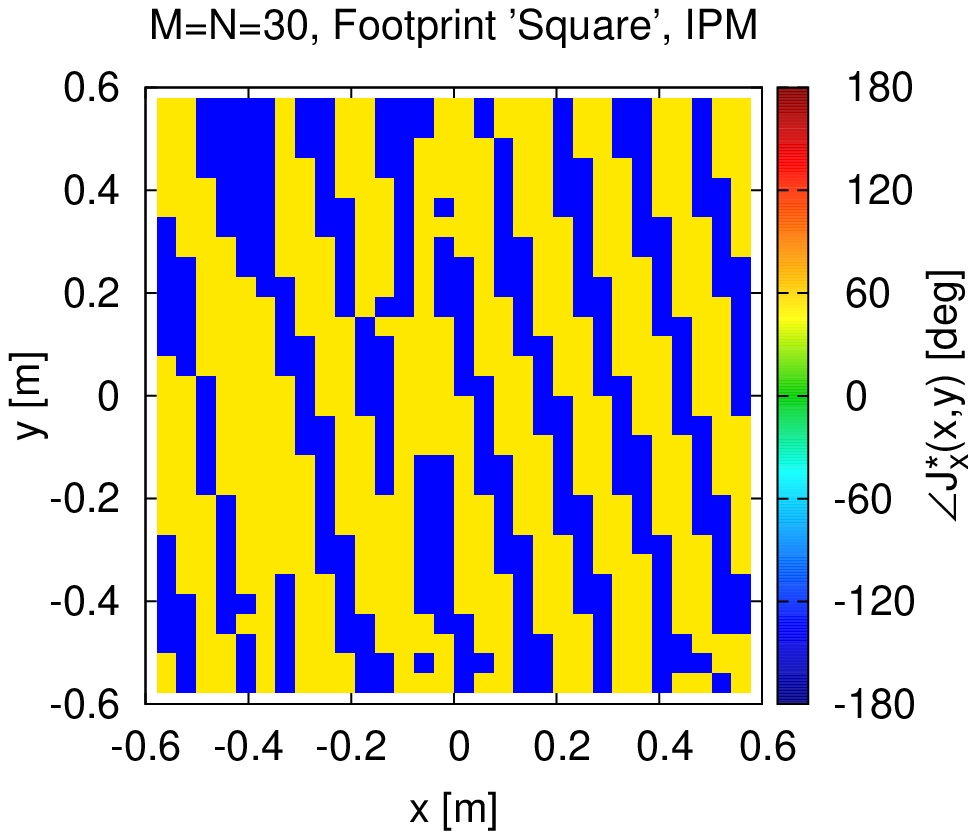}&
\includegraphics[%
  clip,
  width=0.48\columnwidth,
  keepaspectratio]{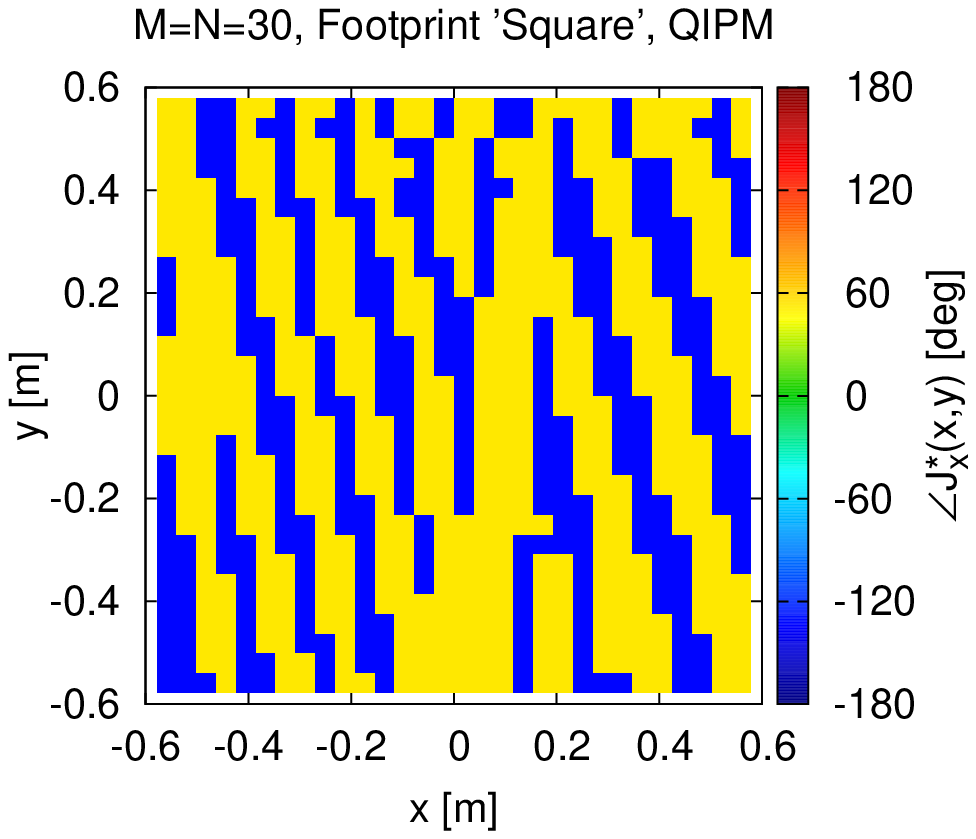}\tabularnewline
(\emph{d})&
(\emph{e})\tabularnewline
\end{tabular}\end{center}

\begin{center}\textbf{Fig. 7 - G. Oliveri et} \textbf{\emph{al.}}\textbf{,}
{}``Multi-Scale Single-Bit \emph{RP-EMS} Synthesis for ...''\end{center}

\newpage
\begin{center}\begin{tabular}{cc}
\multicolumn{2}{c}{\includegraphics[%
  clip,
  width=0.75\columnwidth,
  keepaspectratio]{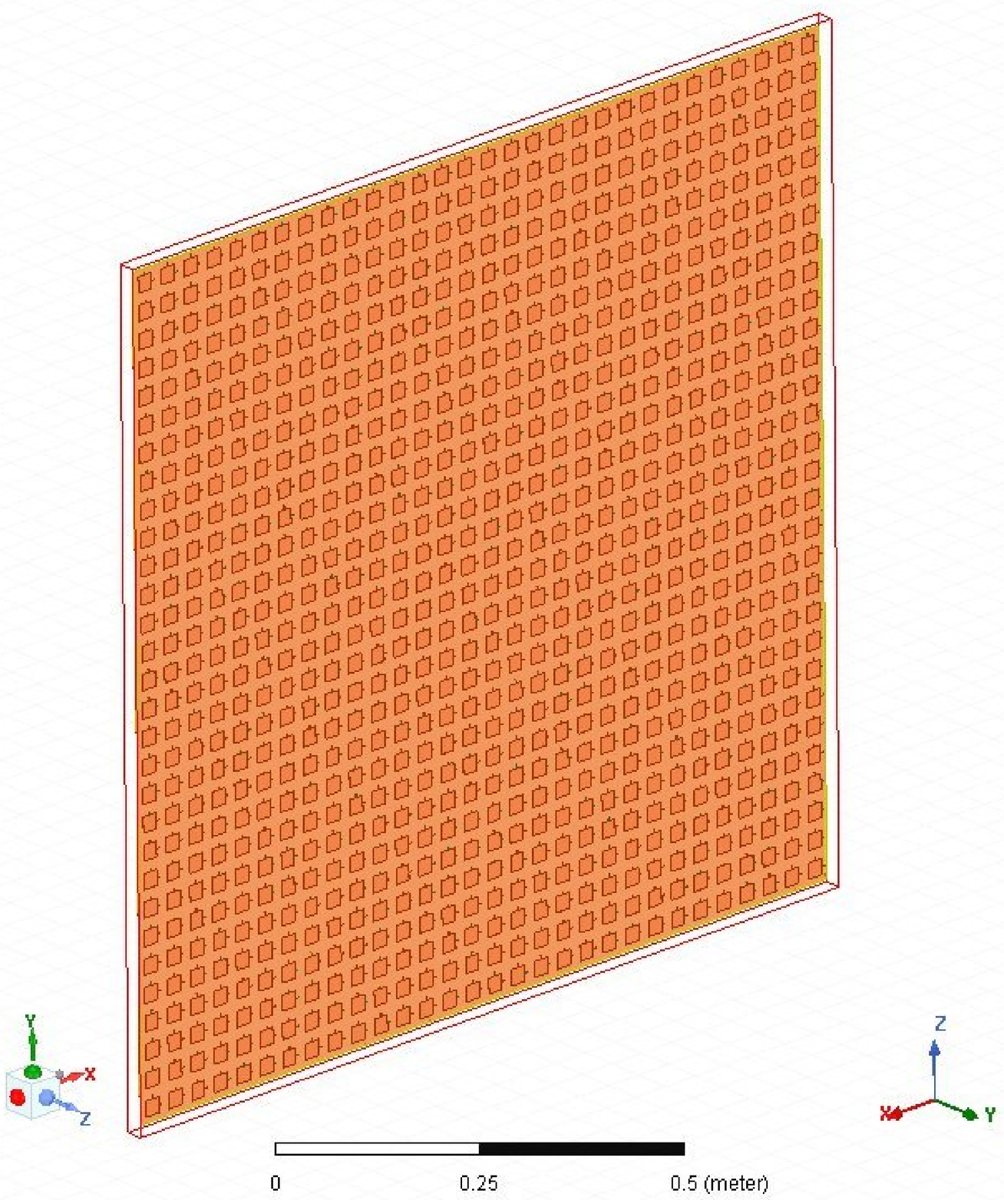}}\tabularnewline
\multicolumn{2}{c}{(\emph{a})}\tabularnewline
\includegraphics[%
  clip,
  width=0.48\columnwidth,
  keepaspectratio]{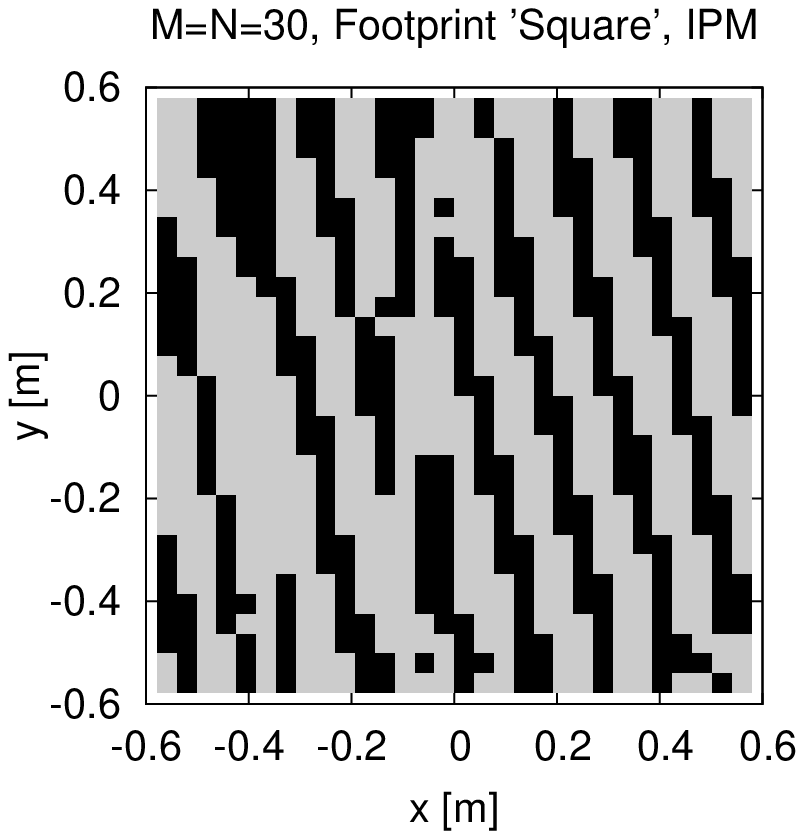}&
\includegraphics[%
  clip,
  width=0.48\columnwidth,
  keepaspectratio]{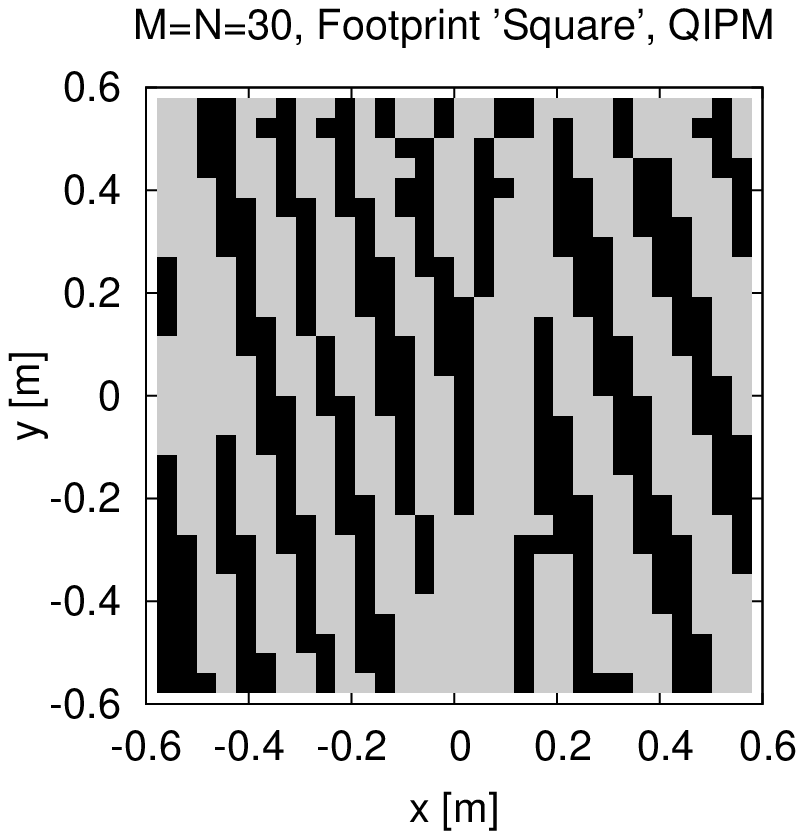}\tabularnewline
(\emph{b})&
(\emph{c})\tabularnewline
\end{tabular}\end{center}

\begin{center}\textbf{Fig. 8 - G. Oliveri et} \textbf{\emph{al.}}\textbf{,}
{}``Multi-Scale Single-Bit \emph{RP-EMS} Synthesis for ...''\end{center}

\newpage
\begin{center}~\vfill\end{center}

\begin{center}\begin{tabular}{c}
\includegraphics[%
  clip,
  width=0.85\columnwidth,
  keepaspectratio]{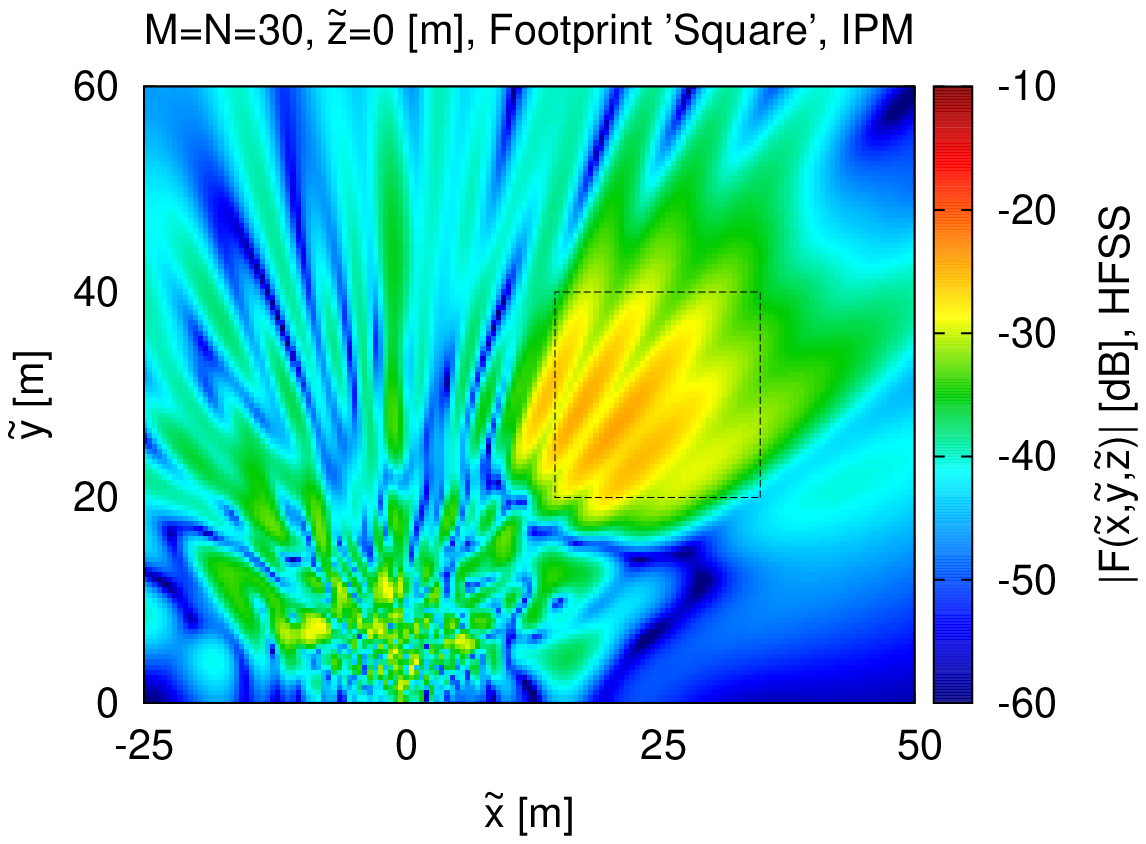}\tabularnewline
(\emph{a})\tabularnewline
\includegraphics[%
  clip,
  width=0.85\columnwidth,
  keepaspectratio]{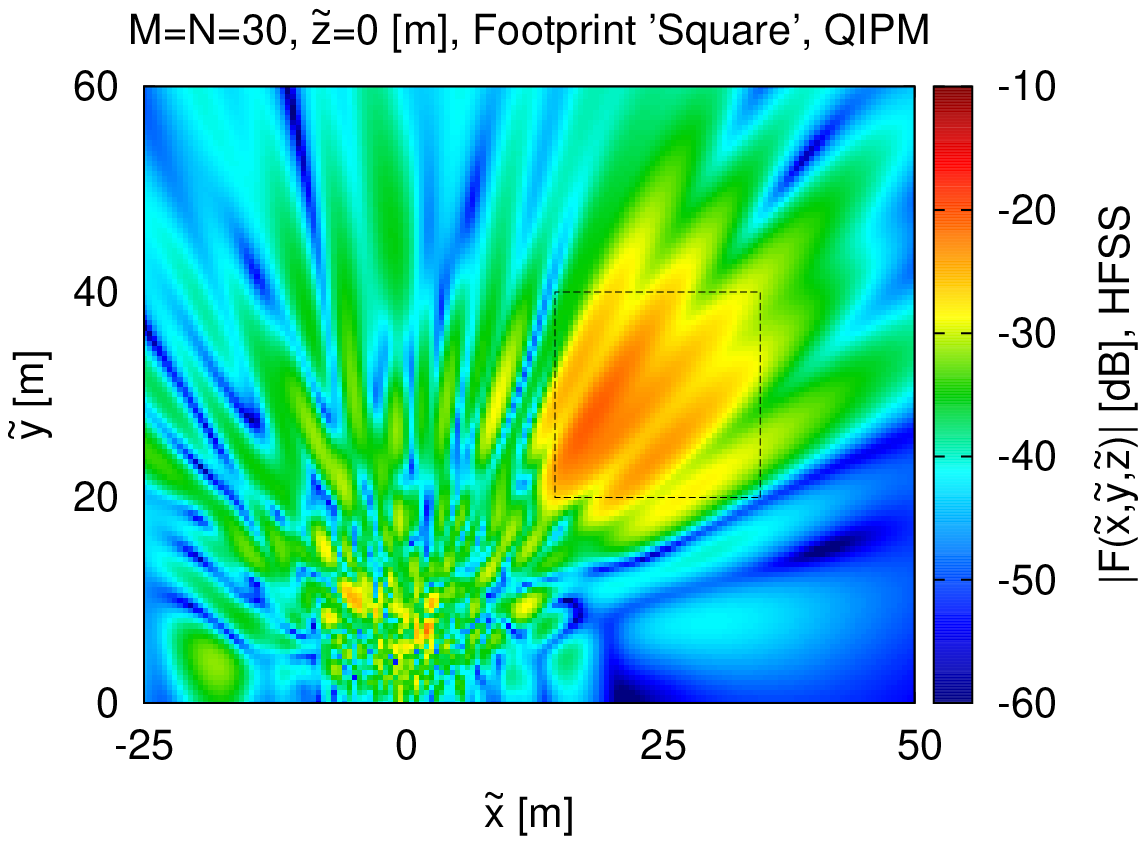}\tabularnewline
(\emph{b})\tabularnewline
\end{tabular}\end{center}

\begin{center}~\vfill\end{center}

\begin{center}\textbf{Fig. 9 - G. Oliveri et} \textbf{\emph{al.}}\textbf{,}
{}``Multi-Scale Single-Bit \emph{RP-EMS} Synthesis for ...''\end{center}

\newpage
\begin{center}\begin{tabular}{cc}
\multicolumn{2}{c}{\includegraphics[%
  clip,
  width=0.85\columnwidth,
  keepaspectratio]{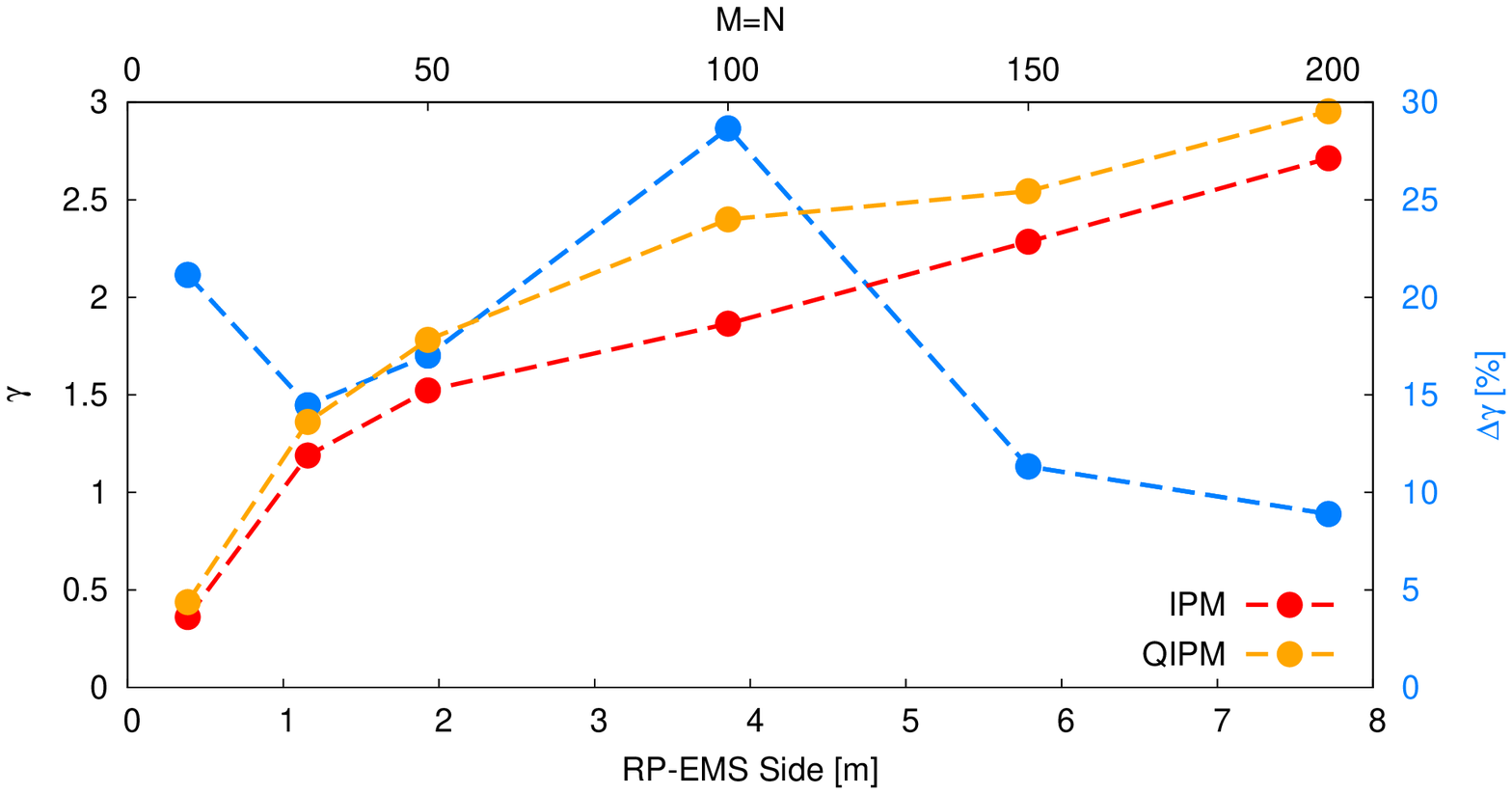}}\tabularnewline
\multicolumn{2}{c}{(\emph{a})}\tabularnewline
\includegraphics[%
  clip,
  width=0.40\columnwidth,
  keepaspectratio]{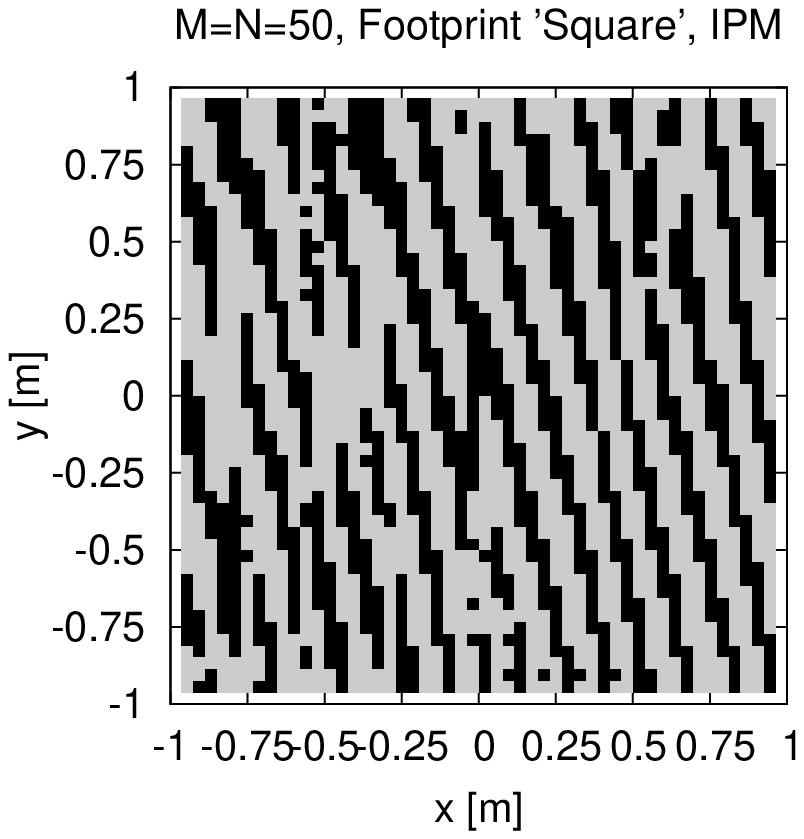}&
\includegraphics[%
  clip,
  width=0.40\columnwidth,
  keepaspectratio]{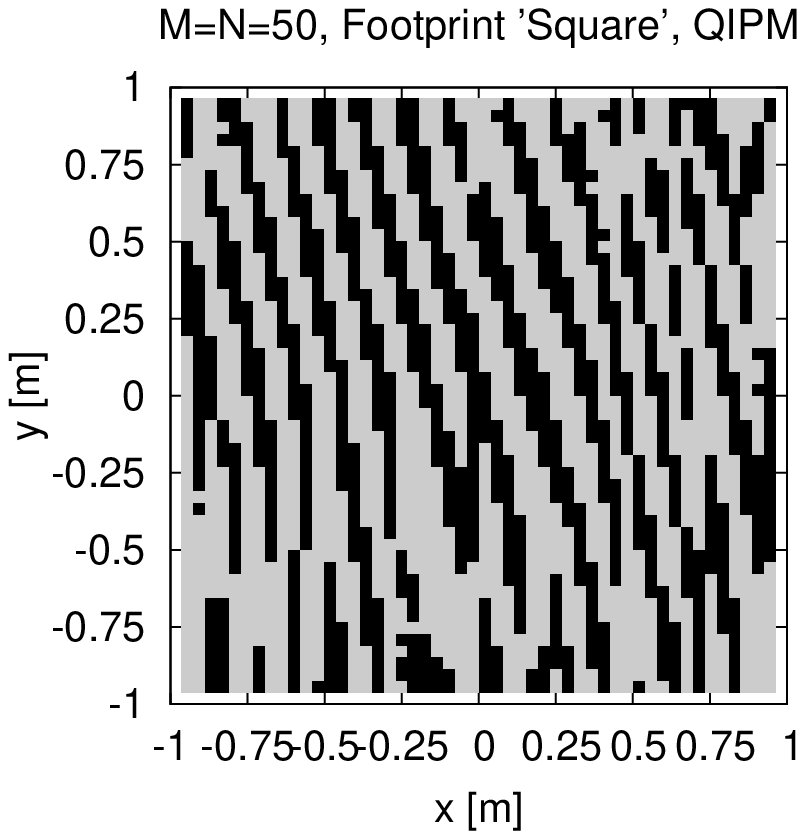}\tabularnewline
(\emph{b})&
(\emph{c})\tabularnewline
\includegraphics[%
  clip,
  width=0.40\columnwidth,
  keepaspectratio]{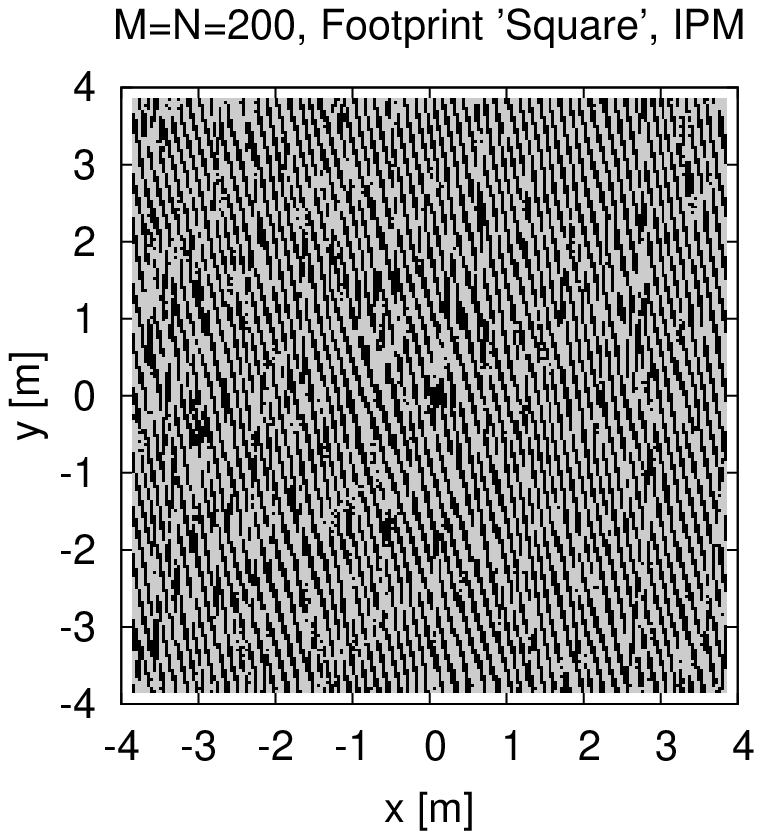}&
\includegraphics[%
  clip,
  width=0.40\columnwidth,
  keepaspectratio]{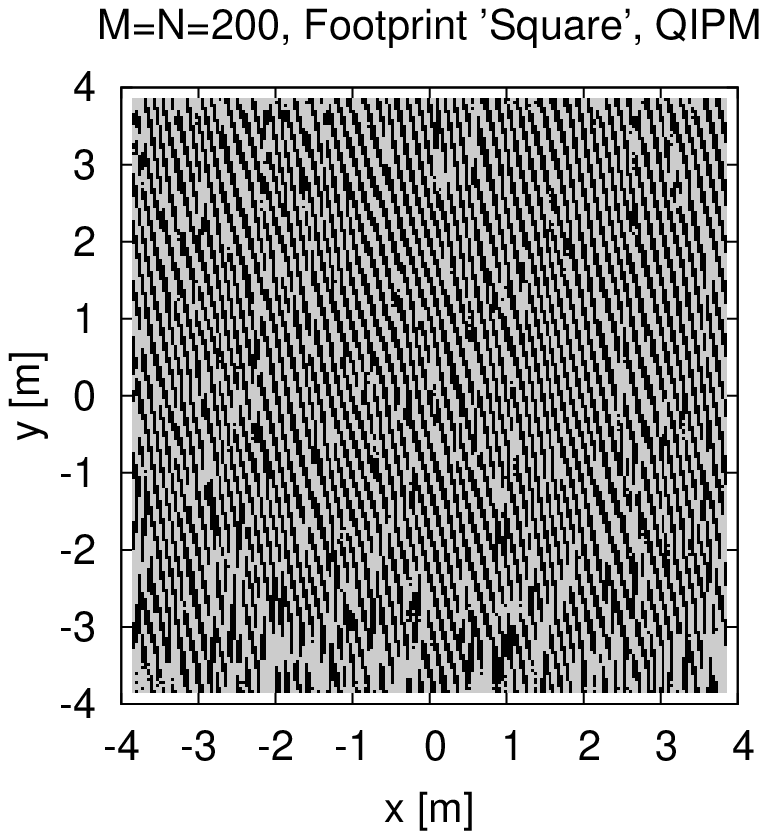}\tabularnewline
(\emph{d})&
(\emph{e})\tabularnewline
\end{tabular}\end{center}

\begin{center}\textbf{Fig. 10 - G. Oliveri et} \textbf{\emph{al.}}\textbf{,}
{}``Multi-Scale Single-Bit \emph{RP-EMS} Synthesis for ...''\end{center}

\newpage
\begin{center}~\vfill\end{center}

\begin{center}\begin{tabular}{cc}
\includegraphics[%
  clip,
  width=0.48\columnwidth,
  keepaspectratio]{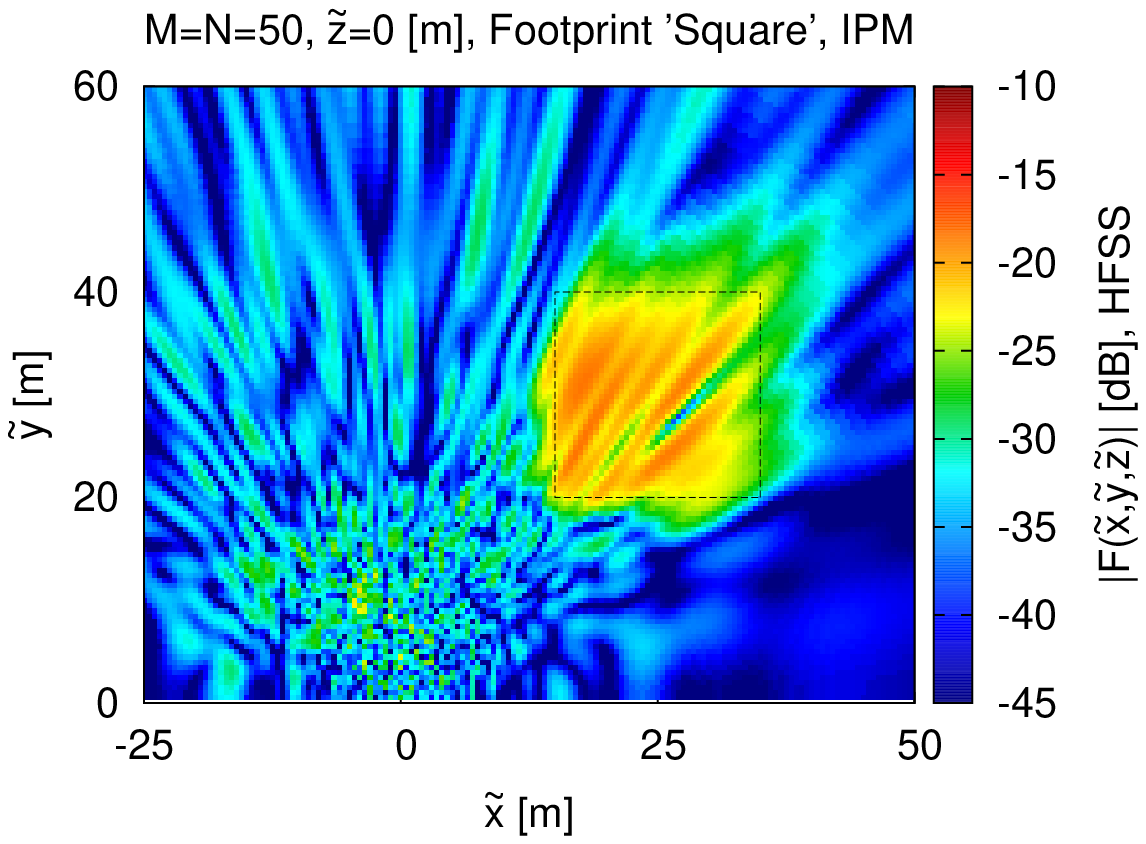}&
\includegraphics[%
  clip,
  width=0.48\columnwidth,
  keepaspectratio]{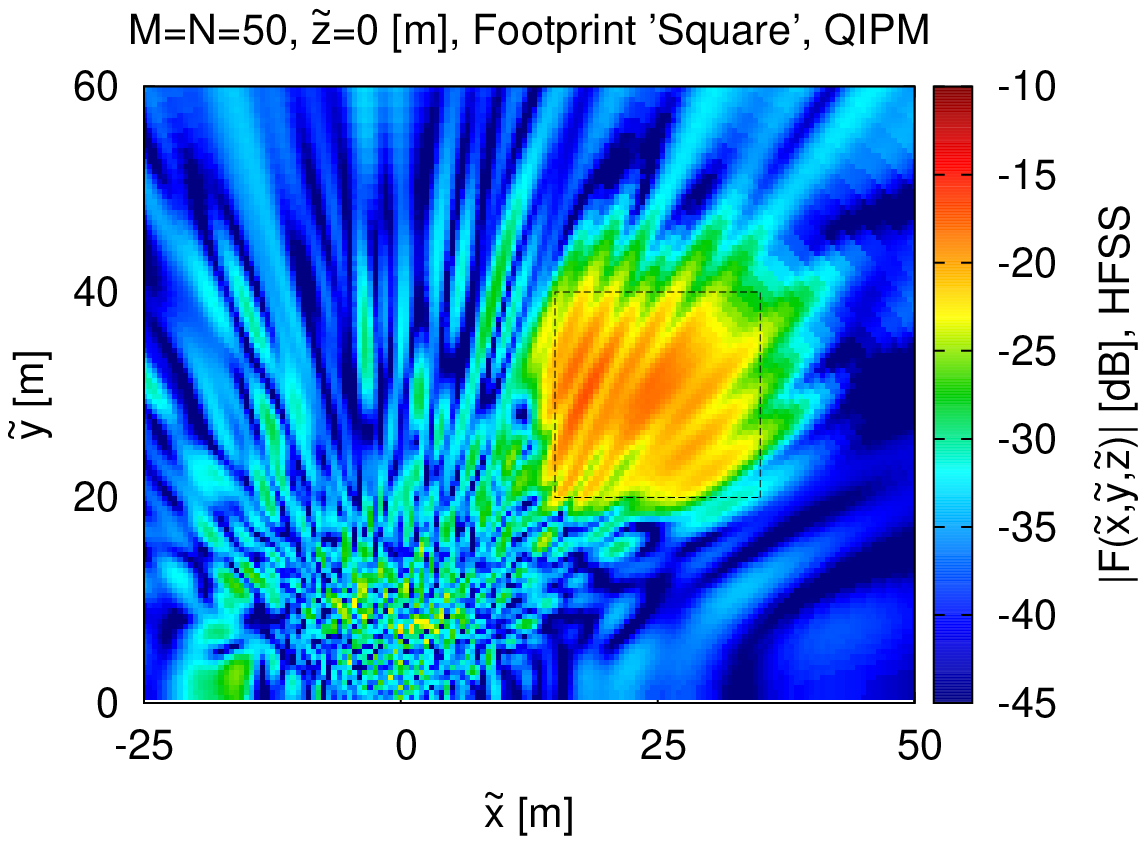}\tabularnewline
(\emph{a})&
(\emph{b})\tabularnewline
\includegraphics[%
  clip,
  width=0.48\columnwidth,
  keepaspectratio]{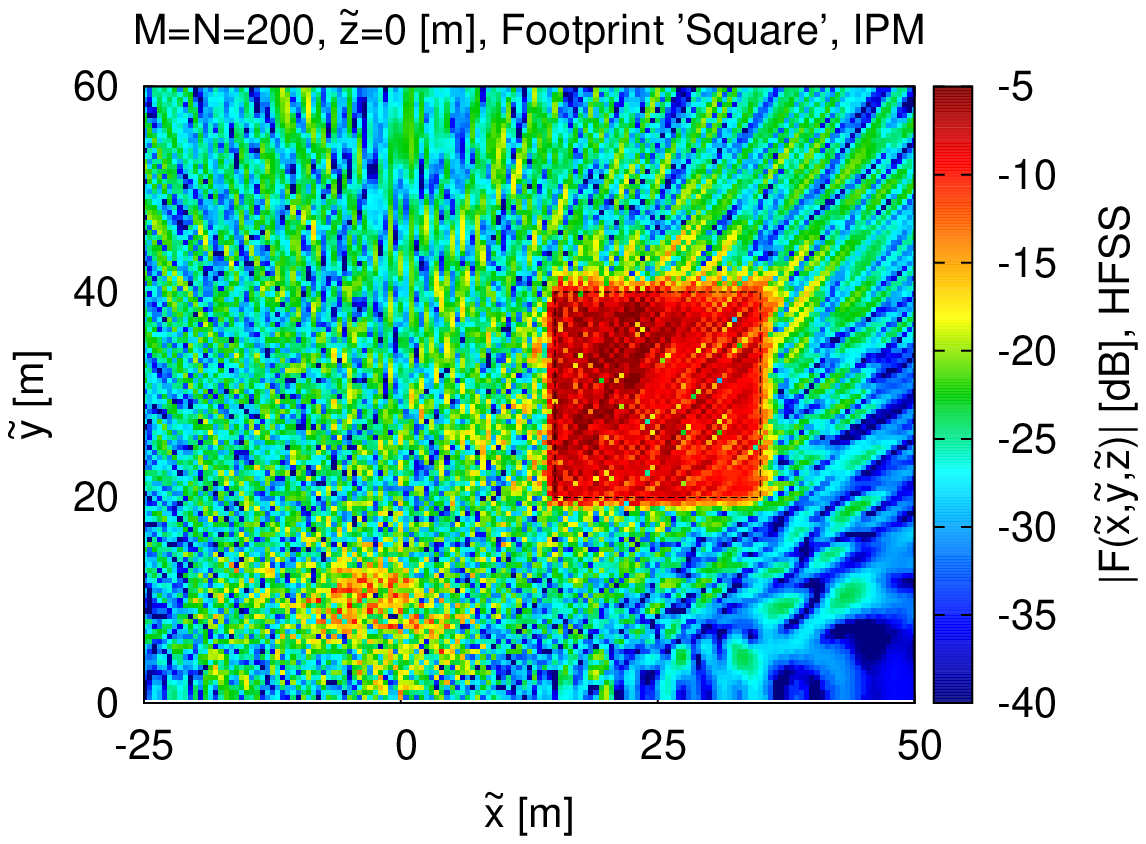}&
\includegraphics[%
  clip,
  width=0.48\columnwidth,
  keepaspectratio]{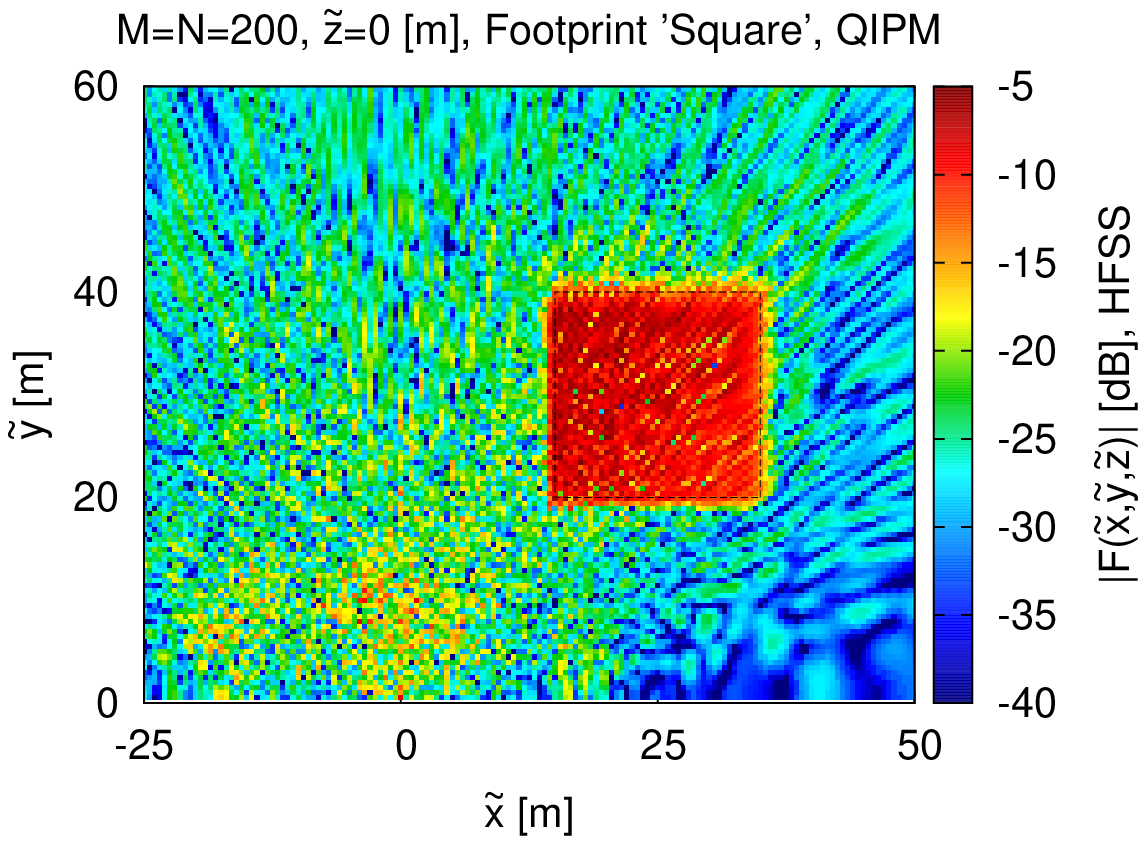}\tabularnewline
(\emph{c})&
(\emph{d})\tabularnewline
\end{tabular}\end{center}

\begin{center}~\vfill\end{center}

\begin{center}\textbf{Fig. 11 - G. Oliveri et} \textbf{\emph{al.}}\textbf{,}
{}``Multi-Scale Single-Bit \emph{RP-EMS} Synthesis for ...''\end{center}

\newpage
\begin{center}~\vfill\end{center}

\begin{center}\begin{tabular}{c}
\includegraphics[%
  clip,
  width=0.70\columnwidth,
  keepaspectratio]{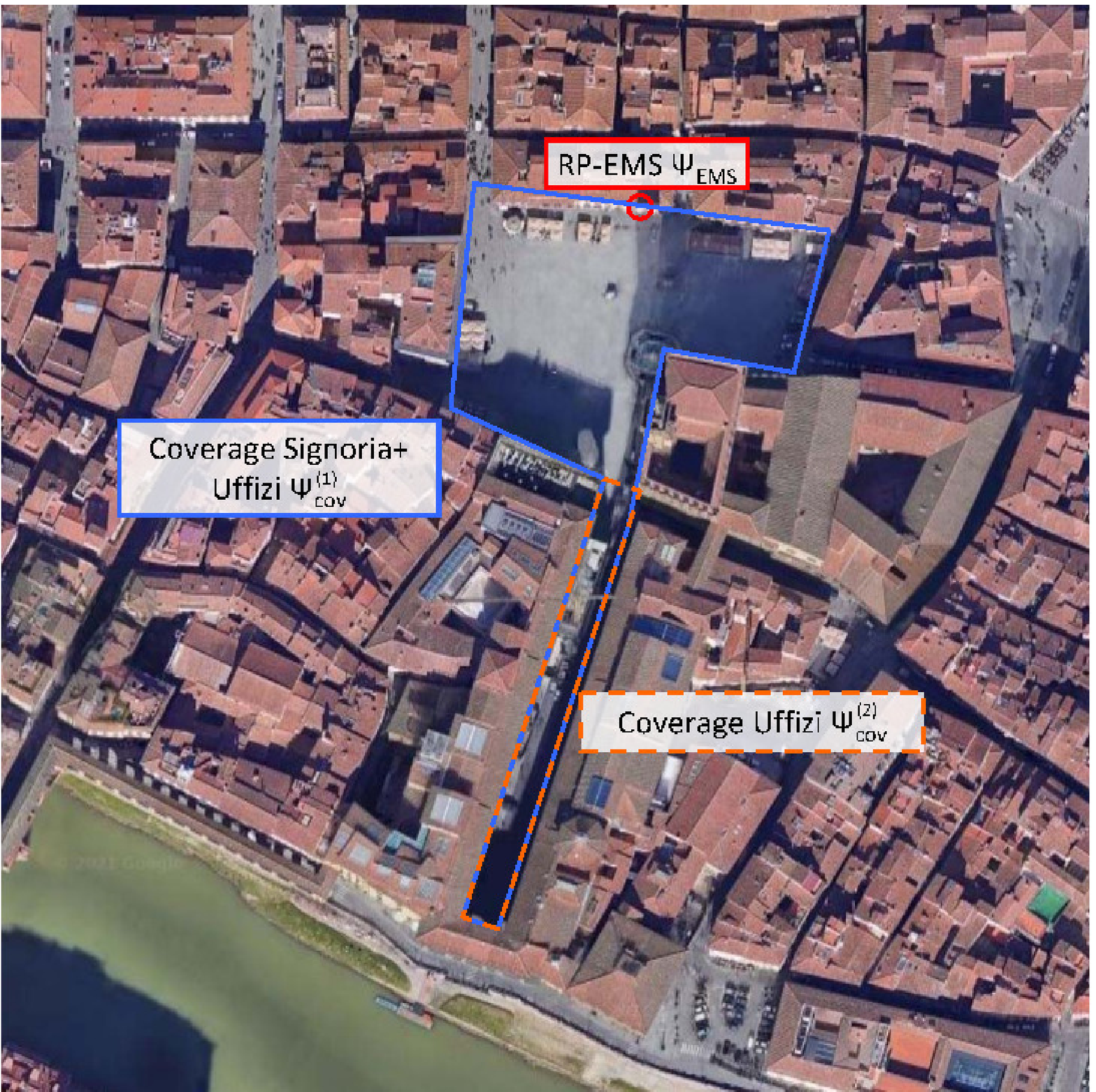}\tabularnewline
(\emph{a})\tabularnewline
\includegraphics[%
  clip,
  width=0.80\columnwidth,
  keepaspectratio]{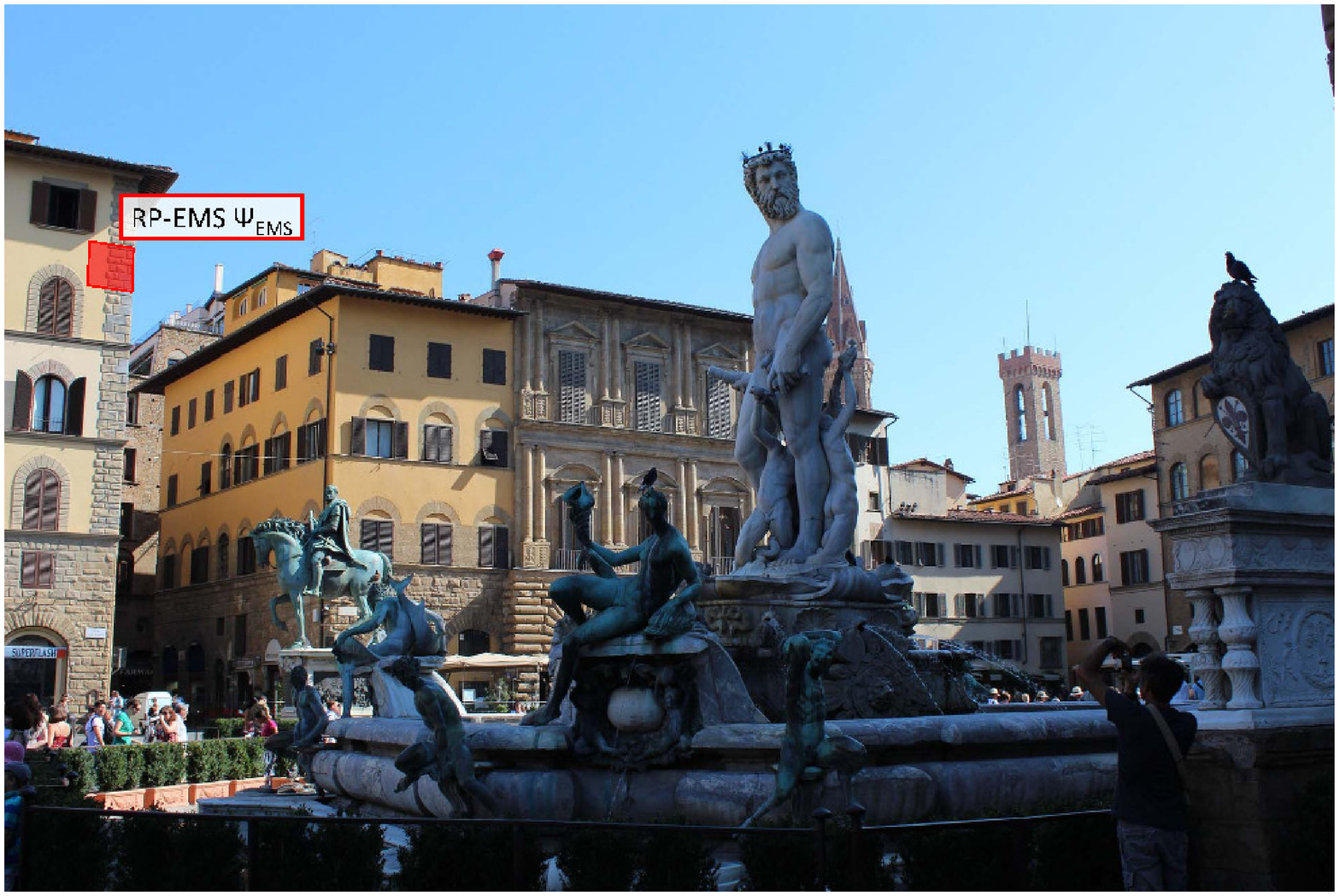}\tabularnewline
(\emph{b})\tabularnewline
\end{tabular}\end{center}

\begin{center}~\vfill\end{center}

\begin{center}\textbf{Fig. 12 - G. Oliveri et} \textbf{\emph{al.}}\textbf{,}
{}``Multi-Scale Single-Bit \emph{RP-EMS} Synthesis for ...''\end{center}

\newpage
\begin{center}~\vfill\end{center}

\begin{center}\begin{tabular}{cc}
\includegraphics[%
  clip,
  width=0.48\columnwidth,
  keepaspectratio]{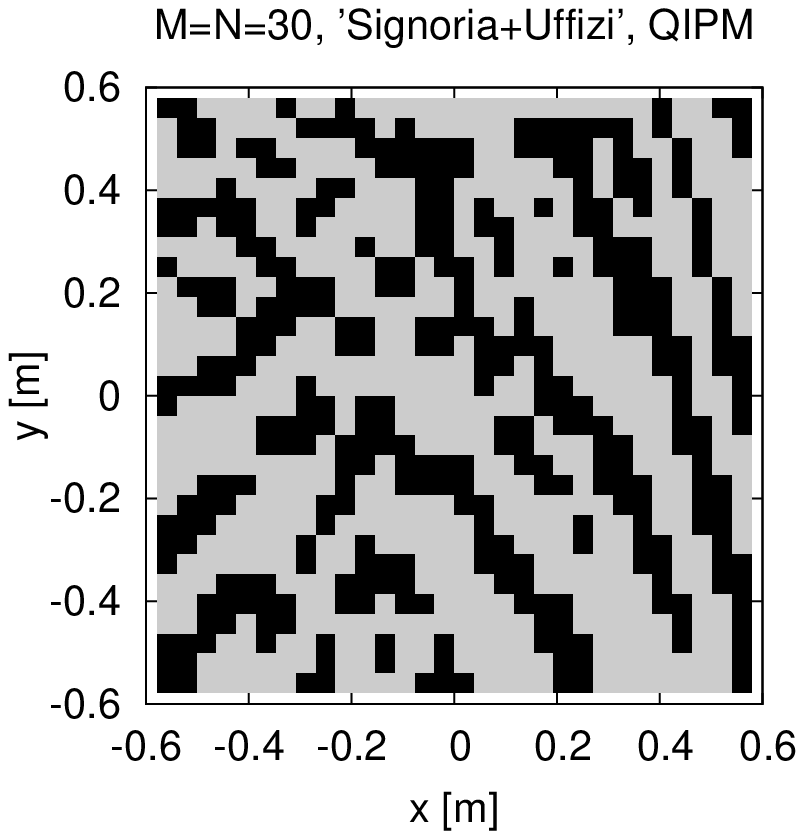}&
\includegraphics[%
  clip,
  width=0.48\columnwidth,
  keepaspectratio]{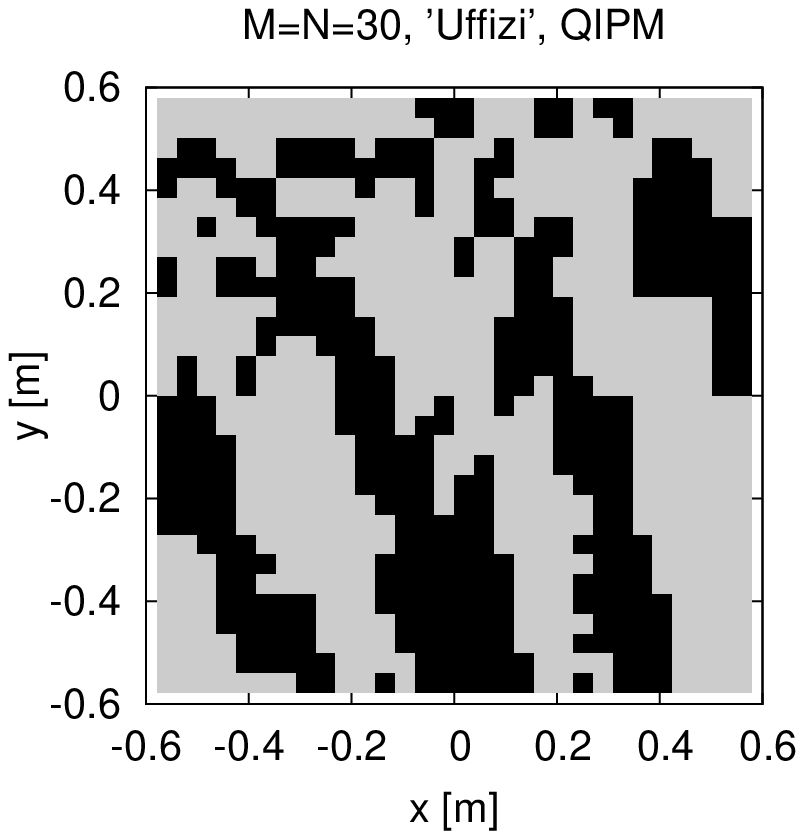}\tabularnewline
(\emph{a})&
(\emph{b})\tabularnewline
\includegraphics[%
  clip,
  width=0.48\columnwidth,
  keepaspectratio]{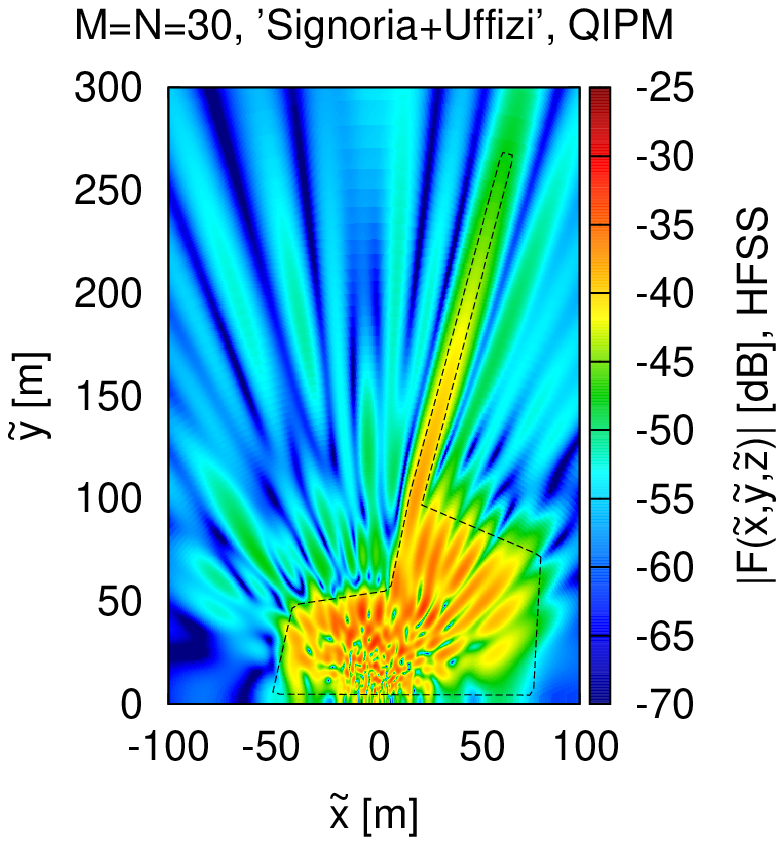}&
\includegraphics[%
  clip,
  width=0.48\columnwidth,
  keepaspectratio]{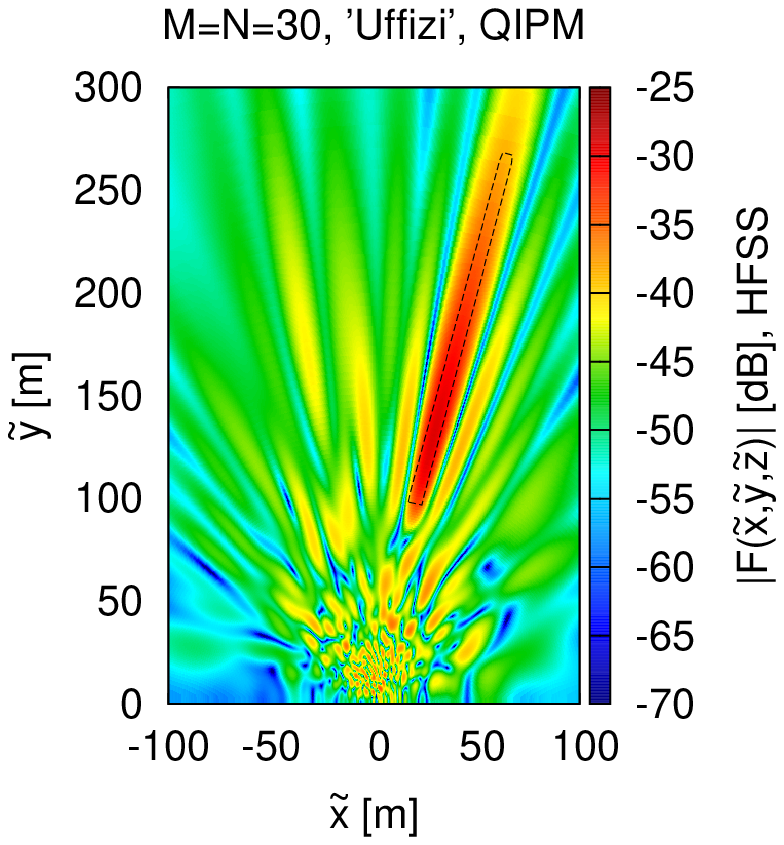}\tabularnewline
(\emph{c})&
(\emph{d})\tabularnewline
\end{tabular}\end{center}

\begin{center}~\vfill\end{center}

\begin{center}\textbf{Fig. 13 - G. Oliveri et} \textbf{\emph{al.}}\textbf{,}
{}``Multi-Scale Single-Bit \emph{RP-EMS} Synthesis for ...''\end{center}

\newpage
\begin{center}~\vfill\end{center}

\begin{center}\begin{tabular}{cc}
\includegraphics[%
  clip,
  width=0.48\columnwidth,
  keepaspectratio]{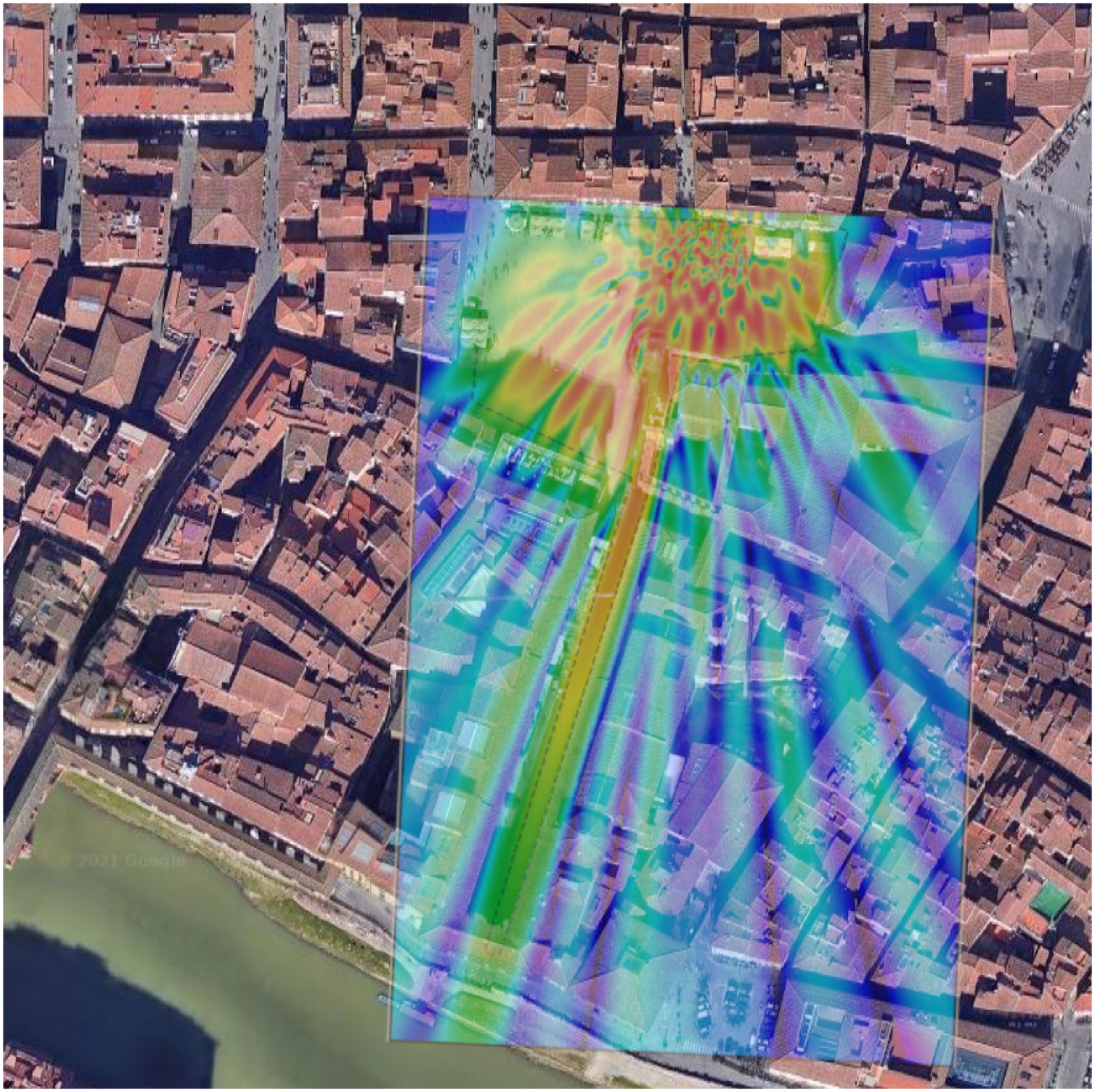}&
\includegraphics[%
  clip,
  width=0.48\columnwidth,
  keepaspectratio]{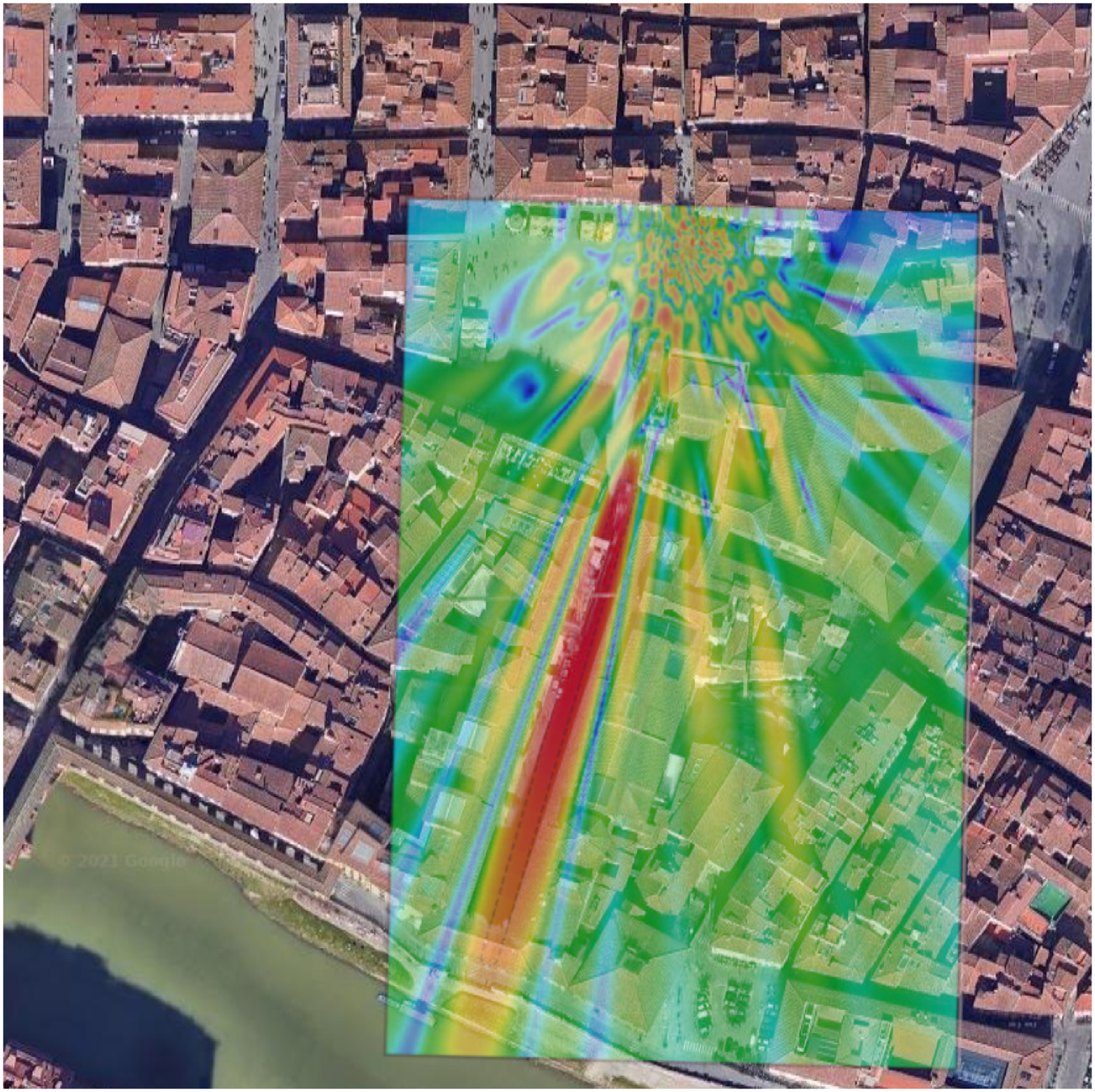}\tabularnewline
(\emph{a})&
(\emph{b})\tabularnewline
\end{tabular}\end{center}

\begin{center}~\vfill\end{center}

\begin{center}\textbf{Fig. 14 - G. Oliveri et} \textbf{\emph{al.}}\textbf{,}
{}``Multi-Scale Single-Bit \emph{RP-EMS} Synthesis for ...''\end{center}

\newpage
\begin{center}~\end{center}

\begin{center}\begin{tabular}{c}
\includegraphics[%
  clip,
  width=0.70\columnwidth,
  keepaspectratio]{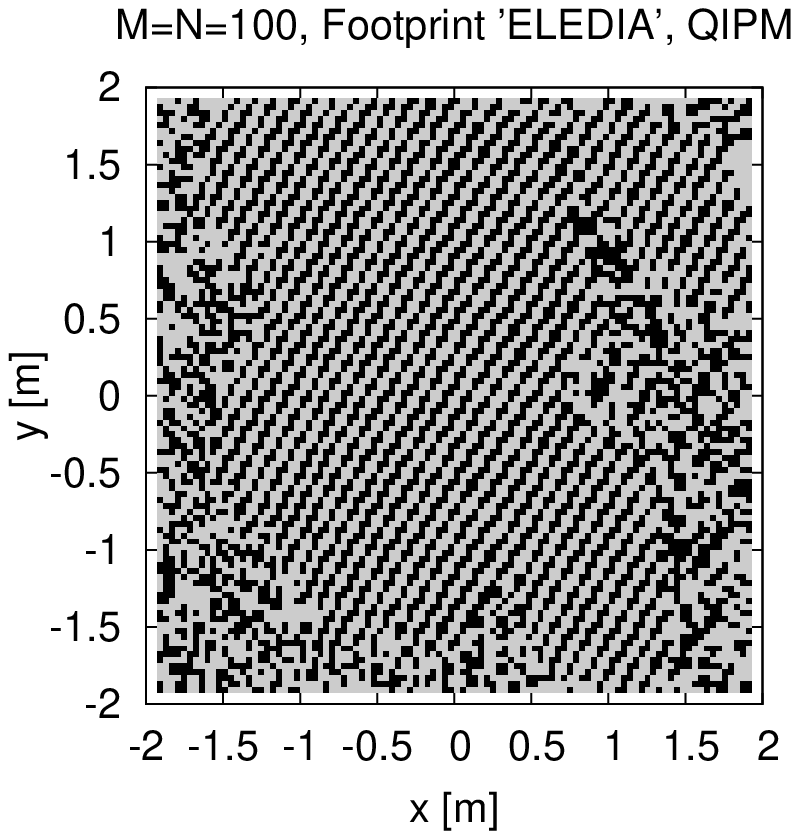}\tabularnewline
(\emph{a})\tabularnewline
\includegraphics[%
  clip,
  width=0.95\columnwidth,
  keepaspectratio]{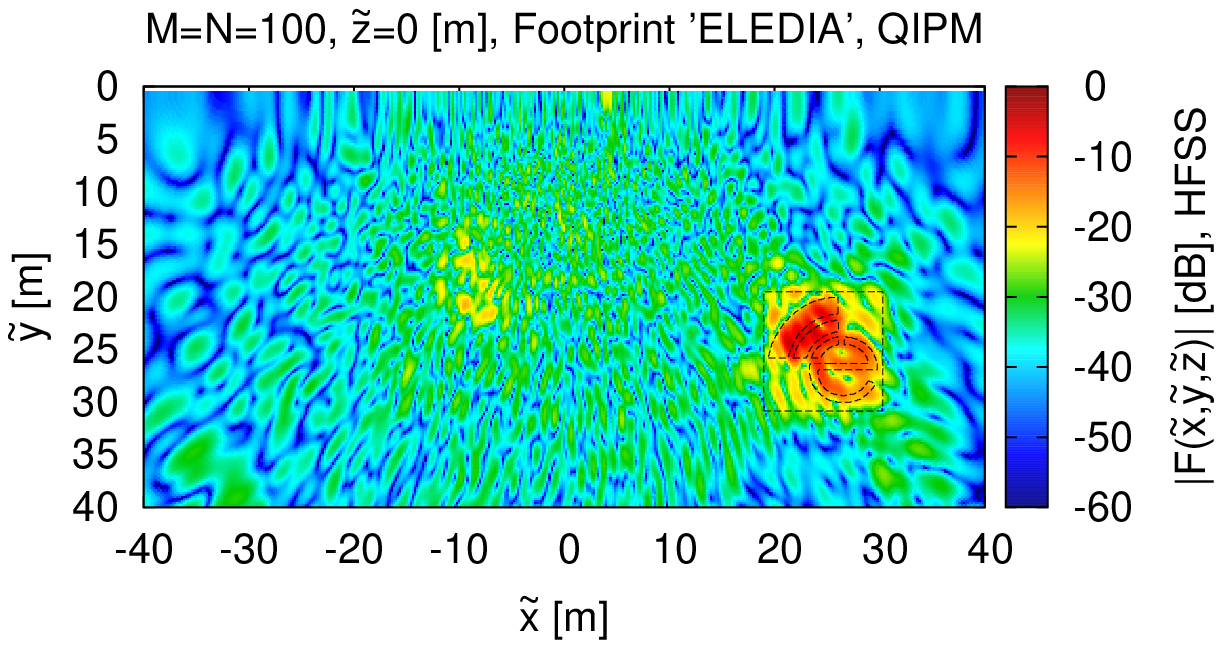}\tabularnewline
(\emph{b})\tabularnewline
\end{tabular}\end{center}

\begin{center}~\vfill\end{center}

\begin{center}\textbf{Fig. 15 - G. Oliveri et} \textbf{\emph{al.}}\textbf{,}
{}``Multi-Scale Single-Bit \emph{RP-EMS} Synthesis for ...''\end{center}

\newpage
\begin{center}~\vfill\end{center}

\begin{center}\begin{tabular}{cccc}
\multicolumn{4}{c}{\includegraphics[%
  clip,
  width=0.70\columnwidth,
  keepaspectratio]{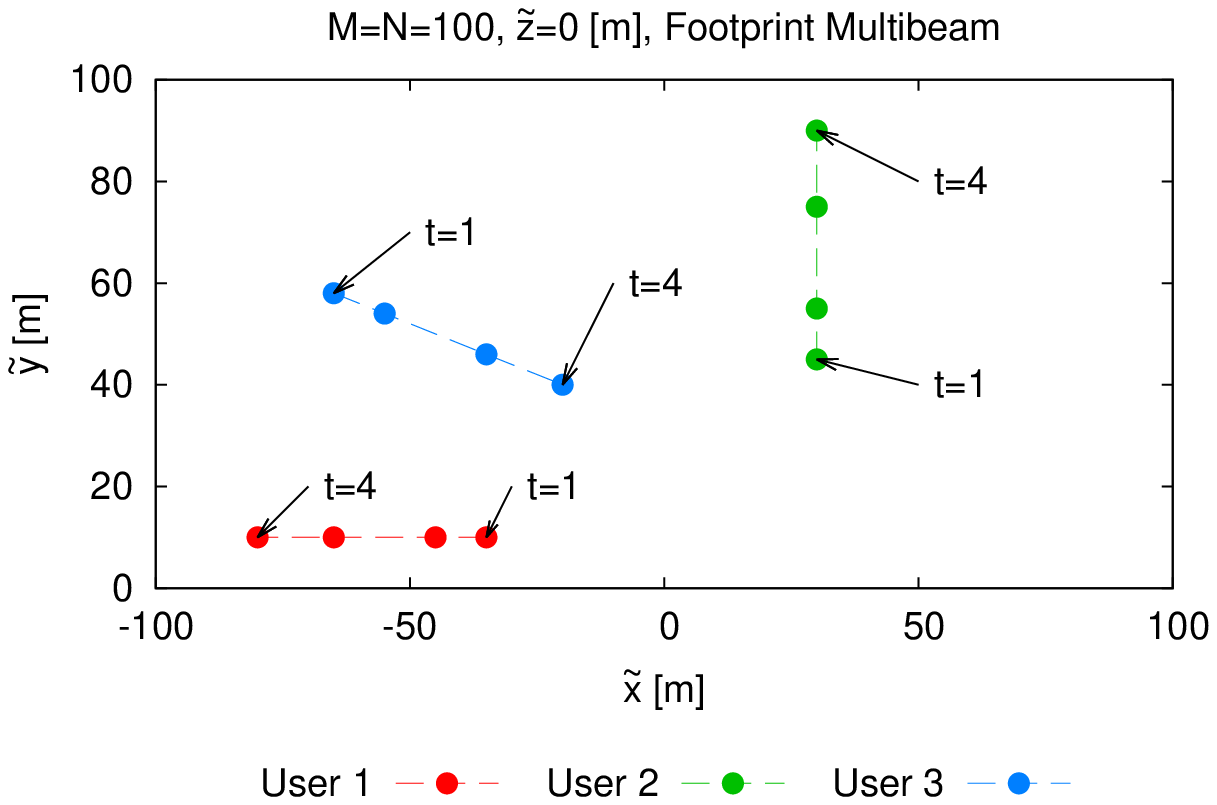}}\tabularnewline
\end{tabular}\end{center}

\begin{center}~\vfill\end{center}

\begin{center}\textbf{Fig. 16 - G. Oliveri et} \textbf{\emph{al.}}\textbf{,}
{}``Multi-Scale Single-Bit \emph{RP-EMS} Synthesis for ...''\end{center}

\newpage
\begin{center}\begin{tabular}{cc}
\includegraphics[%
  clip,
  scale=0.6]{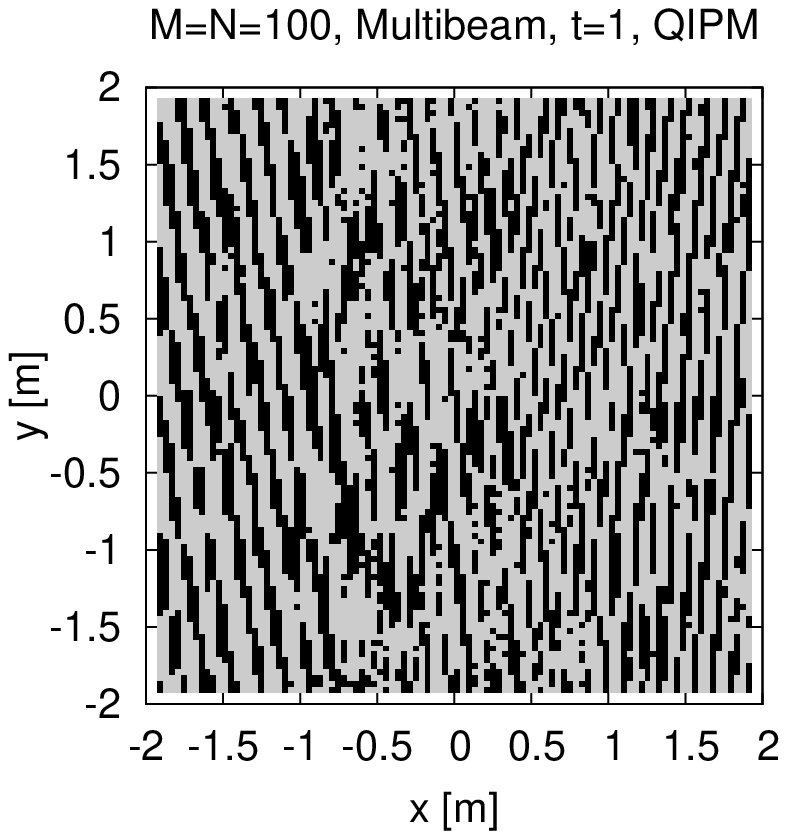}&
\includegraphics[%
  clip,
  scale=0.7]{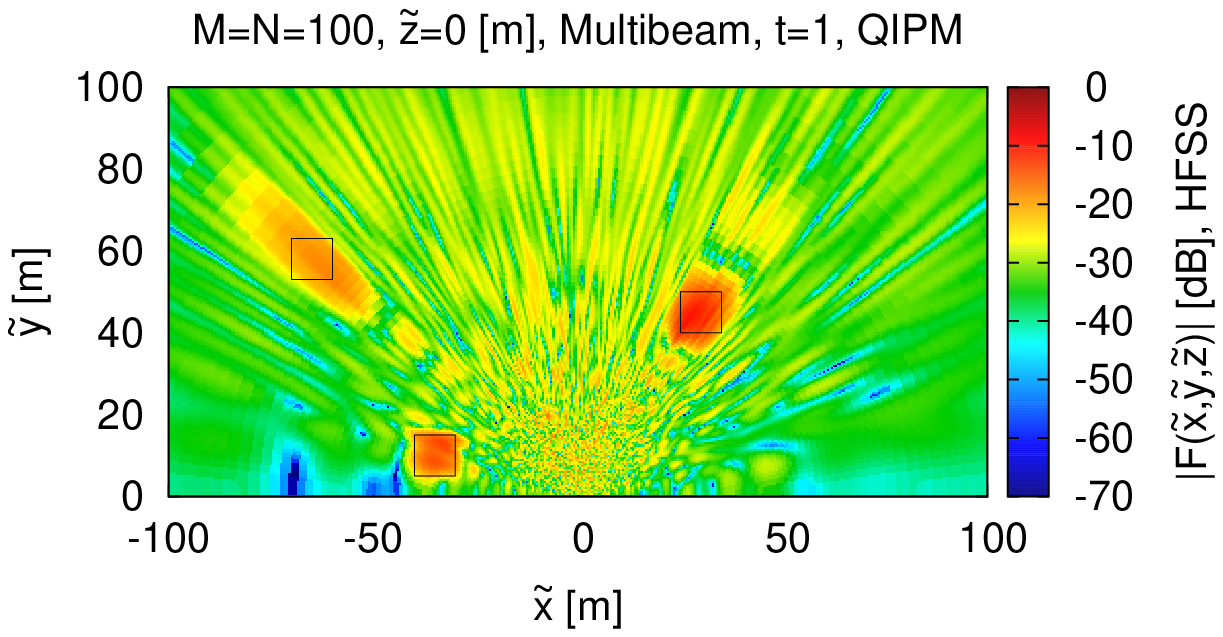}\tabularnewline
(\emph{a})&
(\emph{e})\tabularnewline
\includegraphics[%
  clip,
  scale=0.55]{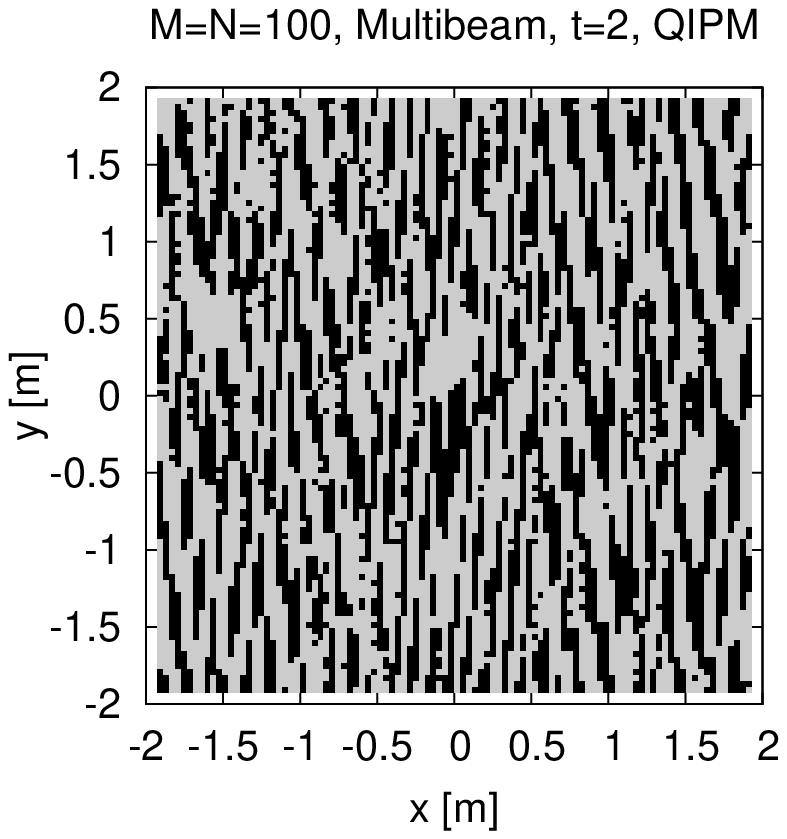}&
\includegraphics[%
  clip,
  scale=0.7]{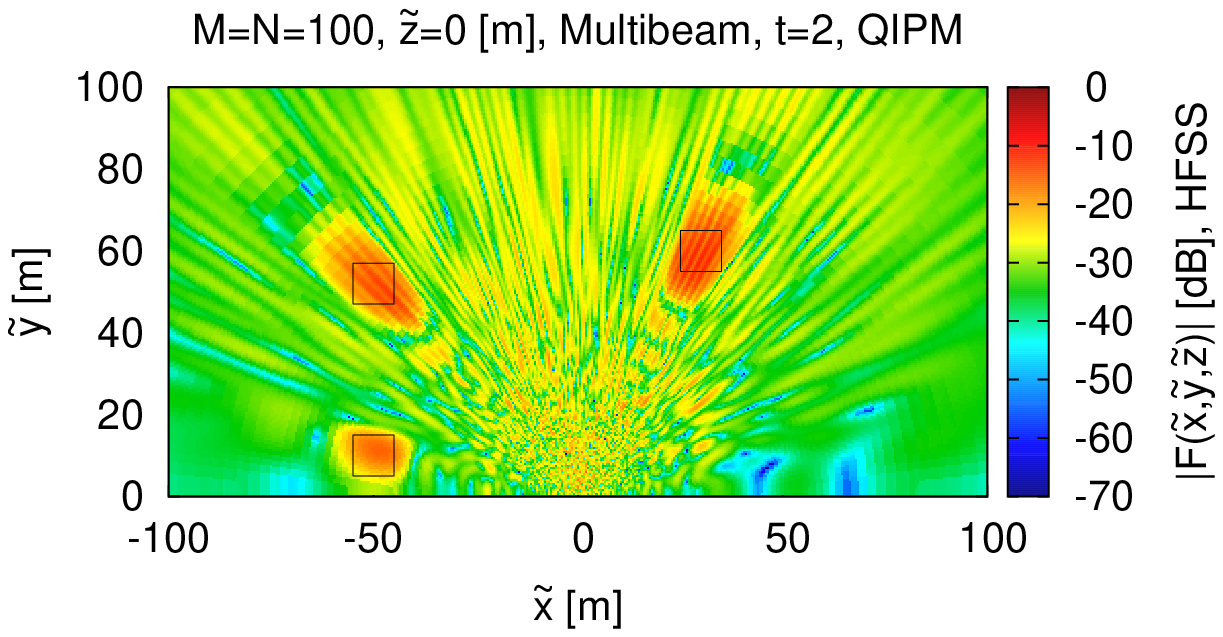}\tabularnewline
(\emph{b})&
(\emph{f})\tabularnewline
\includegraphics[%
  clip,
  scale=0.55]{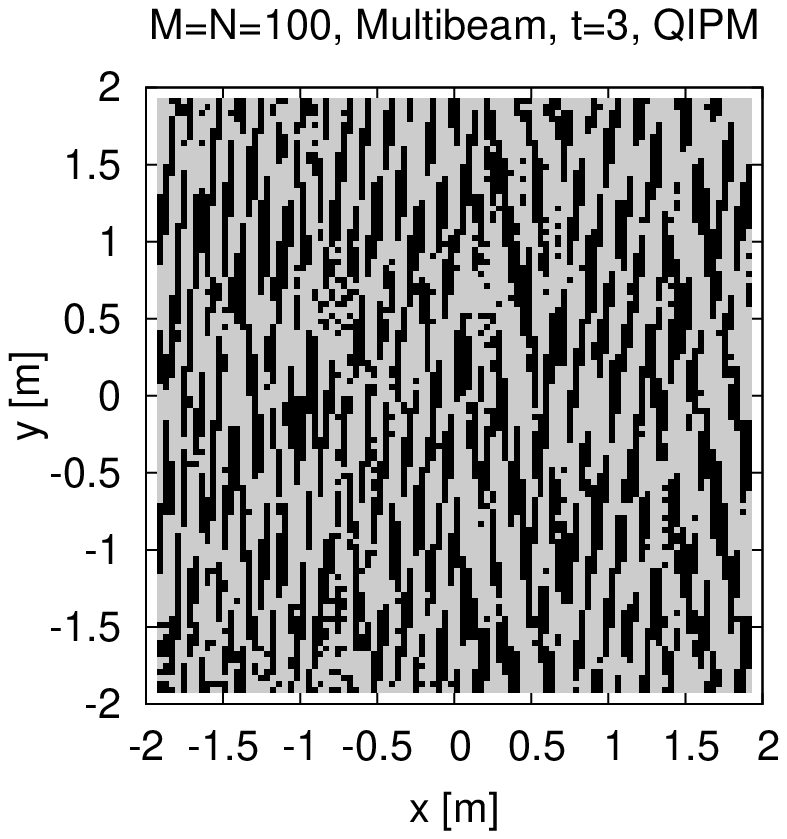}&
\includegraphics[%
  clip,
  scale=0.7]{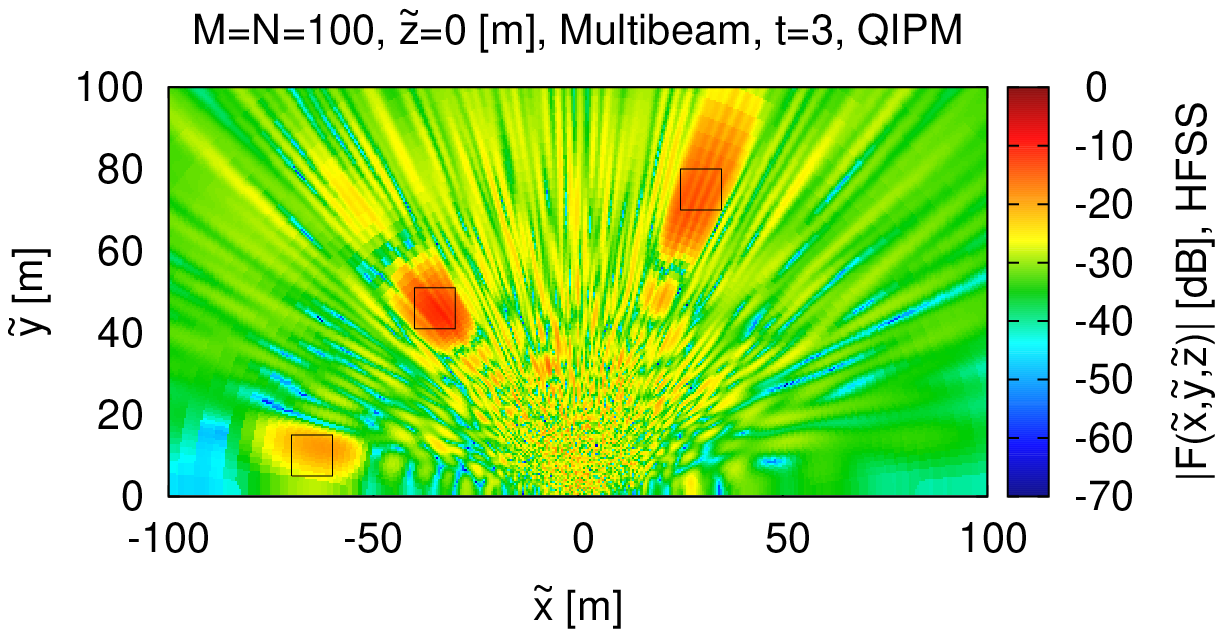}\tabularnewline
(\emph{c})&
(\emph{g})\tabularnewline
\includegraphics[%
  clip,
  scale=0.55]{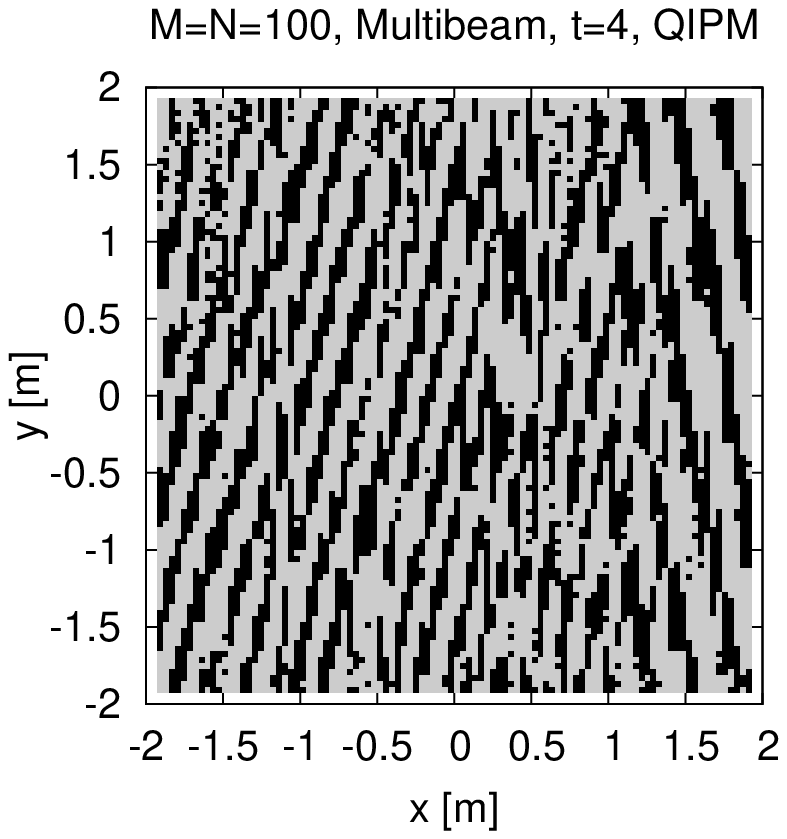}&
\includegraphics[%
  clip,
  scale=0.7]{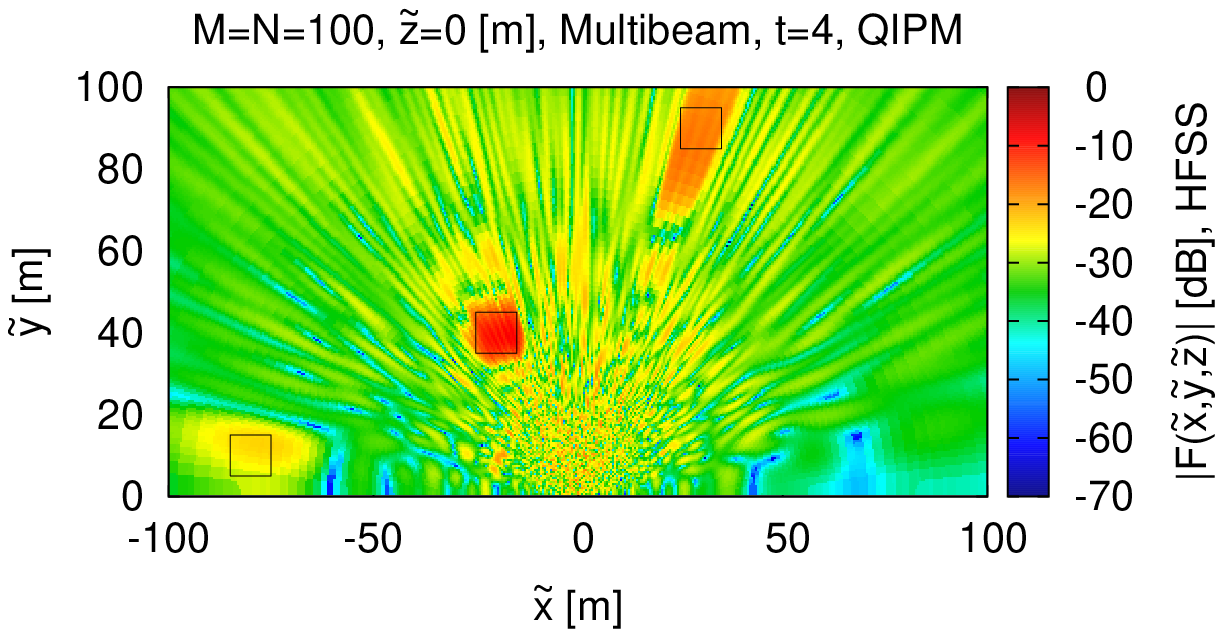}\tabularnewline
(\emph{d})&
(\emph{h})\tabularnewline
\end{tabular}\end{center}

\begin{center}\textbf{Fig. 17 - G. Oliveri et} \textbf{\emph{al.}}\textbf{,}
{}``Multi-Scale Single-Bit \emph{RP-EMS} Synthesis for ...''\end{center}

\newpage
\begin{center}~\vfill\end{center}

\begin{center}\begin{tabular}{|c|c|}
\hline 
\emph{Parameter}&
\emph{Value~}{[}\emph{m}{]}\tabularnewline
\hline
\hline 
$g_{1}^{opt}=g_{2}^{opt}$&
$3.854\times10^{-2}$\tabularnewline
\hline 
$g_{3}^{opt}=g_{4}^{opt}$&
$2.191\times10^{-2}$\tabularnewline
\hline 
$g_{5}^{opt}=g_{11}^{opt}$&
$1.616\times10^{-4}$\tabularnewline
\hline 
$g_{6}^{opt}=g_{12}^{opt}$&
$2.488\times10^{-3}$\tabularnewline
\hline 
$g_{7}^{opt}=g_{13}^{opt}$&
$3.300\times10^{-4}$\tabularnewline
\hline 
$g_{8}^{opt}=g_{14}^{opt}$&
$1.777\times10^{-3}$\tabularnewline
\hline 
$g_{9}^{opt}=g_{15}^{opt}$&
$2.000\times10^{-4}$\tabularnewline
\hline 
$g_{10}^{opt}=g_{16}^{opt}$&
$6.000\times10^{-4}$\tabularnewline
\hline
\end{tabular}\end{center}

\begin{center}\vfill\end{center}

\begin{center}\textbf{Table I - G. Oliveri et} \textbf{\emph{al.}}\textbf{,}
{}``Multi-Scale Single-Bit \emph{RP-EMS} Synthesis for ...''\end{center}
\end{document}